\documentclass[epj-mine,nopacs,numbook]{svjour-mine}
\usepackage{amsmath,amssymb,bbm,graphicx}

\renewcommand{\texttt}{\url}
\usepackage[dvips]{hyperref}
\hyperbaseurl{http://arxiv.org/pdf/}

\newcommand{\emdash}{---}
\newcommand{\tmop}[1]{\operatorname{#1}}
\newcommand{\tmfloatcontents}{}
\newlength{\tmfloatwidth}
\newcommand{\tmfloat}[5]{\bigskip \bigskip\begin{center}
  \renewcommand{\tmfloatcontents}{#4}
  \setlength{\tmfloatwidth}{\widthof{\tmfloatcontents}+1in}
  \ifthenelse{\equal{#2}{small}}
    {\ifthenelse{\lengthtest{\tmfloatwidth > \linewidth}}
      {\setlength{\tmfloatwidth}{\linewidth}}{}}
    {\setlength{\tmfloatwidth}{\linewidth}}  \begin{minipage}[#1]{\tmfloatwidth}
    \begin{center}
      \tmfloatcontents
      \captionof{#3}{#5}
    \end{center}
  \end{minipage}
\end{center}
\bigskip}
\newcommand{\bignone}{}
\newcommand{\mathi}{\mathrm{i}}
\newcommand{\tmem}[1]{{\em #1\/}}
\newcommand{\mathd}{\mathrm{d}}
\newcommand{\mathe}{\mathrm{e}}
\newcommand{\mathpi}{\pi}
\newcommand{\nonesep}{}
\catcode`\à=\active \defà{\`a} \catcode`\À=\active \defÀ{\`A}
\catcode`\á=\active \defá{\'a} \catcode`\Á=\active \defÁ{\'A}
\catcode`\ä=\active \defä{\"a} \catcode`\Ä=\active \defÄ{\"A}
\catcode`\â=\active \defâ{\^a} \catcode`\Â=\active \defÂ{\^A}
\catcode`\å=\active \defå{{\aa}} \catcode`\Å=\active \defÅ{{\AA}}
\catcode`\ç=\active \defç{\c{c}} \catcode`\Ç=\active \defÇ{\c{C}}
\catcode`\è=\active \defè{\`e} \catcode`\È=\active \defÈ{\`E}
\catcode`\é=\active \defé{\'e} \catcode`\É=\active \defÉ{\'E}
\catcode`\ë=\active \defë{\"e} \catcode`\Ë=\active \defË{\"E}
\catcode`\ê=\active \defê{\^e} \catcode`\Ê=\active \defÊ{\^E}
\catcode`\ì=\active \defì{\`{\i}} \catcode`\Ì=\active \defÌ{\`{\I}}
\catcode`\í=\active \defí{\'{\i}} \catcode`\Í=\active \defÍ{\'{\I}}
\catcode`\ï=\active \defï{\"{\i}} \catcode`\Ï=\active \defÏ{\"{\I}}
\catcode`\î=\active \defî{\^{\i}} \catcode`\Î=\active \defÎ{\^{\I}}
\catcode`\ò=\active \defò{\`o} \catcode`\Ò=\active \defÒ{\`O}
\catcode`\ó=\active \defó{\'o} \catcode`\Ó=\active \defÓ{\'O}
\catcode`\ö=\active \defö{\"o} \catcode`\Ö=\active \defÖ{\"O}
\catcode`\ô=\active \defô{\^o} \catcode`\Ô=\active \defÔ{\^O}
\catcode`\ù=\active \defù{\`u} \catcode`\Ù=\active \defÙ{\`U}
\catcode`\ú=\active \defú{\'u} \catcode`\Ú=\active \defÚ{\'U}
\catcode`\ü=\active \defü{\"u} \catcode`\Ü=\active \defÜ{\"U}
\catcode`\û=\active \defû{\^u} \catcode`\Û=\active \defÛ{\^U}
\catcode`\ý=\active \defý{\'y} \catcode`\Ý=\active \defÝ{\'Y}
\catcode`\ÿ=\active \defÿ{\"y} \catcode`\˜=\active \def˜{\"Y}
\catcode`\½=\active \def½{!`}
\catcode`\¾=\active \def¾{?`}
\catcode`\ß=\active \defß{{\ss}}

\begin{document}

\title{The role of spurions in
  Higgs-less electroweak effective theories}
\author{Johannes Hirn\inst{1,2,}\thanks{\email{johannes.hirn@durham.ac.uk}} \and Jan Stern\inst{2,}\thanks{\email{stern@ipno.in2p3.fr}}
}                     
%
%
\institute{Groupe Physique Théorique\thanks{\em{Unit\'e mixte de recherche 8608.}}, IPN Orsay, Université
  Paris-Sud XI, 91406 Orsay, France \and Institute for Particle Physics Phenomenology, University of  Durham, Durham DH1 3LE, UK}
%
%
\abstract{Inspired by recent developments of moose models, we reconsider low-energy
  effective theories of Goldstone bosons, gauge fields and chiral fermions
  applied to low-energy QCD and to Higgs-less electroweak symmetry breaking.
  Couplings and the corresponding reduction of symmetry are introduced via
  constraints enforced by a set of non-propagating covariantly constant
  spurion fields. Relics of the latter are used as small expansion parameters
  conjointly with the usual low-energy expansion. Certain couplings can only
  appear at higher orders of the spurion expansion and, consequently, they
  become naturally suppressed independently of the idea of dimensional
  deconstruction.
  At leading order this leads to a set of generalized Weinberg sum rules and
  to the suppression of non-standard couplings to fermions in Higgs-less EWSB
  models with the minimal particle content. Within the latter, higher spurion 
  terms allow for a fermion mass matrix with the standard CKM structure and
  $C P$ violation. In addition, Majorana masses for neutrinos are possible.
  Examples of non-minimal models are briefly mentioned.
\PACS{
      {PACS-key}{discribing text of that key}   \and
      {PACS-key}{discribing text of that key}
     } 
} 
\maketitle
\section{Introduction} \label{sec:intro} \label{sub:motiv}

In a low-energy effective theory~(LEET)~{\cite{Weinberg:1979kz}}, the
requirement of naturalness~{\cite{'tHooft:1979bh}} plays a central role in
defining a systematic low-energy expansion that is finite order by order
despite the lack of renormalizability and a bad high-energy behavior. LEET
operates with light degrees of freedom, which become massless in a particular
limit, due to chiral and/or gauge symmetries. The lagrangian is constructed
and renormalized order-by-order in a momentum expansion. Naturalness then
amounts to the requirement that, at each order, the lagrangian contains all
terms allowed by the symmetries and by the power counting. The simplest
example is Chiral Perturbation Theory~($\chi$PT)~{\cite{Gasser:1984yg,Gasser:1985gg}} {\emdash}the low-energy effective theory of QCD. It merely
operates with Goldstone bosons of spontaneously broken chiral symmetry. Since
in this case, the high-energy completion of the effective theory~(i.e. QCD) is
known, the~$\mathcal{O} \left( p^4 \right)$ low-energy constants~(LECs) can be
put under partial phenomenological control.

In this article, we are particularly concerned with more complicated effective
theories which, in addition to Goldstone bosons, also contain gauge fields. For
some time this type of LEET was hoped to describe electroweak symmetry
breaking~(EWSB) resulting from the dynamical Higgs mechanism~{\cite{Dobado:1991zh,Espriu:1992vm}}. Especially interesting was the
possibility of EWSB with no scalar particles remaining in the spectrum. The
question whether a natural and phenomenologically viable LEET of EWSB without
scalar relics exists is still of interest today as it would represent an
alternative to the standard interpretation of electroweak precision tests as
constraints on the Higgs mass~{\cite{Group:2002mc}}. Unfortunately, this type
of effective theories lacks predictivity at~$\mathcal{O} \left( p^4 \right)$
order~(i.e. one loop), since its high-energy completion is not known: the
values of the renormalized~$\mathcal{O} \left( p^4 \right)$ LECs can so far
only be discussed within particular models such as ``rescaled QCD''
{\cite{Peskin:1992sw}}, or the gauged linear sigma model in the heavy-mass
limit~{\cite{Appelquist:1980vg,Longhitano:1980iz,Longhitano:1981tm,Nyffeler:1996mb}}. Beyond those models, it seems premature to conclude that a
generic Higgs-less LEET of EWSB is at variance with precision electroweak
tests. On the other hand, already at leading~$\mathcal{O} \left( p^2 \right)$
order, not all couplings one may construct in a generic Higgs-less effective
theory based on the symmetry arguments alone are actually observed.

i) Tree-level~$\mathcal{O} \left( p^2 \right)$ contributions to the
$S$ parameter are in principle possible.

ii) There are non-standard couplings to fermions, which would spoil
universality of the left-handed couplings and introduce right-handed couplings
of the~$W^{\pm}$. This is unacceptable at leading order and the problem we
address is that of suppressing those non-standard couplings naturally.

iii) Last but not least, let us mention the long-standing problem of fermion
masses and flavor symmetry breaking. Indeed, within Higgs-less effective
theories it has always been difficult to have, at leading order,
mass-splittings within a doublet of the same order of magnitude as the masses
themselves.

All these problems suggest that something else should be added to the momentum
expansion based on the low-energy symmetry~$\tmop{SU}\left(2 \right) \times
\mathrm{U} \left( 1 \right)_Y$ in order to construct natural Higgs-less
effective theories of EWSB.

It is interesting to note that a similar problem in principle arises if the
low-energy effective theory of QCD is extended beyond the pure~$\chi$PT
framework towards higher energies, incorporating vector and axial vector
states. If the latter are 
included among protected light states of a LEET, they should be treated 
as weakly coupled gauge particles (although such a scenario need not
be realistic for finite-$N_c$  QCD, we are going to use it as a simple 
theoretical laboratory~\cite{Georgi:1990xy}).  If in the corresponding LEET all terms allowed by the chiral symmetry were kept
{\cite{Meissner:1988ge,Ecker:1989yg}}, the Weinberg sum rules
(WSRs)~{\cite{Weinberg:1967kj}} would be lost. In QCD, the latter do not follow from chiral
symmetry alone, but hinge strongly on short-distance properties of the theory.
Within LEET this leads once more to the problem mentioned in connection with
EWSB: how to naturally suppress unwanted terms in order to recover the two
Weinberg sum rules and obtain finite radiative masses for pseudo-Goldstone
bosons~(PGBs).

Our starting point is therefore a recent discussion of the two Weinberg sum
rules~{\cite{Son:2003et}}, based on moose models~{\cite{Georgi:1986dw}}. The
model considered consists in a chain of Goldstone bosons of spontaneously
broken chiral symmetries coupled in a particular way to gauge fields. Not all
couplings allowed by the symmetries and low-energy power counting are
admitted, leading to a better high-energy behavior than one might expect in
effective theories exclusively based on the low-energy symmetries. The
literature does not relate these properties to a requirement of symmetry.
Indeed, the standard line of thinking has been that of dimensional
deconstruction~{\cite{Arkani-Hamed:2001ca,Cheng:2001vd}}, in which case
unwanted terms in the lagrangian are omitted on a physical basis: the
requirement of locality along the fifth~(discretized) dimension. Only interactions between nearest neighbors along the moose are then allowed.
What happens at higher orders is then unclear: such terms will be generated
and one can question whether it is a licit procedure to omit them altogether
in the first place.

What we would like to demonstrate here is the feasibility and the usefulness
of an alternative bottom-up approach, without reference to a five-dimensional
theory. Instead we will focus on an approach which is closely linked with the
naturalness hypothesis formulated by 't Hooft~{\cite{'tHooft:1979bh}}, that
is, relying on the symmetries of the theory. Our formulation will make
extensive use of non-propagating fields {\emdash}called spurions{\emdash}
within the LEET~{\cite{Weinberg:1979kz,Gasser:1984yg}}, in order to keep
track of the reduction of symmetry when the gauge couplings are introduced.
These spurions and the constraints applied on them serve the purpose of
introducing couplings by restricting the configuration space of gauge
connections, while at the same time allowing the use of a lagrangian which is
invariant under the full symmetry. Spurions have been employed in the past in
various contexts: explicit chiral symmetry breaking due to quark masses~{\cite{Weinberg:1979kz,Gasser:1984yg}}, introduction of electromagnetic
couplings in~$\chi$PT~{\cite{Ecker:1989te,Urech:1995hd}}, radiative masses of
PGBs~{\cite{Moussallam:1997xx}}, and also linear moose models~{\cite{Arkani-Hamed:2001nc}}. In~{\cite{Arkani-Hamed:2001nc}}, spurions were used to 
formally count the occurrences of coupling constants in connection with 
radiative corrections to PGBs masses once the unwanted couplings have 
been dismissed. At the end, the spurions of reference~{\cite{Arkani-Hamed:2001nc}} are set to one.
In this paper we go one step further: we consider covariantly constant 
spurions as small expansion parameters which are a genuine part of the
effective theory, allowing for a natural suppression of unwanted couplings
between Goldstone bosons and gauge fields. The LEET will be constructed 
as a simultaneous expansion in powers of momenta, (gauge) couplings and 
spurions~{\cite{Weinberg:1979kz,Gasser:1984yg,Gasser:1985gg,Urech:1995hd}}: unwanted couplings will be suppressed since they 
only appear at higher orders of the spurion expansion. The spurion formalism developed in this paper represents a
general device to couple~$\tmop{SU}\left( 2 \right) \times \tmop{SU} \left( 2 \right)$
Goldstone bosons and~$\tmop{SU} \left( 2 \right)$ gauge fields. As such, it
applies both to effective theories of QCD and of Higgs-less EWSB. This is the
main reason why these apparently distinct subjects are discussed jointly
within the same paper. The paper is organized as follows.

In section~\ref{sec:spurions-and-mooses} we introduce real
covariantly-constant spurions in the general linear moose model based on
$\tmop{SU} \left( 2 \right)$ groups. We then perform appropriate field
redefinitions to study the spectrum of the theory, which consists in a
multiplet of Goldstone bosons and a tower of massive vectors.

In section~\ref{sec:generalized-WSRs}, we consider the left-right two-point
correlator and show that, at leading order in the spurion expansion, and in
the tree-level approximation~(cf. the large-$N_c$ limit of QCD), it
automatically satisfies~$K$ generalized Weinberg sum rules, where~$K$ is the
length of the moose. Higher spurion terms introduce corrections to these WSRs.
We give the expression for~$L_{10}$, which is related to the~$S$ parameter of
the electroweak sector.

In section~\ref{sub:EWSB-minimal-case}, we consider in detail the Higgs-less
effective EWSB theory with a minimal particle content: electroweak bosons
and fermions. There are no scalar particles below the scale~$4 \mathpi f$~($f
\simeq 250$~GeV) where the effective theory breaks down. The minimal set of
spurions needed to select the reduced symmetry~$\tmop{SU} \left( 2 \right)
\times \mathrm{U} \left( 1 \right)_Y$ out of a larger~$\tmop{SU} \left( 2
\right)^4 \times \mathrm{U} \left( 1 \right)_{B-L}$ natural symmetry is discussed.
It is shown how a complex spurion selects the correct~$\mathrm{U} \left( 1
\right)$ subgroup and describes weak isospin breaking effects. The suppression
of all non-standard couplings to fermions is demonstrated at the leading order
of the spurion expansion. It is further shown that spurions allow the
introduction of fermion mass-matrices with a general CKM structure and $C P$
violation.

In section~\ref{sec:EWSB-extended-cases} we describe two non-minimal models of
EWSB based on extension of the moose of section~\ref{sub:EWSB-minimal-case}.
The first of these two models contains~$W'$ and~$Z'$ excited vectors. Our
purpose is to exhibit what the first corrections to the~$S, T, U$ parameters
might be. Our last model does not contain excited vectors in the low-energy
sector, but rather a triplet of PGBs. We illustrate the role of the Weinberg
sum rules discussed in section~\ref{sec:generalized-WSRs}, in relation with the
radiative masses of these PGBs. This last model provides an example of light
scalars which do not play the same role as the standard Higgs boson. Again, we
introduce fermions in the model and discuss the consequences of the spurion
power counting.

We summarize our findings in section~\ref{sec:concl}.

\section{Spurions and mooses} \label{sec:spurions-and-mooses}

In this section, the open linear moose is studied at lowest order in the
momentum expansion and lowest order in the spurion power counting:
interactions between Goldstone bosons and gauge fields are introduced via
constraints imposed on spurions embodying the principle of naturalness. Indeed, we start with a theory with Goldstone
multiplets and gauge fields. When the parameters are sent to appropriate
limits, the separate theories do not interact anymore. Consequently, it is
sensible to consider an expansion where the parameters are close to these
limits: this is both a well-defined expansion from the practical point of view
and a meaningful one in the sense that the symmetry is increased when the
parameters are set to be equal to the aforementioned limits. This in turn
makes it logically plausible that an underlying theory produces this set of
values for the parameters without fine-tuning because the symmetry preserves
this situation from obtaining large corrections. We want to build a
perturbative expansion around the limit where the Goldstone bosons do not
interact with the gauge fields: the moose is then disconnected.

In order to disentangle the disconnected limit from the case when the gauge
coupling constants are taken to zero, we are compelled to introduce a set of
additional parameters, which can be taken to zero independently. These
parameters will in fact derive from spurion fields, which  render the theory invariant
under a larger symmetry group: the group of transformations where the chiral
transformations on the Goldstone fields are independent of the gauge
transformations. From this we may deduce how the spurions will have to
transform. We find that the connections corresponding to the chiral
transformations on the Goldstone fields and the gauge connections become
identified, and that the spurions reduce to constants, which will become our
small expansion parameters.

In this section, we focus on spurions satisfying, in addition to the
constraint of covariant constancy, a reality condition. This will allow for
each~$\tmop{SU} \left( 2 \right)^2$ symmetry under which the spurion
transforms to be restricted to~$\tmop{SU} \left( 2 \right)$. Later on in
sections \ref{sub:EWSB-minimal-case} and \ref{sec:EWSB-extended-cases}, we
will add complex spurions in order to reduce the symmetry further to
$\mathrm{U} \left( 1 \right)$.

\subsection{Goldstone bosons and Gauge fields} \label{sub:disconnected}

\begin{figure*}\centering
\includegraphics{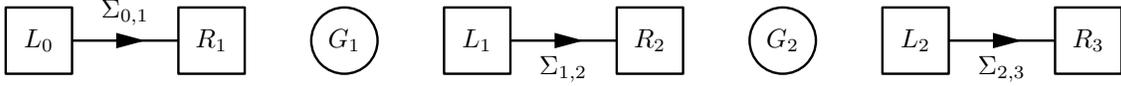}
\caption{Three non-linear sigma models and two Yang-Mills theories.}
\label{fig:3-nlsm-indep}
\end{figure*}

We consider~$K + 1$ independent non-linear sigma models, each describing the
spontaneous breakdown~$\tmop{SU} \left( 2 \right) \times \tmop{SU} \left( 2
\right) \longrightarrow \tmop{SU} \left( 2 \right)$. In addition, we consider
$K$ independent Yang-Mills theories with gauge group~$\tmop{SU} \left( 2
\right)$. We will describe the general case, leaving~$K$ as a free parameter,
but will pick the particular case with~$K  = 2$ for definiteness in order
to draw diagrams in this section, such as Fig.~\ref{fig:3-nlsm-indep}.
This diagram describes the symmetries of the model: for this particular case,
we really have five completely independent theories at this stage. The three
non-linear sigma models are depicted as oriented links between two squares
representing the transformation operating on the Goldstone multiplets. For
instance the left-most item in the figure represents the fact that the unitary
matrix~$\Sigma_{0, 1}$ transforms as
\begin{eqnarray}
  \Sigma_{0, 1} & \longmapsto & L_0 \Sigma_{0, 1} R^{\dag}_1,
\end{eqnarray}
with~$\left( L_0, R_1 \right) \in \tmop{SU}( 2 ) \times \tmop{SU} \left( 2
\right)$. The two gauge groups are represented by circles, and in fact the
circle labeled~$G_1$ means that the gauge transformation on the hermitian
gauge field~$G_{1 \mu}$ is given by
\begin{eqnarray}
  G_{1 \mu} & \longmapsto & G_1 G_{1 \mu} G_1^{\dag} + \frac{\mathi}{g_1} G_1
  \partial_{\mu} G_1^{\dag},
\end{eqnarray}
with~$G_1 \in \tmop{SU} \left( 2 \right)$. The positioning and naming of the
various elements in the diagram is irrelevant at this stage, although the
different pieces will be connected later on to form a chain.

Coming back to the general case, we point out that the model we have just
introduced possesses the symmetry
\begin{eqnarray}
  S_{\tmop{natural}, K} & = & \prod_{l = 1}^K \tmop{SU} \left( 2 \right)_{G_l}\nonumber\\
  &\times& \prod_{k = 0}^K \bignone \tmop{SU} \left( 2 \right)_{L_k} 
\times
  \tmop{SU} \left( 2 \right)_{R_{k + 1}}, \label{Snat}
\end{eqnarray}
under which the fields transform as
\begin{eqnarray}
  \Sigma_{k, k + 1} & \longmapsto & L_k \Sigma_{k, k + 1} R_{k + 1}^{\dag},
  \hspace{.2em} \tmop{for} \hspace{.2em} k = 0, \cdots, K, \\
  G_{k \mu} & \longmapsto & G_k G_{k \mu} G_k^{\dag} + \frac{\mathi}{g_k} G_k
  \partial_{\mu} G_k^{\dag}, \nonumber\\
&\tmop{for}& k = 1, \cdots, K,
\end{eqnarray}
where we have introduced the~$\tmop{SU} \left( 2 \right)$ transformations
$L_k$, $R_k$, $G_k$. The standard practice
consists in considering {\tmem{local}} chiral transformations~$L_k$, $R_k$, in
order to define the generating functional for the Noether currents of the
corresponding symmetries~{\cite{Gasser:1984yg,Gasser:1985gg}}. We will rely heavily on this in the following and we
therefore introduce the following covariant derivatives
\begin{eqnarray}
  D_{\mu} \Sigma_{k, k + 1} & = & \partial_{\mu} \Sigma_{k, k + 1} - \mathi
  L_{k \mu} \Sigma_{k, k + 1} + \mathi \Sigma_{k, k + 1} R_{k + 1 \mu},\nonumber\\
& \tmop{for}& \hspace{.2em} k = 0, \cdots, K,
\end{eqnarray}
where the~$L_{k \mu}$ and~$R_{k \mu}$ are sources.

Now, we want to build a LEET characterized by an expansion in powers of
momenta, the momentum scale being set by the momenta of light external
particles. Therefore, we assume slowly varying external sources and consider an
expansion in powers of derivatives. In order to write down the most general
lagrangian consistent with the symmetry~$S_{\tmop{natural}, K}$ of the
problem, we note that if one performs the following transformations on the
sources under the full group~$S_{\tmop{natural}, K}$
\begin{eqnarray}
  L_{k \mu} & \longmapsto & L_k L_{k \mu} L_k^{\dag} + \mathi L_k
  \partial_{\mu} L_k^{\dag}, \\
  R_{k \mu} & \longmapsto & R_k R_{k \mu} R_k^{\dag} + \mathi R_k
  \partial_{\mu} R_k^{\dag},
\end{eqnarray}
then the effective lagrangian must be invariant in order to reproduce the Ward
identities of the theory~{\cite{Leutwyler:1994iq}}, up to possible fermion
anomalies~{\cite{Wess:1971yu}}, which only impact the next-to-leading order in
the momentum expansion~{\cite{Gasser:1984yg,Gasser:1985gg}}. We use the
appropriate power counting for covariant derivatives and gauge fields~{\cite{Urech:1995hd,Wudka:1994ny}}
\begin{eqnarray}
  g_k & = & \mathcal{O} \left( p^1 \right),  \label{eq:pow-count-gk}\\
  G_{k \mu} & = & \mathcal{O} \left( p^0 \right), \\
  L_{k \mu}, R_{k \mu} & = & \mathcal{O} \left( p^1 \right),
  \label{eq:pow-count-sources}
\end{eqnarray}
defined to have the connections and derivatives counted on the same footing,
in order to obtain a covariant expansion order by order, and to have the
kinetic terms for the dynamical gauge fields appear at the same order as that
for the Goldstone bosons. Alternatively, one may consider the normalization of
states for this last step. With these rules we get at lowest order in the
expansion in powers of momenta, that is to say at~$\mathcal{O} \left( p^2
\right)$, the following lagrangian
\begin{eqnarray}
  \mathcal{L} & = & \frac{1}{4}  \sum^K_{k = 0} f_k^2  \left\langle D_{\mu}
  \Sigma_{k, k + 1} D^{\mu} \Sigma_{k, k + 1}^{\dag} \right\rangle \nonumber\\
&-&  \frac{1}{2}  \sum_{k = 1}^K \left\langle G_{k \mu \nu} G_k^{\mu \nu}
  \right\rangle.  \label{eq:Op2-lag}
\end{eqnarray}
In the above, the usual definition for the field-strength applies
\begin{eqnarray}
  G_{k \mu \nu} & = & \partial_{\mu} G_{k \mu} - \partial_{\nu} G_{k \mu} -
  \mathi g_k  \left[ G_{k \mu}, G_{k \nu} \right], \nonumber\\
&\tmop{for}&
 k = 1, \ldots, K,
\end{eqnarray}
and the symbol~$\left\langle \cdots \right\rangle$ denotes the trace of
the two-by-two matrices.

\subsection{Couplings along the chain}

We now wish to introduce couplings between the Goldstone multiplets via
interactions with gauge fields. However, the lagrangian~(\ref{eq:Op2-lag})
does not describe interactions between the disconnected elements of Fig.~\ref{fig:3-nlsm-indep}, nor does it suggest any line-ordering either, except for the relative positions we have
chosen for the purpose of the diagram. To remedy this, we now introduce
couplings between the independent theories, not by adding new interaction
terms, but by restricting fields in the lagrangian through constraints
{\footnote{This is quite standard: for instance the non-linear sigma model
itself introduces interactions in this manner.}}.

With this objective in mind, we now consider spurions, which we introduce as
non-propagating fields satisfying some constraints. These constraints will
restrict the allowed space of gauge configurations. In other words, they imply identifications
between various symmetry transformations operating on the model, while
enabling us to keep the full invariance group~$S_{\tmop{natural}, K}$
of the original theory.

\subsubsection{Real spurions} \label{sub:SU2-case}

In this section, we introduce spurions as two-by-two matrix-valued fields
$X_k, Y_k$ for~$k = 1, \cdots, K,$ subject to a reality condition. The
constraint of covariant constancy is then imposed on them. We show that one
can find a gauge in which the spurions reduce to real constants times the
identity matrix
\begin{eqnarray}
  \left. X_{k} \right|_{\tmop{const.}} & = & \xi_k 
  \mathbbm{1}_{2 \times 2},  \label{eq:before-last-claim}\\
  \left. Y_k \right|_{\tmop{const.}} & = & \eta_k  \mathbbm{1}_{2 \times
  2} .  \label{eq:last-claim}
\end{eqnarray}
The constants~$\xi_k, \eta_k$ are assumed to be small, and taken as expansion
parameters. The constraints are solved in this gauge to yield
\begin{eqnarray}
  \left. R_{k \mu}^a \right|_{\tmop{const.}} & = & g_k G_{k \mu}^a,
  \hspace{.2em} \tmop{for} \hspace{.2em} a = 1, 2, 3,  \label{eq:Rkmu=gkGkmu}\\
  \left. L^a_{k \mu} \right|_{\tmop{const.}} & = & g_k G^a_{k \mu},
  \hspace{.2em} \tmop{for} \hspace{.2em} a = 1, 2, 3,  \label{eq:Lkmu=gkGkmu}
\end{eqnarray}
implying that the connections are identified, thereby introducing the
couplings of Goldstone fields to gauge fields. The~$G_k$ transformation is
unconstrained.

\begin{figure*}\centering
\includegraphics{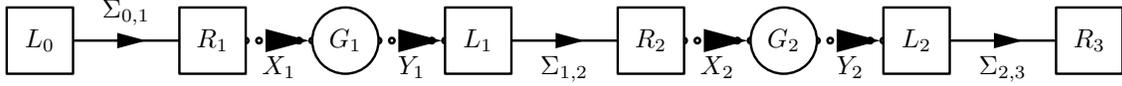}
\caption{Introduction of spurions in the model of section~\ref{sub:disconnected}.}
\label{fig:intr-gauge-coupl-K=2}
\end{figure*}

We now proceed to describe the spurions and the constraints, before solving
these: in the setting of section~\ref{sub:disconnected}, we introduce the
matrices~$X_k, Y_k$ with the assumed transformation properties
\begin{eqnarray}
  X_{k \text{}} & \longmapsto & R_k X_k G_k^{\dag}, \hspace{.2em} \tmop{for}
  \hspace{.2em} k = 1, \ldots, K,  \label{eq:transf-Xk}\\
  Y_k & \longmapsto & G_k Y_k L_k^{\dag}, \hspace{.2em} \tmop{for} \hspace{.2em} k = 1,
  \ldots, K,  \label{eq:transf-Yk}
\end{eqnarray}
as depicted in Fig.~\ref{fig:intr-gauge-coupl-K=2}~(compare figure
\ref{fig:3-nlsm-indep}). These spurions serve the purpose of relating the
gauge transformations with the chiral transformations operating on the
Goldstone multiplets. In relation with the hypothesis of naturalness and the
limit in which the Goldstone bosons and the gauge fields do not interact with each other, we will consider the case where the
entries in these spurions are functions with a small modulus, and therefore
consider
\begin{eqnarray}
  X_k, Y_k & = & \mathcal{O} \left( \epsilon \right), \hspace{.2em} \tmop{for}
  \hspace{.2em} k = 1, \ldots, K,  \label{eq:first-time-epsilon}
\end{eqnarray}
where~$\epsilon$ is by assumption a small parameter. We have assumed the
simplest situation here, considering all spurions to be of the same order of magnitude.
This is however not necessarily the case and at this point we have no way of
deciding what the appropriate counting is. We will therefore continue with
this simple assumption.

In order to introduce the gauge couplings to the Goldstone bosons via the
identification of connections, we  now demand that the spurions satisfy a constraint. Since we want the spurions to reduce to constants once the constraints are enforced while maintaining a covariant formulation and at the same time identifying the connections, it seems natural to impose the condition of covariant constancy~{\footnote{Note that such a
constraint could be enforced using Lagrange multipliers. For this simple
application, we find it more convenient to impose it by hand.}}
\begin{eqnarray}
  D_{\mu} X_k & = & 0, \hspace{.2em} \tmop{for} \hspace{.2em} k = 1, \ldots, K,
  \label{eq:covariant-constancy-Xk}\\
  D_{\mu} Y_k & = & 0, \hspace{.2em} \tmop{for} \hspace{.2em} k = 1, \ldots, K. 
  \label{eq:covariant-constancy-Yk}
\end{eqnarray}
From~(\ref{eq:transf-Xk}) and~(\ref{eq:transf-Yk}) we deduce that the
covariant derivatives in the above equations are given by
\begin{eqnarray}
  D_{\mu} X_k & = & \partial_{\mu} X_k - \mathi R_{k \mu} X_k + \mathi g_k X_k
  G_{k \mu}, \\
  D_{\mu} Y_k & = & \partial_{\mu} Y_k - \mathi g_k G_{k \mu} Y_k + \mathi Y_k
  L_{k \mu} . 
\end{eqnarray}
Notice that the constraints~(\ref{eq:covariant-constancy-Xk}) and~(\ref{eq:covariant-constancy-Yk}) restrict both the gauge connections contained in the covariant derivatives and also the non-propagating fields.
Now, the nature of the spurion fields itself impacts the way the full symmetry
is reduced: we consider in this section~the simplest case, imposing the
following reality condition on the spurions
\begin{eqnarray}
  X_k^c & = & X_k, \hspace{.2em} \tmop{for} \hspace{.2em} k = 1, \ldots, K,
  \label{eq:constraint-Xk}\\
  Y_k^c & = & Y_k, \hspace{.2em} \tmop{for} \hspace{.2em} k = 1, \ldots, K,
  \label{eq:constraint-Yk}
\end{eqnarray}
where we have defined the conjugate~$X^c$ of any two-by-two matrix~$X$ as
\begin{eqnarray}
  X^c & = & \tau^2 X^{\ast} \tau^2 . 
\end{eqnarray}
Note that the reality conditions~(\ref{eq:constraint-Xk}) and~(\ref{eq:constraint-Yk}) are compatible with the transformations
(\ref{eq:transf-Xk}) and~(\ref{eq:transf-Yk}). This fact, as well as the
reality condition itself, is specific to the group~$\tmop{SU} \left( 2
\right)$. The reality conditions~(\ref{eq:constraint-Xk}) and~(\ref{eq:constraint-Yk}) are in fact equivalent to the statement that we may
write
\begin{eqnarray}
  X_k & = & \xi_k U_k, \hspace{.2em} \tmop{for} \hspace{.2em} k = 1, \ldots, K,
  \label{eq:decomp-Xk}\\
  Y_k & = & \eta_k V_k, \hspace{.2em} \tmop{for} \hspace{.2em} k = 1, \ldots, K,
  \label{eq:radial-times-SU}
\end{eqnarray}
where~$\xi_k, \eta_k$ are real functions and~$U_k$ and~$V_k$ are~$\tmop{SU}
\left( 2 \right)$ matrices. From~(\ref{eq:first-time-epsilon}) we see that the
small parameter is the magnitude of the functions~$\xi_k, \eta_k$
\begin{eqnarray}
  \xi_k, \eta_k & = & \mathcal{O} \left( \epsilon \right), \hspace{.2em}
  \tmop{for} \hspace{.2em} k = 1, \ldots, K. 
\end{eqnarray}
In order to demonstrate our claims
(\ref{eq:before-last-claim}-\ref{eq:Lkmu=gkGkmu}), we perform an~$\tmop{SU}
\left( 2 \right)_{R_k}$ transformation with the following parameter
\begin{eqnarray}
  R_k & = & U_k^{\dag} .  \label{eq:Rk=Ukcross}
\end{eqnarray}
Identifying the components in the constraint~(\ref{eq:covariant-constancy-Xk}),
one then obtains that~$X_k$ is written as in~(\ref{eq:before-last-claim}),
with
\begin{eqnarray}
  \partial_{\mu} \xi_k & = & 0,
\end{eqnarray}
as well as the result announced in~(\ref{eq:Rkmu=gkGkmu}). Following the same
steps for~$Y_k$, with the~$\tmop{SU} \left( 2 \right)_{L_k}$ transformation given by
\begin{eqnarray}
  L_k & = & V_k,  \label{eq:Lk=Vk}
\end{eqnarray}
to write~$Y_k$ in the form of~(\ref{eq:last-claim}), we find that the
constraint results in
\begin{eqnarray}
  \partial_{\mu} \eta_k & = & 0,
\end{eqnarray}
together with~(\ref{eq:Lkmu=gkGkmu}). Thus, the constraints can be solved
easily in the gauge reached by the transformations~(\ref{eq:Rk=Ukcross}) and~(\ref{eq:Lk=Vk}), where they imply the identification of the connections with
the gauge fields.

Note that we did not have to use the gauge transformation~$G_k$ corresponding
to the dynamical gauge field. This~$\tmop{SU} \left( 2 \right)$ gauge field by
itself is therefore free of any constraints. On the other hand, we have
performed~$R_k$ and~$L_k$ transformations, and the corresponding
parameters are no more independent. Hence, we see that the real spurions select
in the original~$S_{\tmop{natural}, K}$ group the following subgroup which we
denote by~$S_{\tmop{reduced}, K}$
\begin{eqnarray}
  S_{\tmop{reduced}, K} & = & \bignone \tmop{SU} \left( 2 \right)_{L_0} \times
  \tmop{SU} \left( 2 \right)_{R_{K + 1}} \nonumber\\
&\times& \prod_{k = 1}^K \tmop{SU}
  \left( 2 \right)_{R_k+G_k+L_k} . 
\end{eqnarray}
A subgroup of this,
\begin{eqnarray}
  S_{\tmop{dynamical}, K} & = & \prod_{k = 1}^K \tmop{SU} \left( 2
  \right)_{R_k+G_k+L_k},
\end{eqnarray}
corresponds to dynamical gauge fields which propagate: this is the subgroup
which is `gauged'.  $S_{\tmop{reduced},K}$ coincides with the invariance group of 
the solution~(\ref{eq:before-last-claim}-\ref{eq:Lkmu=gkGkmu}) to the constraints~(\ref{eq:covariant-constancy-Xk}) and~(\ref{eq:covariant-constancy-Yk}) . Defining~$S_{\tmop{natural},K}$~(\ref{Snat}) as the maximal symmetry
of a given set of uncoupled Goldstone bosons and Yang Mills fields~(c.f. Fig.~\ref{fig:3-nlsm-indep} ), the list of spurions needed to operate the reduction of~$S_{\tmop{natural},K}$ to~$S_{\tmop{reduced},K}$ appears to be fixed essentially uniquely.

For the moment, we have not introduced any spurions
transforming under either~$L_0$ or~$R_{K + 1}$ and an additional dynamical
gauge group. We will consider this in sections \ref{sub:EWSB-minimal-case} and
\ref{sec:EWSB-extended-cases}, and these additional gauge groups will then be
identified with the weak gauge groups. To achieve the proper gauge group
reduction, we will be interested in generic spurions for which the reality
condition is dropped. This will leave us with the~$\mathrm{U} \left( 1
\right)$ subgroup of~$\tmop{SU} \left( 2 \right)$, but for this section~and
section~\ref{sec:generalized-WSRs}, we limit ourselves to cases were the
symmetry groups at the end of the moose are not dynamical.

\subsubsection{Leading-order lagrangian}

\begin{figure*}\centering
\includegraphics{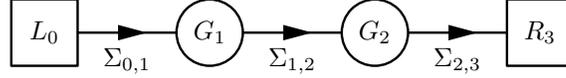}
\caption{The `reduced' diagram for the model depicted in Fig.~\ref{fig:intr-gauge-coupl-K=2}, after imposing the constraints.}
\label{fig:K=2-condense}
\end{figure*}

Imposing the constraints can be diagrammatically represented by the
modification of Fig.~\ref{fig:intr-gauge-coupl-K=2} into a new
{\emdash}reduced{\emdash} notation describing the same model, as shown in
Fig.~\ref{fig:K=2-condense}. This representation of the moose is closer to
the standard one found in the literature~{\cite{Arkani-Hamed:2001ca,Son:2003et}}: it describes the moose after application of the constraints,
which we will call the `constrained' moose. Note that some information is lost
with this diagram as there is no trace of the spurions, which will nonetheless
play a prominent role in the sequel.

We also introduce a new notation for the covariant derivative acting on the
Goldstone multiplets when the solution to the constraints in the standard
gauge as defined in section~\ref{sub:SU2-case} is injected, for~$k = 1,
\cdots, K - 1$
\begin{eqnarray}
  \nabla_{\mu} \Sigma_{k, k + 1} & = & \left. D_{\mu} \Sigma_{k, k + 1}
  \right|_{\tmop{const.}} \nonumber\\
  & = & \partial_{\mu} \Sigma_{k, k + 1} - \mathi g_k G_{k \mu} \Sigma_{k, k
  + 1} \nonumber\\
&+& \mathi g_{k + 1} \Sigma_{k, k + 1} G_{k + 1 \mu},
  \label{eq:nabla-first}
\end{eqnarray}
and for the ends of the moose
\begin{eqnarray}
  \nabla_{\mu} \Sigma_{0, 1} & = & \left. D_{\mu} \Sigma_{0, 1}
  \right|_{\tmop{const.}} \nonumber\\
  & = & \partial_{\mu} \Sigma_{0, 1} - \mathi L_{0 \mu} \Sigma_{0, 1} \nonumber\\
&+&
  \mathi g_1 \Sigma_{0, 1} G_{1 \mu}, \\
  \nabla_{\mu} \Sigma_{K, K + 1} & = & \left. D_{\mu} \Sigma_{K, K + 1}
  \right|_{\tmop{const.}} \nonumber\\
  & = & \partial_{\mu} \Sigma_{K, K + 1} - \mathi g_K G_{K \mu} \Sigma_{K, K
  + 1}\nonumber\\
& +& \mathi \Sigma_{K, K + 1} R_{K + 1 \mu} .  \label{eq:nabla-last}
\end{eqnarray}
Now, we may write the leading-order lagrangian, that is the~$\mathcal{O}
\left( p^2 \epsilon^0 \right)$ lagrangian. In fact, all the terms of this
order are already collected in~(\ref{eq:Op2-lag}). This~$\mathcal{O} \left(
p^2 \epsilon^0 \right)$ lagrangian, in which the spurions do not appear
explicitly, is reproduced here
\begin{eqnarray}
  \mathcal{L}^{\left( 2, 0 \right)} & = & \frac{1}{4}  \sum^K_{k = 0} f_k^2 
  \left\langle D_{\mu} \Sigma_{k, k + 1} D^{\mu} \Sigma_{k, k + 1}^{\dag}
  \right\rangle\nonumber\\
& -& \frac{1}{2}  \sum_{k = 1}^K \left\langle G_{k \mu \nu}
  G_k^{\mu \nu} \right\rangle . 
\end{eqnarray}
The point of this spurion formalism is the following: before we impose the
constraints~(\ref{eq:covariant-constancy-Xk}) and~(\ref{eq:covariant-constancy-Yk}), the Goldstone bosons do not know about the
gauge fields. It is only upon injection of the solution to these constraints
that we find interactions between the Goldstone bosons and the gauge fields.
With the help of the
definitions~(\ref{eq:nabla-first}-\ref{eq:nabla-last}) for the~$\nabla_{\mu}$
covariant derivatives, we obtain the following `constrained' $\mathcal{O} \left( p^2 \epsilon^0 \right)$ lagrangian
\begin{eqnarray}
  \left. \mathcal{L}^{\left( 2, 0 \right)} \right|_{\tmop{const.}} & = &
  \frac{1}{4}  \sum^K_{k = 0} f_k^2  \left\langle \nabla_{\mu} \Sigma_{k, k +
  1} \nabla^{\mu} \Sigma_{k, k + 1}^{\dag} \right\rangle \nonumber\\
&-& \frac{1}{2} 
  \sum_{k = 1}^K \left\langle G_{k \mu \nu} G_k^{\mu \nu} \right\rangle . 
  \label{eq:Op2x0-lag}
\end{eqnarray}
Here, the~$G_{k \mu}$ connections are dynamical fields while~$L_{0 \mu}$ and
$R_{K + 1 \mu}$~(appearing in the~$\nabla_{\mu}$ operator acting on
$\Sigma_{0, 1}$ and~$\Sigma_{K, K + 1}$) are the sources enabling us to define
the Noether currents of the chiral~$\tmop{SU} \left( 2 \right)_{L_0} \times
\tmop{SU} \left( 2 \right)_{R_{K + 1}}$ symmetry of this model. We thus get at
this order exactly the same lagrangian as is assumed in the case of
dimensional deconstruction, and the property of locality along the linear
moose is also evident. The interest of this approach has to do with the additional
terms at non-leading order in powers of~$\epsilon$: using the spurion
expansion, we are in a position to determine which terms will appear next.
Indeed the number of spurions required to turn the terms allowed by the
$S_{\tmop{reduced}, K}$ symmetry into terms invariant under the full~$S_{\tmop{natural},
K}$ symmetry gives us the order at which the corresponding term should appear.
We now briefly describe these additional terms.

\subsubsection{Structure of the effective lagrangian}

Having described the spurion formalism to be used in the context of effective
theories, we now go back to the problem of writing down the most general
lagrangian satisfying our requirements of symmetry: invariance under
$S_{\tmop{reduced}, K}$. If we now add the requirement of naturalness, our
previous discussions show that this implies invariance under
$S_{\tmop{natural}, K}$ and therefore involves spurions.

In order to write the most general effective lagrangian, we will use the fact
that all terms which are invariant under the full~$S_{\tmop{natural}, K}$
symmetry qualify as terms in our effective lagrangian. Such terms may be found
by forming suitable combinations of building blocks which are covariant under
the chiral transformations~$\prod_{k = 0}^K \bignone \tmop{SU} \left( 2
\right)_{L_k} \times \tmop{SU} \left( 2 \right)_{R_{k + 1}}$ acting on the
Goldstone fields. One can then also apply covariant derivatives. Examples of
such building blocks are
\begin{eqnarray}
  X_k Y_k & \longmapsto & R_k X_k Y_k L_k^{\dag}, \\
  X_k G_{k \mu \nu} Y_k & \longmapsto & R_k X_k G_{k \mu \nu} Y_k L_k^{\dag} .
\end{eqnarray}
We are also interested in classifying the terms in this lagrangian, following
a double expansion: the usual momentum expansion and simultaneously an
expansion in powers of spurions. After we have written down the full
lagrangian, we may `reduce' the terms by going to our standard gauge as
described in section~\ref{sub:SU2-case} and then inject the solution to the
constraints to see what the dynamical content of the new terms is. One then
obtains the constrained lagrangian, which does not involve the full spurions
but rather the constants~$\xi_k$ and~$\eta_k$. The power counting is then
provided, in addition to the expansion in momenta and gauge
coupling-constants, by the powers of~$\xi_k$ and $\eta_k$, that is, by the powers
of~$\epsilon$.

We have already written down in~(\ref{eq:Op2x0-lag}) the leading-order terms
in the lagrangian, which are the terms of order~$\mathcal{O} \left( p^2
\epsilon^0 \right)$. Those terms  do not involve spurions. We now discuss the next terms: once the
constraints are applied, the terms that would appear at~$\mathcal{O} \left(
p^2 \epsilon^2 \right)$ are found to be constants or simple renormalizations
of~$\mathcal{O} \left( p^2 \epsilon^0 \right)$ terms since we have
\begin{eqnarray}
  X_k X_k^{\dag}  & = & \xi_k^2  \mathbbm{1}_{2 \times 2},  \label{eq:detXk-identity}
\end{eqnarray}
and likewise for~$Y_k$. On the other hand, there are new terms at~$\mathcal{O}
\left( p^2 \epsilon^4 \right)$. These yield the following terms in the
standard gauge used in section~\ref{sub:SU2-case}
\begin{eqnarray}
  &  &  \left\langle D_{\mu} \Sigma_{k, k + 1} X_{k + 1} Y_{k + 1}\right. \nonumber\\
&\times &
\left. \left. D^{\mu} \Sigma_{k + 1, k + 2} \Sigma_{k + 1, k + 2}^{\dag} Y_{k + 1}^{\dag}
  X_{k + 1}^{\dag} \Sigma_{k, k + 1}^{\dag} \right\rangle
  \right|_{\tmop{const.}} \nonumber\\
  & = & \xi_{k + 1}^2 \eta_{k + 1}^2  \left\langle \nabla_{\mu} \Sigma_{k, k
  + 1} \right.\nonumber\\
&\times&
 \left. \nabla^{\mu} \Sigma_{k + 1, k + 2} \Sigma_{k + 1, k + 2}^{\dag}
  \Sigma_{k, k + 1}^{\dag} \right\rangle,  \label{eq:add-int-1}
\end{eqnarray}
and
\begin{eqnarray}
  &  & \left. \left\langle G_{k \mu \nu} Y_k \Sigma_{k, k + 1} X_{k + 1} G_{k
  + 1}^{\mu \nu} X_{k + 1}^{\dag} \Sigma_{k, k + 1}^{\dag} Y_k^{\dag}
  \right\rangle \right|_{\tmop{const.}} \nonumber\\
  & = & \xi_{k + 1}^2 \eta_k^2  \left\langle G_{k \mu \nu} \Sigma_{k, k + 1}
  G_{k + 1}^{\mu \nu} \Sigma_{k, k + 1}^{\dag} \right\rangle . 
  \label{eq:add-int-2}
\end{eqnarray}
There are in fact other terms that would involve only products of spurions
and Goldstone boson matrices without derivatives. Such terms would be
non-leading in~$\epsilon$, but of order~$\mathcal{O} \left( p^0 \right)$.
However, upon going to the above-mentioned gauge, such terms yield constant
numbers as a consequence of the identity~(\ref{eq:detXk-identity}).

As we will see, the terms in~(\ref{eq:add-int-1}) and~(\ref{eq:add-int-2})
have important implications regarding the properties of the model: they yield
corrections to the WSRs. Regarding symmetries and naturalness, we emphasize
once again that the constrained lagrangian~(\ref{eq:Op2x0-lag}) is not
invariant under the full original symmetry~$S_{\tmop{natural}, K}$, but only
under the~$S_{\tmop{reduced}, K}$ subgroup, since the transformations~$L_k$
and~$R_k$ have been identified with~$G_k$ for~$k = 1, \cdots, K$. Thus, there
is no reason why the terms in the right-hand side of~(\ref{eq:add-int-1}) and~(\ref{eq:add-int-2}) should not be included. Since they are invariant under this reduced symmetry, they will in fact be required as counter-terms if we
consider loops. Our formalism involving spurions shows how these terms may be
consistently treated as being of higher order in an expansion around the
lagrangian~(\ref{eq:Op2x0-lag}).

\subsection{Field redefinitions and unitary gauge} \label{sub:field-redef}

The model is the stage of multiple Higgs mechanisms:~$K$ gauge fields get
masses and~$K$ multiplets of Goldstone bosons disappear from the spectrum. The
mass matrix for the gauge fields has a very specific structure, due to the
nearest-neighbor interactions, and this will have important consequences later
on. Counting the number of scalar fields, one notices that there remains a
physical Goldstone multiplet in the spectrum. In order to describe the Higgs
mechanism, we will perform a gauge-independent change of variables. We will
refer to these field redefinitions by the phrase `going to the unitary gauge',
due to the similarity in the resulting lagrangian, but as a matter of fact,
the procedure does not involve gauge-fixing~{\cite{Gross:1972pv,Grosse-Knetter:1993nn}}. We will give the appropriate field redefinitions to
be performed in the full lagrangian but will then only write the resulting
constrained lagrangian in order to avoid unnecessarily complicated equations.

We first define the unitary matrix~$U$ describing the Goldstone bosons
remaining in the spectrum, the normalization being fixed by the requirement
that it be unitary. This implies that the multiplicative factors~$\xi_k$ and
$\eta_k$~{\footnote{We remind the reader that~$\xi_k$ and~$\eta_k$ are at this
stage functions: we are writing down the fields redefinitions in a general
gauge and independently of the constraints.}} drop out from the definition
\begin{eqnarray}
  U & = & \Sigma_{0, 1} U_1  \left( \prod^{K - 1}_{j = 1} V_j \Sigma_{j, j +
  1} U_{j + 1} \right) \nonumber\\
&\times& V_K \Sigma_{K, K + 1},
\end{eqnarray}
which merely contains the angular parts~$U_k$ and~$V_k$ of the spurion fields as introduced in~(\ref{eq:decomp-Xk}) and~(\ref{eq:radial-times-SU}). We then perform the
change of variable from~$\Sigma_{0, 1}$ to~$U$. Next, we define the vector
fields~$W_k^{\mu}$ by the following relation, for~$k = 1, \cdots, K$
\begin{eqnarray}
  g_k W_k^{\mu} & = & \mathi \left( \Sigma_{0, 1} U_1  \prod^{k - 1}_{j = 1}
  V_j \Sigma_{j, j + 1} U_{j + 1} \right)\nonumber\\
&\times& D^{\mu} \left( \left( \prod^{k -
  1}_{j = 1} U_{j + 1}^{\dag} \Sigma_{j, j + 1}^{\dag} V_j^{\dag} \right)
  U_1^{\dag} \Sigma_{0, 1}^{\dag} \right) \nonumber\\
  & - &  \mathi \alpha_k UD^{\mu} U^{\dag},  \label{eq:redef-Wks}
\end{eqnarray}
and perform the change of variables from the~$G_k^{\mu}$ fields to the
$W_k^{\mu}$ fields. From the transformation properties of~$U_k$ and~$V_k$, we
know that the covariant derivatives appearing in this equation are given by
\begin{eqnarray}
  D_{\mu} U_k & = & \partial_{\mu} U_k - \mathi R_{k \mu} U_k + \mathi g_k U_k
  G_{k \mu}, \\
  D_{\mu} V_k & = & \partial_{\mu} V_k - \mathi g_k G_{k \mu} V_k + \mathi V_k
  L_{k \mu},
\end{eqnarray}
which identically vanish when the constraints are imposed on the spurions. The
coefficients~$\alpha_k$ in~(\ref{eq:redef-Wks}) are defined by
\begin{eqnarray}
  \alpha_k & = & f_{\pi}^2  \sum_{j = 0}^{k - 1} \frac{1}{f_j^2} \bignone,
  \hspace{.2em} \tmop{for} \hspace{.2em} k = 1, \cdots, K,  \label{eq:alphak}
\end{eqnarray}
where we have introduced
\begin{eqnarray}
  \frac{1}{f_{\pi}^2} & = & \sum_{j = 0}^K \frac{1}{f_j^2} . 
  \label{eq:def-fpi}
\end{eqnarray}
The transformation properties of the newly defined fields are as follows
\begin{eqnarray}
  U & \longmapsto & L_0 UR_{K + 1}^{\dag},  \label{eq:transf-U}\\
  W_{k \mu} & \longmapsto & L_0 W_{k \mu} L_0^{\dag}, \hspace{.2em} \tmop{for}
  \hspace{.2em} k = 1, \cdots, K,
\end{eqnarray}
and when we then perform the change of variables according to
\begin{eqnarray}
  &&\left\{ G_{1 \mu}, \cdots, G_{K \mu}, \Sigma_{0, 1}, \cdots, \Sigma_{K, K +
  1} \right\}\nonumber\\
 & \longrightarrow & \left\{ W_{1 \mu}, \cdots, W_{K \mu}, U,
  \Sigma_{1, 2}, \cdots, \Sigma_{K, K + 1} \right\} . 
\end{eqnarray}
We find that, as a consequence of the symmetries, the lagrangian written in
terms of these variables does not depend on~$\Sigma_{1, 2}, \ldots, \Sigma_{K,
K + 1}$: this is the Higgs mechanism. In addition, the choice
(\ref{eq:alphak}) for the coefficients~$\alpha_k$ eliminates all mixing
between the Goldstone bosons~$U$ and the vector fields~$W_{k \mu}$.

It is in fact sufficient at this stage to work directly with the constrained
form of the lagrangian. With this in mind, we give the constrained form of the
field redefinitions in order to display the field content in a more intuitive
way
\begin{eqnarray}
  \left. U \right|_{\tmop{const.}} & = & \prod^K_{j = 0} \Sigma_{j, j +
  1},  \label{eq:def-U}\\
  \left. g_k W_k^{\mu} \right|_{\tmop{const.}} & = & \mathi \left(
  \prod^{k - 1}_{j = 0} \bignone \Sigma_{j, j + 1} \right) D^{\mu} \left(
  \prod^{k - 1}_{j = 0} \bignone \Sigma_{j, j + 1} \right)^{\dag} \nonumber\\
  & - & \left.  \mathi \alpha_k UD^{\mu} U^{\dag}
  \right|_{\tmop{const.}}, \nonumber\\
& \tmop{for}& \hspace{.2em} k = 1, \cdots,
  K.  \label{eq:redef-champs}
\end{eqnarray}
Upon rewriting the lagrangian in terms of the new variables, one then
recognizes~$f_{\pi}$, defined in~(\ref{eq:def-fpi}), as the Goldstone boson
decay constant.

In addition, the entry in the~$k$-th row and~$k'$-th column of the
mass-squared matrix~$\mathcal{M}^2$ for the~$W_k^{\mu}$ fields are given by
\begin{eqnarray}
  \left[ \mathcal{M}^2 \right]_{k, k'} & = & \frac{1}{4}   \delta_{k,
  k'} g_k^2  \left( f_{k - 1}^2 + f_k^2 \right) \nonumber\\
&-&\frac{1}{4}\left(  \delta_{k, k' + 1} +
  \delta_{k + 1, k'} \right) g_k g_{k + 1} f_k^2 ,
  \label{eq:coeff-matrice-masse}
\end{eqnarray}
for~$k, k' = 1, \cdots, K$. The elements of this mass matrix are of order
$\mathcal{O} \left( p^2 \epsilon^0 \right)$ since they contain two powers of
gauge couplings. This entails that the masses of the vector bosons have to be
considered small in some sense in our expansion. In practice, this has to be
understood as follows: the corresponding fields really belong to the LEET only
if their masses are smaller than the scale at which the low-energy
perturbation expansion breaks down and at which other resonances have to be
taken into account. We will not discuss the issue of estimating the value of
this scale. Suffices to say here that the information that can be gleaned from
the LEET containing the Goldstone modes only~{\cite{Manohar:1984md,Georgi:1985kw,Georgi:1993dw}}, points to a value in the vicinity of~$4 \mathpi
f_{\pi}$, but that there is a debate over these matters~\cite{Chang:2003vs}, due to the restoration of tree unitarity from Kaluza-Klein
excitations of gauge fields in the extra-dimensional case
{\cite{SekharChivukula:2001hz,Csaki:2003dt}}, or from the very resonances we
are considering here or in the dimensional deconstruction view.
Coming back to our main line of discussion, we notice that the only non-zero
entries in the mass-squared matrix~(\ref{eq:coeff-matrice-masse}) are on the
diagonal and just above or just below it. This structure entails that the
first powers of this matrix have zero coefficient in the off-diagonal corners,
that is
\begin{eqnarray}
  \left[ \left( \mathcal{M}^2 \right)^{l - 1} \right]_{1, K} & = & 0,
  \hspace{.2em} \tmop{for} \hspace{.2em} l = 1, \cdots, K - 1.  \label{eq:powers-of-M2}
\end{eqnarray}
On the other hand, the first inverse power of this matrix is such that it
satisfies the relation
\begin{eqnarray}
  \frac{g_1 g_K f_0^2 f_K^2}{4}  \left[ \left( \mathcal{M}^2 \right)^{- 1}
  \right]_{1, K} - f_{\pi}^2 & = & 0,  \label{eq:1ere-WSR-par-transversite}
\end{eqnarray}
as can be calculated explicitly by inference on~$K$, for~$K \geqslant 1$. We
will make use of relations~(\ref{eq:powers-of-M2}) and~(\ref{eq:1ere-WSR-par-transversite}) later on in section~\ref{sub:WSRs}. Note
that in sections \ref{sub:EWSB-minimal-case} and
\ref{sec:EWSB-extended-cases}, we will be coupling gauge fields at the ends of
the chain, and also connecting the two ends. This has no influence on the mass
matrix for the~$W_{k \mu}$ fields, but only on the kinetic terms for these
fields. In order to bring the lagrangian to the canonical form, further field
redefinitions involving rescaling of fields will then be required.

\section{The left-right two-point correlator} \label{sec:generalized-WSRs}

The constrained moose shown in Fig.~\ref{fig:K=2-condense} represents a
chain of~$K$ gauge fields~$G_{k \mu}$ interacting with~$K + 1$ Goldstone boson
multiplets~$\Sigma_{k, k + 1}$, where~$K = 2$ in the figure. As a consequence
of the dynamical Higgs mechanism, the latter disappear from the spectrum but
for one combination parametrized by~$U$. This gives the mass matrix~(\ref{eq:coeff-matrice-masse}) to the vectors.

The characteristic long-distance feature of the model is its global chiral
symmetry~$\tmop{SU} \left( 2 \right)_{L_0} \times \tmop{SU} \left( 2
\right)_{R_{K + 1}}$, which is spontaneously broken, generalizing the usual
case of the non-linear sigma model~($K = 0$). The Noether currents generating
this symmetry may be obtained by taking the functional derivative with respect
to the sources~$L_{0 \mu}$ and~$R_{K + 1 \mu}$ in~(\ref{eq:Op2x0-lag}).
However, the lagrangian~(\ref{eq:Op2x0-lag}) is endowed with additional
interesting short-distance properties which are reminiscent of QCD, despite
the model bearing no obvious resemblance to any QCD-like theory.

Indeed, in QCD, the spontaneously broken chiral symmetry combines with the
operator product expansion~(OPE) and the operator content of the theory to
yield the two Weinberg sum rules~{\cite{Weinberg:1967kj}}: the two-point
correlator of left-handed and right-handed Noether currents transforms
according to the~$( 3, 3 )$ representation of the chiral~$\tmop{SU} \left( 2
\right) \times \tmop{SU} \left( 2 \right)$ group and there are no
corresponding local operators in the theory with mass-dimension strictly lower
than six~{\cite{Shifman:1979bx}}. Consequently, the left-right correlator
behaves smoothly at short distances, leading among other things to the finite
electroweak mass of the pseudo-Goldstone bosons~(pions)~{\cite{Das:1967it}}.

It has already been pointed out~{\cite{Son:2003et}} that the lagrangian
(\ref{eq:Op2x0-lag}) also leads to the two Weinberg sum rules, rendering the
theory smoother at high energies than naively expected: we stress that this is
in fact at the origin of the little Higgs phenomenon. However, as is the case
in QCD, this feature is {\tmem{not}} a consequence of low-energy symmetries
characteristic of the lagrangian~(\ref{eq:Op2x0-lag}). What is crucial in
addition is the property of locality along the chain: the fact that only
nearest-neighbors interact. This property is emphasized when the lagrangian
(\ref{eq:Op2x0-lag}) is rewritten in the form
\begin{eqnarray}
  \left. \mathcal{L}^{( 2, 0 )} \right|_{\tmop{const.}} & = &
  \mathcal{L}_0 \left( L_{0 \mu}, \Sigma_{1, 0}, G_{1 \mu} \right)\nonumber\\
 & +&\mathcal{L}_K \left( G_{K \mu}, \Sigma_{K, K + 1}, R_{K + 1 \mu} \right)
  \nonumber\\
  & + & \sum_{k = 1}^{K - 1} \mathcal{L}_k \left( G_{k \mu}, \Sigma_{k, k +
  1}, G_{k + 1 \mu} \right) .  \label{eq:locality}
\end{eqnarray}
We then see that, in order to connect the left and right sources~$L_{0 \mu}$
and~$R_{K + 1 \mu}$, one has to perform at least~$K$ contractions, pushing the
left-right correlator to higher orders. This property of the theory indeed
does not follow from the symmetry~$S_{\tmop{reduced}, K}$ of
(\ref{eq:Op2x0-lag}) alone: one  needs either to invoke dimensional
deconstruction and locality in the fifth dimension, or to use the
higher symmetry~$S_{\tmop{natural}, K}$ and the corresponding spurion
formalism. Terms which are not of the form~(\ref{eq:Op2x0-lag}) and which
break the Weinberg sum rules are indeed part of the theory, but are of higher
order in~$\epsilon$.

In this section, these statements are further clarified and extended: we are
going to show that, at order~$\mathcal{O} \left( p^2 \epsilon^0 \right)$,
the theory actually contains~$K$ Weinberg sum rules generalizing those known
from QCD.

\subsection{Generalized Weinberg sum rules} \label{sub:WSRs}

The claims made above concerning the left-right correlator pertain to the
two-point function of the Noether currents  corresponding to the symmetry
under which the Goldstone multiplet remaining in the spectrum transforms.
Recalling the transformation properties of~$U$~(\ref{eq:transf-U}), we now
focus on the second derivative of the generating functional with respect to
the sources~$L_{0 \mu}$ and~$R_{K + 1 \mu}$: this study is performed at
lowest-order and consequently at tree-level. We define the function~$\Pi_{L
R}$ through the relation
\begin{eqnarray}
&&  4 \mathi \int \mathd x \mathe^{\mathi q \cdot x}  \bignone \left\langle 0
  \left| TJ_{L_0}^{a \mu}( x ) J_{R_{K + 1}}^{b \nu} \left( 0 \right) \right|
  0 \right\rangle\nonumber\\
 & = & - \delta^{a b}  \left( \eta^{\mu \nu} q^2 - q^{\mu}
  q^{\nu} \right) \Pi_{L R} \left( - q^2 \right) .  \label{eq:def-PiLR}
\end{eqnarray}
One may evaluate this by taking the second derivative of the path integral with respect to the
sources and then evaluating the tree-level diagrams in the
basis of vector fields which diagonalizes the mass matrix
(\ref{eq:coeff-matrice-masse}). We will however follow another route, which
displays in a more transparent fashion the origin of the relations we are
interested in, using the~$W_k^{\mu}$ fields.

We first extract the expressions for the currents in terms of the~$W_{k \mu}$
fields: this is easily done by first taking the functional derivatives of the
original expression for the lagrangian~(\ref{eq:Op2x0-lag}), and then
performing the change of variables implied by the definitions
(\ref{eq:redef-champs}) and~(\ref{eq:def-U}). Again, we work with the
constrained lagrangian in order to keep the equations as simple as possible.
We find the expected field-current identities~{\cite{Sakurai:1969,Craigie:1986qf,Meissner:1988ge}}, with an additional term involving the
remaining Goldstone bosons
\begin{eqnarray}
  J_{L_0}^{a \mu} & = & \left. \frac{\delta \mathcal{L}_2}{\delta L_{0 \mu}^a}
  \right|_{L_{0 \rho} = 0, R_{K + 1 \sigma} = 0} \nonumber\\
  & = & - \frac{g_1 f_0^2}{4} W_1^{a \mu} - \mathi \frac{f_{\pi}^2}{4} 
  \left\{ U \partial^{\mu} U^{\dag} \right\}^a,  \label{eq:JL}
\end{eqnarray}
\begin{eqnarray}
  J_{R_{K + 1}}^{a \mu} & = & \left. \frac{\delta \mathcal{L}_2}{\delta R_{K +
  1 \mu}^a} \right|_{L_{0 \rho} = 0, R_{K + 1 \sigma} = 0} \nonumber\\
  & = & - \frac{g_K f_K^2}{4}  \left\{ U^{\dag} W_K^{\mu} U \right\}^a \nonumber\\
&-&  \mathi \frac{f_{\pi}^2}{4}  \left\{ U^{\dag} \partial^{\mu} U \right\}^a . 
  \label{eq:JR}
\end{eqnarray}
The presence of the matrix~$U$ in the second line of
(\ref{eq:JR}) follows from the lack of symmetry of our definitions
(\ref{eq:redef-champs}) relative to the center of the chain. This is a minor
flaw, and we stick to these definitions based on their simplicity.

Only the terms linear in fields in the above expressions for the currents are
relevant for the evaluation of the two-point function at tree level. The only contributions in this model come from Goldstone boson and massive resonance exchange, and we find at this order
\begin{eqnarray}
  &  & 4 \mathi \int \mathd x \mathe^{\mathi q \cdot x}  \bignone
  \left\langle 0 \left| TJ_{L_0}^{a \mu}( x ) J_{R_{K + 1}}^{b \nu} \left( 0
  \right) \right| 0 \right\rangle \nonumber\\
  & = & \delta^{a b}  \frac{g_1 g_K f_0^2 f_K^2}{4}  \nonumber\\
&\times& \left[ \left(
  \eta^{\mu \nu}  \mathbbm{1} - q^{\mu} q_{}^{\nu}  \left( \mathcal{M}^2
  \right)^{- 1} \right)  \left( q^2  \mathbbm{1} -\mathcal{M}^2 \right)^{- 1}
  \right]_{1, K} \nonumber\\
&+& \delta^{a b} f_{\pi}^2  \frac{q^{\mu} q^{\nu}}{q^2} .
  \label{eq:PiLR}
\end{eqnarray}
Expanding this for large euclidean momenta~$Q^2 = - q^2 \rightarrow + \infty$
yields
\begin{eqnarray}
  \Pi_{L R} \left( Q^2 \right) & = & -\frac{g_1 g_K f_0^2 f_K^2}{4} \nonumber\\
&\times& \sum_{l =
  1}^{\infty}  \frac{\left[ \left(- \mathcal{M}^2
  \right)^{l - 1} \right]_{1, K}}{Q^{2 \left( l + 1 \right)}} . 
  \label{eq:PiLR-expansion}
\end{eqnarray}
Note that the~$l = 0$ term in this equation is absent for~$K \geqslant 1$, due
to the relation~(\ref{eq:1ere-WSR-par-transversite})~(compare expressions
(\ref{eq:PiLR}) and~(\ref{eq:PiLR-expansion})). This is the first Weinberg sum rule, which is valid here at tree-level: the
first moment in the expansion of~$\Pi_{L R}$ at large~$Q^2$ is zero. In this model where the contributions to
the correlator come from infinitely narrow resonances, the Weinberg sum
rules are simply relations between the masses of the resonances and the decay constants of the Goldstone bosons and resonances as we shall see
in section~\ref{sub:other-aspects}. Note that relation~(\ref{eq:1ere-WSR-par-transversite})  can also be deduced from
expression~(\ref{eq:PiLR}) and the transversity of the two-point function,
which is itself a consequence of current-conservation. From the relations~(\ref{eq:powers-of-M2}), we see that the terms with~$l = 1, \cdots, K -
1$ in the infinite series in the right-hand-side of~(\ref{eq:PiLR-expansion})
are also zero. This gives us additional Weinberg sum rules. The case~$l = 1$
is recognized as the second Weinberg sum rule. Thus, the case of a chain with
two intermediate sites ($K=2$) reproduces exactly two Weinberg sum rules.

In the general case, equation~(\ref{eq:PiLR-expansion}) shows that the two-point function~$\Pi_{L R}$
decreases for large euclidean momenta as~$1 / Q^{2 \left( K + 1 \right)}$ and
not as~$1 / Q^2$, where~$K$ is the number of internal sites in the chain. This
in turn implies that the~$K$ first operators appearing as coefficients in the
expansion~(\ref{eq:PiLR-expansion}), which are order parameters of the chiral
symmetry, are found to have a vanishing vacuum expectation value in such models at this order. We
have thus related the smooth high-energy behavior of moose models to an
extension of a well-known property of QCD.

\subsection{Other aspects} \label{sub:other-aspects}

In this section, we point out other relevant aspects of the left-right
correlator which require a closer inspection of the algebra involved with the
diagonalization of the mass matrix. In fact, it turns out that we can learn
more about the left-right correlator, even without explicitly diagonalizing
the mass matrix, but just using the orthogonality properties of the
transformation matrix to the diagonal basis.

As already mentioned, an alternative writing for~$\Pi_{L R}$ uses the basis in
which the mass matrix is diagonal: this proceeds through the definition of the
$A_n^{\mu}$ fields~{\cite{Son:2003et}}
\begin{eqnarray}
  A_n^{\mu} & = & \sum_{k = 1}^K b_n^k W_k^{\mu}, \hspace{.2em} \tmop{for}
  \hspace{.2em} n = 1, \cdots, K,
\end{eqnarray}
where the orthogonal matrix~$b$ is as yet unknown. We denote by~$M_n^2$ the mass squared of the
field~$A_n^{\mu}$. These masses as well as the~$b_n^k$
coefficients may be calculated explicitly only in particular cases
{\footnote{We will work out an example explicitly in section
\ref{sub:particular-cases}.}}. Taking the second functional derivative with
respect to the sources~$L_{0 \mu}, R_{K + 1 \nu}$ of the generating functional
defined by the path-integral over the lagrangian~(\ref{eq:Op2x0-lag}), we get
at tree-level
\begin{eqnarray}
  \Pi_{L R} \left( Q^2 \right) & = & - \frac{f_{\pi}^2}{Q^2} + \sum_{n = 1}^K
  \frac{F_n^2}{Q^2 + M_n^2} \bignone,  \label{eq:PiLR-Fn2}
\end{eqnarray}
where the coupling constant~$F_n^2$ of the massive vector field~$A_n^{\mu}$ is
given by
\begin{eqnarray}
  F_n^2 & = & \frac{1}{4} g_1 g_K f_0^2 f_K^2  \frac{b_n^1 b_n^K}{M_n^2} . 
  \label{eq:Fn2}
\end{eqnarray}
Explicit calculation of the quantities
\begin{eqnarray}
  \sum_{n = 1}^K \bignone F_n^2  - f_{\pi}^2 &=& 0 , \label{WSR1}\\
 \sum_{n =
  1}^K \bignone F_n^2 M_n^{2 \left( l - 1 \right)} & = & 0,\hspace{.2em} \text{for} \hspace{.2em} l=2, \cdots, K, \label{WSR2}
\end{eqnarray}
using~(\ref{eq:Fn2}) then gives an alternative derivation of the~$K$ Weinberg
sum rules of section~\ref{sub:WSRs}.

Depending on the sign of the constant~$F_n^2$, one may call the corresponding
$n$-th vector field a {\tmem{vector}} or an {\tmem{axial}} resonance even
though the names do not apply in the strict sense since the chain is not
necessarily symmetric under reflection with respect to its center
{\emdash}`parity'~{\footnote{It is possible to show quite
generally that the signs for the~$F_n^2$ alternate, starting with a vector
resonance~($F_1^2 > 0$). This was already noticed in~{\cite{Son:2003et}} for the parity-invariant or symmetric moose. The authors of~{\cite{Knecht:1998ts}} also point out that this is the case in general whatever
the model for resonances when one has exactly as many resonances as Weinberg
sum rules.}}. In this manner, one may determine the first non-zero moment in
the expansion~(\ref{eq:PiLR-expansion})
\begin{eqnarray}
  \Pi_{L R} \left( Q^2 \right) & = & - \left( \prod_{i = 1}^K g_i^2 \bignone
  \right)  \left( \prod_{j = 0}^K \frac{f_j^2}{4} \bignone \right) 
  \frac{1}{Q^{2 \left( K + 1 \right)}} \nonumber\\
&+&\mathcal{O} \left( \frac{1}{Q^{2
  \left( K + 2 \right)}} \right) .  \label{eq:Witten}
\end{eqnarray}
Thus, in the limit of large~$Q^2$ we have~$Q^2 \Pi_{L R} \left(
Q^2 \right) < 0$, as implied in the case of QCD by the inequality derived by
Witten~{\cite{Witten:1983ut}}. This result once again confirms that of
{\cite{Knecht:1998ts}} which considered resonances independently of a
lagrangian model, but assuming invariance under parity.

We briefly turn to the low-energy consequences of our model: here again, we do
not need to know the diagonalization for the mass matrix explicitly, just its
properties.

By integration of the massive resonances, one may in principle derive the
values of the LECs for the chiral lagrangian corresponding to this model.
Integration of the resonances require that they be parametrically heavier than
the remaining particles. In our case, even though the masses of the resonances
are small parameters in the sense of our expansion~(they are counted as
$\mathcal{O} \left( p^2 \epsilon^0 \right)$), the remaining particles are the
massless Goldstone bosons, and the procedure is thus meaningful. As a simple
application, one may determine the constant~$L_{10}$ defined by Gasser and
Leutwyler~{\cite{Gasser:1985gg}}, by making use of the equations of motion for
the resonances~{\cite{Ecker:1989te}}, or directly through the following
relation~{\cite{Gasser:1984yg}}
\begin{eqnarray}
  L_{10} & = & - \frac{1}{4}  \int_0^{+ \infty} \mathd s \left(
  \frac{1}{\mathpi} \tmop{Im} \left( \Pi_{L R} \left( s \right) \right) +
  f_{\pi}^2 \delta \left( s \right) \right) \bignone \bignone \nonumber\\
  & = & - \frac{1}{4}  \sum_{n = 1}^K \bignone \frac{F_n^2 }{M_n^2} . 
\end{eqnarray}
We find
\begin{eqnarray}
  L_{10} & = & - \sum^K_{k = 1} \frac{\alpha_k  \left( 1 - \alpha_k
  \right)}{g_k^2} .  \label{eq:L10}
\end{eqnarray}
Note that the constants~$\alpha_k$ appearing in this equation are known
explicitly from~(\ref{eq:alphak}) in the generic case. If we take the view
that our moose model is to be used as a model for the low-energy sector of a
QCD-type theory, then this constant~$L_{10}$ is a non-local order parameter of
chiral symmetry breaking: we observe that~$L_{10} < 0$ in this model. This
fact will have repercussions later on in section~\ref{sec:EWSB-extended-cases}
when we use these moose models as a basis for electroweak symmetry
breaking: the constant~$L_{10}$ then gives us the value of the~$S$ parameter
{\cite{Espriu:1992vm}} since we are restricting ourselves to tree-level. We
have seen that the sign of the parameter is fixed~{\footnote{Changing the sign
would require a model where there are fewer Weinberg sum rules than
resonances, as can be seen from~{\cite{Knecht:1998ts}}.}}, however, its
magnitude can be made arbitrarily small if we set one of the decay constants
at the end of the chain to be much smaller than the others. In practice, we
would choose~$f_K \ll f_k$ for~$k = 0, \cdots, K - 1$ with the conventions to
be adopted in section~\ref{sub:EWSB-minimal-case} and
\ref{sec:EWSB-extended-cases}. One may also choose the gauge coupling
constants to be large enough in order to avoid a large value of~$L_{10}$ at
tree-level but the freedom in this respect is limited given the power counting
(\ref{eq:pow-count-gk}) for the gauge coupling constants~$g_k$.

\subsection{Particular cases} \label{sub:particular-cases}

\subsubsection{Case with all decay constants equal and all gauge couplings
equal}

Up to this point the derivations have been general and the results apply in
all cases. Still, we are limited for practical applications as soon as~$K$ is
larger than a few units, since we are then unable to diagonalize the mass
matrix and find the explicit values of the~$b_n^k$ coefficients. One may want
to derive further explicit results to get a feeling of what happens. We will do
this in the simplest possible case while keeping~$K$ undetermined: we pick the
particular case where all decay constants are equal, and all gauge
coupling constants are equal
\begin{eqnarray}
  f_k & = & f, \hspace{.2em} \tmop{for} \hspace{.2em} k = 0, \cdots, K, \\
  g_k^{} & = & g, \hspace{.2em} \tmop{for} \hspace{.2em} k = 1, \cdots, K. 
\end{eqnarray}
We then derive the following masses for the~$K$ resonances
\begin{eqnarray}
  M_n^2 & = & g^2 f^2 \sin^2 \left( \frac{\mathpi n}{2 \left( K + 1 \right)}
  \right), \nonumber\\
& \tmop{for}& n = 1, \cdots, K,
  \label{eq:Mn2-cas-simple}
\end{eqnarray}
and the following explicit expressions for the~$F_n^2$
\begin{eqnarray}
  \hspace{.2em} F_n^2 & = & \frac{2}{K + 1}  \left( - 1 \right)^{n + 1} f^2
  \cos^2 \left( \frac{\mathpi n}{2 \left( K + 1 \right)} \right), \nonumber\\
&\tmop{for}& \hspace{.2em} n = 1, \cdots, K . 
\end{eqnarray}
With this formula, we can explicitly verify that we have a tower of
alternating vector and axial resonances~{\footnote{This time, the names apply
in the strict sense since we are considering a symmetric moose.}}. The expression
for~$L_{10}$ in this particular case is found to be
\begin{eqnarray}
  L_{10} & = & - \frac{1}{6 g^2}  \frac{K \left( K + 2 \right)}{K + 1} . 
\end{eqnarray}

\subsubsection{Case with~$K = 1$} \label{sub:K=1-moose}

We give a short summary of relevant results for this simple case as a
preparation for section~\ref{sub:radcor-mPGBs}. The study of the~$K = 1$
linear moose model is performed as in the general case presented above: there
is only one triplet of massive vector fields~$W_{1 \mu}^a$,~$a = 1, 2, 3$,
with masses
\begin{eqnarray}
  M_{W_1}^2 & = & g_1^2  \frac{f_0^2 + f_1^2}{4} .  \label{eq:MW2}
\end{eqnarray}
The decay constant~$f_{\pi}$ of the remaining Goldstone bosons is related to
the original parameters in the lagrangian through relation~(\ref{eq:def-fpi}),
which becomes
\begin{eqnarray}
  f_{\pi}^2 & = & \frac{f_0^2 f_1^2}{f_0^2 + f_1^2} . 
\end{eqnarray}
For the~$K = 1$ linear moose model, we have exactly one Weinberg sum rule, and
(\ref{eq:PiLR-Fn2}) is rewritten
\begin{eqnarray}
  \Pi_{L R} \left( Q^2 \right) & = & - f_{\pi}^2 M_{W_1}^2  \frac{1}{Q^2 
  \left( Q^2 + M_{W_1}^2 \right)} .  \label{eq:PiLR-1-WSR}
\end{eqnarray}
\subsection{Non-leading interactions and corrections to the WSRs}
\label{sub:corrections}

We now briefly describe the effect of the non-leading terms in~$\epsilon$
mentioned in~(\ref{eq:add-int-1}) and~(\ref{eq:add-int-2}). These additional
interactions are of order~$\mathcal{O}\left(p^2 \epsilon^4\right)$, and thus only a limited number of
them may occur at a given order in the expansion. For most of this paper we
will not fix the relation between the counting in powers of~$p$ and that in
powers of~$\epsilon$, since we want to be as general as possible.

In the dimensional deconstruction approach the terms we are considering here
are omitted on the basis that they correspond to non-local interactions with
respect to the fifth~(deconstructed) dimension. However, we want to consider
their effect since they are not forbidden by the symmetries of the problem
once we have introduced the gauge interactions, but only suppressed in our
power counting scheme. In any case, they will be produced by radiative
corrections.

We first consider the following lagrangian, where~$\mathcal{O}\left(p^2 \epsilon^4\right)$ terms of the type~(\ref{eq:add-int-1}) have been included
\begin{eqnarray}
  \left. \mathcal{L}' \right|_{\tmop{const.}} & = & \frac{1}{4} 
  \sum^K_{k = 0} \bignone f_k^2  \left\langle \nabla_{\mu} \Sigma_{k, k + 1}
  \nabla^{\mu} \Sigma_{k, k + 1}^{\dag} \right\rangle \nonumber\\ 
&-& \frac{1}{2}  \sum_{k =
  1}^K \left\langle G_{k \mu \nu} G_k^{\mu \nu} \right\rangle \nonumber\\
  & + &  \frac{1}{2}  \sum_{k = 0}^{K - 1} \xi_{k + 1}^2 \eta_{k + 1}^2 f_{k,
  k + 1} \left\langle \nabla_{\mu} \Sigma_{k, k + 1} \nabla^{\mu} \Sigma_{k +
  1, k + 2} \right.\nonumber\\
&\times&  \left.\Sigma_{k + 1, k + 2}^{\dag} \Sigma_{k, k + 1}^{\dag}
  \right\rangle .  \label{eq:second-neighbor}
\end{eqnarray}
If we derive the expression for the two-point left-right function implied
by this lagrangian at tree-level and up to order~$\mathcal{O}\left(p^2 \epsilon^4\right)$, we find that the last Weinberg sum rule obtains a correction,~(the $l=K$ equation in~(\ref{WSR2}) then includes an~$\mathcal{O}\left(p^{2 \left(K-1\right)}\epsilon^4\right)$ term in the right-hand side whereas individual terms in left-hand side are of order~$\mathcal{O}\left(p^{2 \left(K-1\right)}\epsilon^0\right)$). This is expected from the naive
observation that the effective length of the chain is reduced: we indeed have
interactions involving two neighboring links. This may be rephrased along the
same lines as the reasoning following~(\ref{eq:locality}): the number of gauge
interactions required to have a non-zero correlation between left and right
currents is reduced. Indeed, linear combinations of terms of the
type
\begin{eqnarray}
&&  \left\langle \nabla_{\mu} \Sigma_{k, k + 1} \nabla^{\mu} \Sigma_{k, k +
  1}^{\dag} \right\rangle ,
\end{eqnarray}
and
\begin{eqnarray}
 &&\left\langle \nabla_{\mu} \Sigma_{k,
  k + 1} \nabla^{\mu} \Sigma_{k + 1, k + 2} \Sigma_{k + 1, k + 2}^{\dag}
  \Sigma_{k, k + 1}^{\dag} \right\rangle,
\end{eqnarray}
can be recast as combinations of
\begin{eqnarray}
 && \left\langle \nabla_{\mu} \Sigma_{k, k + 1} \nabla^{\mu} \Sigma_{k, k +
  1}^{\dag} \right\rangle ,
\end{eqnarray}
and
\begin{eqnarray}
&& \left\langle \nabla_{\mu} \left(
  \Sigma_{k, k + 1} \Sigma_{k + 1, k + 2} \right)\right.\nonumber\\
&\times& \left. \nabla^{\mu} \left(
  \Sigma_{k + 1, k + 2}^{\dag} \Sigma_{k, k + 1}^{\dag} \right) \right\rangle
  . 
\end{eqnarray}
The point is that the right transformation of~$\Sigma_{k, k + 1}$
is identified with the left transformation of~$\Sigma_{k + 1, k + 2}$, and the
corresponding interactions come in without an additional power of gauge coupling
constant. This description is reminiscent of the standard motivation for
little Higgs models~{\cite{Arkani-Hamed:2001nc}}.

Coming back to our discussion of the low-energy constant~$L_{10}$~(\ref{eq:L10}), we note that, provided the spurion term in
(\ref{eq:second-neighbor}) represents a small correction, one expects
\begin{eqnarray}
  \xi_{k + 1}^2 \eta_{k + 1}^2  \left| f_{k, k + 1} \right| & \ll & f_j^2,
\end{eqnarray}
which in turn implies that~$L_{10}$ remains negative.

If one now considers the additional terms involving gauge field curvatures as
shown in~(\ref{eq:add-int-2}), then all Weinberg sum rules but the first one
are affected by the presence of~$\mathcal{O}\left(p^2 \epsilon^4\right)$ terms in the lagrangian: the particular sums of~$\mathcal{O}\left(p^{2 \left(l-1\right)} \epsilon^0\right)$ constants on the left-hand side of~(\ref{WSR2}) are evaluated to be of order~$\mathcal{O}\left(p^{2 \left(l-1\right)} \epsilon^4\right)$. This comes about because the diagonalization of the kinetic terms for the vector fields is modified. The first Weinberg sum rule on the other hand, being related to the
transversity of the two-point left-right function remains valid. In fact, it
is possible to destroy this sum rule as well, provided we introduce terms
similar to~(\ref{eq:add-int-1}) involving the sources~{\footnote{This does
not conflict with the transversity argument just alluded to because in this
case, there are additional local contributions to the two-point left-right
function which destroy the link between the first Weinberg sum rule and the
transversity of this two-point function. }}
\begin{eqnarray}
  & \left\langle L_{0 \mu \nu} \Sigma_{0, 1} X_1 G_1^{\mu \nu} X_1^{\dag}
  \Sigma_{0, 1}^{\dag} \right\rangle, & \\
  & \left\langle G_{K \mu \nu} Y_K \Sigma_{K, K + 1} R_{K + 1}^{\mu \nu}
  \Sigma_{K, K + 1}^{\dag} Y_K^{\dag} \right\rangle . & 
\end{eqnarray}
However, such terms only appear at order~$\mathcal{O} \left( p^3 \epsilon^2
\right)$: for them to appear at a lower order, they would need to be divided
by a gauge coupling constant, making them ill-defined in the limit~$g_k \longrightarrow
0$. The same would then be true of the Noether currents of the theory, even
though the contribution to the charge would be zero, being a surface term. Therefore, the requirement that the off-shell Noether currents of the theory be
well-defined in this limit forbids these terms at~$\tmop{order} \mathcal{O}
\left( p^2 \epsilon^2 \right)$ and implies that the first Weinberg sum rule
could at most be violated starting at order~$\mathcal{O} \left( p^3 \epsilon^2
\right)$.

In this section, we have seen that the WSRs get modified by terms of higher powers in the expansion. To the extent that the~$\xi_k$ and~$\eta_k$ constants are small parameters, these modifications represent small corrections: the right-hand side of~(\ref{WSR2}) no more vanishes, but is suppressed by a factor of order~$\mathcal{O}\left(p^0 \epsilon^4\right)$ relative to individual terms in the left-hand side.
This is similar to the case of QCD, where the second WSR is corrected due to non-zero quark masses, which in $\chi$PT are also viewed as spurions.

\subsection{Discussion}

We have seen in section~\ref{sub:WSRs} that imposing naturalness {\emdash}in
the sense of the spurion formalism of section~\ref{sec:spurions-and-mooses}{\emdash} on an open linear moose model with~$K$ internal sites implies~$K$
Weinberg sum rules at lowest order. At next order, these sum rules receive
corrections as discussed in section~\ref{sub:corrections}. This means that in
the underlying model whose LEET we have been constructing, the order
parameters which give the coefficients in the expansion
(\ref{eq:PiLR-expansion}) are naturally small. Now the particular case where a
given number of sum rules are valid has been studied in
{\cite{Knecht:1998ts}}, where the coupling constants and masses of the vector
and axial resonances are studied assuming parity. What we have displayed here is not
the full theory either, but a possible lagrangian for the low-energy sector,
which means that we are in a position to discuss other Green's functions, not
only two-point ones. Our model is only one particular instance realizing a
given number of Weinberg sum rules within some approximation, but it might
nonetheless be useful in conjunction with the large-$N_c$ expansion of QCD
{\cite{'tHooft:1974jz,Rossi:1977cy,Witten:1979kh}}. In view of such applications, one would
only be interested in the case~$K = 1$ or~$K = 2$.

Other cases with a longer moose would give us low-energy descriptions for
theories where the chiral symmetry breaking pattern is different from that of
QCD, suppressing the order parameters appearing in the expansion
(\ref{eq:PiLR-expansion}). Even though such models do not seem to be relevant
to QCD, they might be helpful in understanding the various patterns of chiral
symmetry breaking. In particular, one would be interested in knowing in which
respects the underlying theory would differ from QCD if such a comparison is
possible. However, such an endeavor is to be carried out outside the domain of
applicability of the effective theory we have built.

\section{Minimal model of EWSB} \label{sub:EWSB-minimal-case}

As already mentioned in section~\ref{sec:generalized-WSRs}, the case of the
linear moose with~$K = 0$ corresponds to the non-linear sigma model and can be
viewed as the minimal symmetry-breaking sector for the electroweak gauge
theory: it only contains the three Goldstone modes required to give masses to
the~$W^{\pm}$ and~$Z^0$ vector bosons and no additional particles, in
particular no physical Higgs boson. Even though the corresponding lowest-order
lagrangian has been used to tackle the heavy-Higgs limit of the SM in
{\cite{Appelquist:1980vg,Longhitano:1980iz,Longhitano:1981tm}} and more
systematically in~{\cite{Nyffeler:1996mb}}, we wish to emphasize that in
general there need not be an underlying scalar resonance to replace the Higgs
and that the approach is indeed much more general. In this section, we focus
on the simplest LEET for the electroweak sector where the Higgs mechanism
occurs dynamically, that is to say, where there is no physical Higgs boson in
the spectrum as a remnant of a complex doublet~{\footnote{Or equivalently, as
the radial component of a two-by-two matrix~$\Sigma$ satisfying a reality
condition~$\Sigma^c = \Sigma$, see section~\ref{sub:SU2-case}.}}.

The point of view of the LEET framework is to construct the effective
lagrangian based on the assumed symmetries of the underlying theory
{\emdash}which we will call `techni-theory'{\emdash} with as little reference
as possible to its details. The denomination techni-theory stems from the
known fact that the effective lagrangian~{\cite{Espriu:1992vm}} for
technicolor~{\cite{Susskind:1979ms}} and other alternative symmetry-breaking
sectors is built around the non-linear sigma model whereas the generalizations
we will be dealing with are based on extensions of the moose. In fact, the
authors of~{\cite{Chivukula:2003wj}} already noticed that moose models
encompass many cases of dynamical electroweak symmetry breaking.

A typical difficulty of technicolor theories~(requiring the introduction of
extended technicolor) is the problem of fermion mass generation
{\cite{Dimopoulos:1979es}} and of the presence of anomalous couplings for
these fermions~{\cite{Appelquist:1985rr,Peccei:1990kr}}. This last point is
also true of BESS models~{\cite{Casalbuoni:1987vq,Casalbuoni:1989xm}}. We
will see that, although the LEET does not deal with the origin of the
constraints applied on the space of gauge configurations, it provides a
rationale for the classification of their effects through the use of spurions,
enabling us to discuss whether or not a given term may appear in the
lagrangian at leading order. It also shows that any theory of dynamical EWSB
{\tmem{a priori}} involves enough weak isospin breaking to provide
mass-splittings within fermion doublets. In this section~we discuss how the
expansion in powers of spurions does address some of the long-standing
questions we have mentioned.

We now describe the leading-order properties of the simplest model one can
imagine for EWSB based on mooses: the low-energy description then only
includes the electroweak gauge fields and the three Goldstone modes that are
required to give masses to the~$W^{\pm}$ and~$Z^0$ bosons. In other words, the
spectrum of the LEET does not include any scalars. This is a well-known case,
but the interest of our formalism here mostly has to do with fermions:
possible anomalous couplings are suppressed by powers of~$\epsilon$. This is a
situation where those terms allowed by the reduced symmetry of the theory
are not allowed by the larger natural symmetry, and can therefore be
consistently treated as being of higher order, the order being given by the
number of spurions involved. Another prediction from our spurion formalism is
that mass-splittings within the doublets automatically appear at the same
order as the mean masses in the doublets, whereas these splittings have always
been difficult to account for in traditional approaches to dynamical
symmetry breaking. In addition, we note that the operators giving a tree-level
$S$ parameter do not appear at leading order. All these results are accounted
for by the same formalism, embodied by the spurion expansion, which was
originally introduced with a different aim: to restore naturalness in a given
limit.

This first model is based on the~$K = 0$ linear moose {\emdash}which has an
$\tmop{SU} \left( 2 \right) \times \tmop{SU} \left( 2 \right)$
symmetry{\emdash} but adds two dynamical~$\tmop{SU}( 2 )$ groups.
Furthermore, we will introduce a~$\mathrm{U} \left( 1 \right)_{B - L}$ group
when we consider fermions. This means that the natural symmetry for this
model, before it is reduced by the constraints imposed on the spurions, is
\begin{eqnarray}
  S_{\tmop{natural}} & = & \tmop{SU} \left( 2 \right)^4 \times \mathrm{U}
  \left( 1 \right)_{B - L} .  \label{eq:Snatural-minimal}
\end{eqnarray}
Applying constraints on the spurions will enable us to reduce this symmetry to
\begin{eqnarray}
  S_{\tmop{reduced}} & = & \tmop{SU} \left( 2 \right) \times \mathrm{U} \left(
  1 \right)_Y . 
\end{eqnarray}
The details of this reduction, as well as the physical consequences are the
subject of this section. As we shall see later on in sections
\ref{sub:WZprimes} and \ref{sub:closed-moose}, many of the results still
hold when we consider extensions of this minimal case. For this reason, the
current section, being central to the paper, is fairly detailed.

\subsection{Complex spurion and~$\mathrm{U} \left( 1 \right)_{\tau^3}$}
\label{sub:one-complex-spurion}

\begin{figure*}\centering
\includegraphics{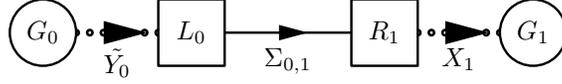}
\caption{Coupling the electroweak gauge fields to the~$K = 0$ moose via spurions.}
\label{fig:EWSB-minimal}
\end{figure*}

The model is obtained from the~$K = 0$ linear moose by coupling gauge fields
to the only Goldstone multiplet as shown in Fig.~\ref{fig:EWSB-minimal}
{\footnote{The notations~$L_0$ and~$R_1$ are just names given to the groups,
and are in no way related to the chirality of fermions to be introduced
later.}}. The two~$\tmop{SU} \left( 2 \right)~$gauge fields transform as
\begin{eqnarray}
  G_{0 \mu} & \longmapsto & G_0 G_{0 \mu} G_0^{\dag} + \frac{\mathi}{g_0} G_0
  \partial_{\mu} G_0^{\dag}, \\
  G_{1 \mu} & \longmapsto & G_1 G_{1 \mu} G_1^{\dag} + \frac{\mathi}{g_1} G_1
  \partial_{\mu} G_1^{\dag},
\end{eqnarray}
and the corresponding transformations commute with the chiral symmetry generated by~$L_0$ and~$R_1$.

The final selection of the~$\tmop{SU} \left( 2 \right) \times \mathrm{U}
\left( 1 \right)$ gauge group that becomes dynamical follows from the
constraints applied on the spurions. It turns out that the two natural
possibilities allowed for the introduction of a two-by-two matrix transforming
under unitary symmetries {\emdash}namely, the case satisfying the reality
condition used before, and the generic case{\emdash} are exactly what we need.

Here, the~$X_1$ spurion is restricted to satisfy the reality condition~$X_1^c
= X_1$, as before, whereas the~$\tilde{Y}_0$ spurion is arbitrary in this
respect. We require both spurions to satisfy the constraint of covariant
constancy, as we did in section~\ref{sub:SU2-case}. The outcome is then
the following: we can perform suitable~$\tmop{SU} \left( 2 \right)_{R_1}$ and
$\tmop{SU} \left( 2 \right)_{L_0}$ transformations such that the spurions
reduce to three real constants and one phase
\begin{eqnarray}
  X_1 |_{\tmop{const.}} & = & \xi_1  \mathbbm{1}_{2 \times 2},
  \label{eq:X1-id}\\
  \left. \tilde{Y}_0 \right|_{\tmop{const.}} & = & \mathe^{\mathi
  \varphi_Y}  \left(\begin{array}{cc}
    \eta_{0 1} & 0\\
    0 & \eta_{0 2}
  \end{array}\right) .  \label{eq:Ytilde0-const}
\end{eqnarray}
The magnitude of these constants then can serve as expansion parameters, and
in addition,~$\tilde{Y}_0$, being diagonal but not proportional to the
identity, allows for weak isospin breaking. Such diagonal matrices have been
used in the literature~{\footnote{See~{\cite{Arkani-Hamed:2001nc}} among
others.}} in order to select a gauge group, but our point is that {\emdash}at
least for the case of reducing~$\tmop{SU} \left( 2 \right)$ to~$\mathrm{U}
\left( 1 \right)$ {\emdash} these constant diagonal matrices can in fact be
traced back to spurions and that an expansion can be built based on that.
Therefore, the possibility of building a systematic expansion is guaranteed,
since the appropriate terms for the renormalization procedure can be constructed with
the help of the spurions~{\cite{Urech:1995hd,Knecht:1998jw}}. Moreover, the
diagonal and constant matrices descend from fields that are considered as
non-propagating at this level, pointing to a dynamical origin for this
subgroup selection.

In the same gauge where the spurions are diagonalized, we get
\begin{eqnarray}
  \left. R_{1 \mu} \right|_{\tmop{const.}} & = & g_1 G_{1 \mu},
  \label{eq:id-R0mu}\\
  \left. L^{1, 2}_{0 \mu} \right|_{\tmop{const.}} & = & G_{0 \mu}^{1, 2}
  \hspace{.2em} = \hspace{.2em} 0,  \label{eq:id-L0mu1,2}\\
  \left. L^3_{0 \mu} \right|_{\tmop{const.}} & = & g_0 G^3_{0 \mu} . 
  \label{eq:id-L0mu3}
\end{eqnarray}
The~$R_{1 \mu}$ connection and the gauge field~$g_1 G_{1
\mu}$ are identified, whereas, due to the complex nature of the~$\tilde{Y}_0$
spurion, only the~$\mathrm{U} \left( 1 \right)$ subgroup of~$G_0$ can be
gauged. In this case, the identification between connections only concerns the
third component: the spurion has selected the~$\mathrm{U} \left( 1
\right)_{\tau^3}$ subgroup.

In order to arrive at this result, the spurions are introduced as follows:
$X_1$ transforms as before
\begin{eqnarray}
  X_1 & \longmapsto & R_1 X_1 G_1^{\dag},  \label{eq:X1-transf}
\end{eqnarray}
with~$G_1$ the~$\tmop{SU} \left( 2 \right)$ gauge transformation, and is
restricted to satisfy, as in~(\ref{eq:constraint-Xk})
\begin{eqnarray}
  X_1^c & = & X_1 .  \label{eq:constraint-X1}
\end{eqnarray}
In order to identify~$G_0$ with the~$\mathrm{U} \left( 1 \right)$ subgroup of
$\tmop{SU} \left( 2 \right)$, we will consider a spurion~$\tilde{Y}_0$ which
is a generic two-by-two matrix-valued function {\emdash}albeit with small
entries{\emdash} transforming as
\begin{eqnarray}
  \tilde{Y}_0 & \longmapsto & G_0  \tilde{Y}_0 L_0^{\dag} . 
  \label{eq:Ytildeo-transf}
\end{eqnarray}
Before we proceed to solve the constraints, we first write all leading-order
terms for the unconstrained lagrangian, that is the terms of order
$\mathcal{O} \left( p^2 \epsilon^0 \right)$. They are collected in the
following lagrangian
\begin{eqnarray}
  \mathcal{L}_{\tmop{bosons}}^{( 2, 0 )} & = & \frac{f^2_0}{4}  \left\langle
  D_{\mu} \Sigma_{0, 1} D^{\mu} \Sigma_{0, 1}^{\dag} \right\rangle -
  \frac{1}{2}  \left\langle G_{1 \mu \nu} G_1^{\mu \nu} \right\rangle \nonumber\\
&-&
  \frac{1}{2}  \left\langle G_{0 \mu \nu} G_0^{\mu \nu} \right\rangle,
  \label{eq:L2-EWSB-min-orig}
\end{eqnarray}
where the covariant derivative operator~$D_{\mu}$ involves both sources~$L_{0
\mu}$ and~$R_{1 \mu}$, but not the dynamical gauge fields~$G_{0 \mu}$ and $G_{1
\mu}$
\begin{eqnarray}
  D_{\mu} \Sigma_{0, 1} & = & \partial_{\mu} \Sigma_{0, 1} - \mathi L_{0 \mu
  \nonesep} \Sigma_{0, 1} + \mathi \Sigma_{0, 1} R_{1 \mu} . 
\end{eqnarray}

We now turn to the constraints: interactions with the gauge fields are
introduced by requiring the spurions to satisfy the usual constraints of
covariant constancy
\begin{eqnarray}
  D_{\mu} X_1 & = & 0, \\
  D_{\mu} \tilde{Y}_0 & = & 0,  \label{eq:DmuYtild0-itself}
\end{eqnarray}
where the covariant derivatives are defined in accordance with
(\ref{eq:X1-transf}) and~(\ref{eq:Ytildeo-transf}).

The constraint is solved as usual for the~$X_1$ spurion, using the
decomposition
\begin{eqnarray}
  X_1 & = & \xi_1 U_1,  \label{eq:decomp-X1}
\end{eqnarray}
and performing an~$\tmop{SU} \left( 2 \right)_{R_1}$ transformation with
\begin{eqnarray}
  R_1 & = & U_1^{\dag},
\end{eqnarray}
to arrive at the expected results~(\ref{eq:X1-id}) and then~(\ref{eq:id-R0mu})
when the constraint is explicitly written. The~$\tmop{SU} \left( 2 \right)$
transformation~$G_1$ remains unconstrained.

As for the~$\tilde{Y}_0$ spurion, we decompose the generic two-by-two matrix
according to
\begin{eqnarray}
  \tilde{Y}_0 & = & \mathe^{\mathi \varphi_Y} G_Y^{\dag} D_Y L_Y,
  \label{eq:decomp-Ytilde0}
\end{eqnarray}
where~$\varphi_Y$ is real and~$G_Y, L_Y$ are elements of~$\tmop{SU} \left( 2
\right)$ and~$D_Y$ is diagonal and real. Since the~$\tau^3$ matrix commutes
with~$D_Y$, the couple~$\left( G_Y, L_Y \right)$ is unique only up to a
diagonal~$\mathrm{U} \left( 1 \right)$ transformation
\begin{eqnarray}
  \left( G_Y, L_Y \right) & \in & \frac{\tmop{SU} \left( 2 \right) \times
  \tmop{SU} \left( 2 \right)}{\mathrm{U} \left( 1 \right)_{\tau^3,
  \tmop{diagonal}}} . 
\end{eqnarray}
One can check that the number of independent real parameters in the right-hand
side of~(\ref{eq:decomp-Ytilde0}) is indeed eight, as it should be. Performing
the following~$\tmop{SU} \left( 2 \right)_{G_0}$ and~$\tmop{SU} \left( 2
\right)_{L_0}$ transformations
\begin{eqnarray}
  G_0 & = & \mathe^{- \mathi f \frac{\tau^3}{2}} G_Y,
  \label{eq:first-G0-transf}\\
  L_0 & = & \mathe^{- \mathi f \frac{\tau^3}{2}} L_Y,
  \label{eq:first-L0-transf}
\end{eqnarray}
we get
\begin{eqnarray}
  \tilde{Y}_0 & = & \mathe^{\mathi \varphi_Y} D_Y,
\end{eqnarray}
independently of the gauge function~$f$. Now, as a consequence of
(\ref{eq:DmuYtild0-itself}) we have
\begin{eqnarray}
  D_{\mu} \left( \tilde{Y}_0  \tilde{Y}_0^{\dag} \right) & = & 0,
\end{eqnarray}
which is explicitly written in the gauge reached by performing the transformations~(\ref{eq:first-G0-transf})
and~(\ref{eq:first-L0-transf}) as
\begin{eqnarray}
  \partial_{\mu} \left( D_Y^2 \right) - \mathi g_0  \left[ G_{0 \mu}, D_Y^2
  \right] & = & 0 . 
\end{eqnarray}
From this we deduce that
\begin{eqnarray}
  D_Y & = & \left(\begin{array}{cc}
    \eta_{0 1} & 0\\
    0 & \eta_{0 2}
  \end{array}\right),  \label{eq:DX}
\end{eqnarray}
must be a constant matrix, yielding~(\ref{eq:Ytilde0-const}). Its two entries
will be considered small, and of the same order as those of the real spurions
{\footnote{This is {\tmem{not}} forced on us by the consistency of the
expansion: one could distinguish between this spurion and others, as they play
a different role.\label{foot:pow-count-Ytilde0}}}
\begin{eqnarray}
  \eta_{0 1}, \eta_{0 2} & = & \mathcal{O} \left( \epsilon \right) . 
\end{eqnarray}
Furthermore, we also find that~$\varphi_Y$ must be a constant. In addition, one
concludes in the generic case where~$\eta_{0 1} \neq \eta_{0 2}$, that
\begin{eqnarray}
  G^{1, 2}_{0 \mu} & = & 0,  \label{eq:G0mu-1,2=0}
\end{eqnarray}
in this gauge. Turning to the equation~(\ref{eq:DmuYtild0-itself}) itself, we
get
\begin{eqnarray}
  g_0 G_{0 \mu} D_Y - D_Y L_{0 \mu}  & = & 0 . 
\end{eqnarray}
With the help of~(\ref{eq:DX}) and~(\ref{eq:G0mu-1,2=0}), this last equation
implies the promised result reproduced in details in~(\ref{eq:id-L0mu1,2}) and~(\ref{eq:id-L0mu3}).

Reaching the appropriate gauge this time involved a~$G_0$ transformation, but
it turned out that an abelian subgroup remained unconstrained: the gauge field is
only restricted to lie in the third direction due to the covariant constancy
conditions. Indeed, when the constraints are applied, the only degree of
freedom left of the six initial real gauge functions in the~$\tmop{SU} \left(
2 \right)_{L_0} \times \tmop{SU} \left( 2 \right)_{G_0}$ transformation is the
$\mathrm{U} \left( 1 \right)$ transformation~$f$. We may thus use the notation
\begin{eqnarray}
  \left. G_{0 \mu}^3 \right|_{\tmop{const.}} & = & b^0_{\mu},
\end{eqnarray}
since the connection is an abelian one, giving the corresponding
field-strength
\begin{eqnarray}
  b^0_{\mu \nu} & = & \partial_{\mu} b^0_{\nu} - \partial_{\nu} b^0_{\mu} . 
\end{eqnarray}
The gauge transformation for~$b^0_{\mu}$, is then
\begin{eqnarray}
  b^0_{\mu} & \longmapsto & b^0_{\mu} - \frac{1}{g_0} \partial_{\mu} f,
\end{eqnarray}
which is compatible with the constraints: we have a~$\mathrm{U} \left( 1
\right)$ symmetry left.
One should not be surprised to find that considering a generic spurion instead
of a real one gives more restrictions on the gauge field configurations.
Indeed, since we have a larger number of parameters, the covariant constancy
equation implies a larger number of relations.

Upon injecting the solutions to the constraints involving the spurions, the
lagrangian~(\ref{eq:L2-EWSB-min-orig}) becomes
\begin{eqnarray}
  \left. \mathcal{L}_{\tmop{bosons}}^{( 2, 0 )} \right|_{\tmop{const.}} &
  = & \frac{f^2_0}{4}  \left\langle \nabla_{\mu} \Sigma_{0, 1} \nabla^{\mu}
  \Sigma_{0, 1}^{\dag} \right\rangle - \frac{1}{2}  \left\langle G_{1 \mu \nu}
  G_1^{\mu \nu} \right\rangle\nonumber\\
& -& \frac{1}{4} b^0_{\mu \nu} b^{0 \mu \nu},
  \label{eq:l2-one-link-EW}
\end{eqnarray}
where the following definition applies
\begin{eqnarray}
  \nabla_{\mu} \Sigma_{0, 1} & = & \left. D_{\mu} \Sigma_{0, 1}
  \right|_{\tmop{const.}} \nonumber\\
  & = & \partial_{\mu} \Sigma_{0, 1} - \mathi g_0 b^0_{\mu}  \frac{\tau^3}{2}
  \Sigma_{0, 1} \nonumber\\
&+& \mathi g_1 \Sigma_{0, 1} G_{1 \mu} . 
\end{eqnarray}
The following terms, yielding a non-zero contribution to the~$S$ parameter
\begin{eqnarray}
  \left\langle L_{0 \mu \nu} \Sigma_{0, 1} R_1^{\mu \nu} \Sigma_{0, 1}^{\dag}
  \right\rangle & = & \mathcal{O} \left( p^4 \epsilon^0 \right), \\
  \left\langle L_{0 \mu \nu} \Sigma_{0, 1} X_1 G_1^{\mu \nu} X_1^{\dag}
  \Sigma_{0, 1}^{\dag} \right\rangle & = & \mathcal{O} \left( p^3 \epsilon^2
  \right), \\
  \left\langle G_{0 \mu \nu}  \tilde{Y}_0 \Sigma_{0, 1} R_1^{\mu \nu}
  \Sigma_{0, 1}^{\dag}  \tilde{Y}_0^{\dag} \right\rangle & = & \mathcal{O}
  \left( p^3 \epsilon^2 \right), \\
  \left\langle G_{0 \mu \nu}  \tilde{Y}_0 \Sigma_{0, 1} X_1 G_1^{\mu \nu}
  X_1^{\dag} \Sigma_{0, 1}^{\dag}  \tilde{Y}_0^{\dag} \right\rangle & = &
  \mathcal{O} \left( p^2 \epsilon^4 \right),
\end{eqnarray}
are absent at leading order: they are only corrections to the original
$\mathcal{O} \left( p^2 \epsilon^0 \right)$ lagrangian
(\ref{eq:L2-EWSB-min-orig}).

Note that, whenever a term with~$\tilde{Y}_0$ is possible, the same term with
$\tilde{Y}_0^c$ can also be written down, as~$\tilde{Y}_0^c$ transforms in the
same manner as~$\tilde{Y}_0$, but is independent from it for a complex
spurion.

In summary, we see that the complex spurion now allows us to take into account
weak isospin breaking. This is obviously what they had been introduced for in
the first place: to restrict the symmetries of the model, and the space of
gauge configurations. However, they turn out to be constant diagonal matrices
in the gauge in which we solve the constraints. The spurions are therefore a
way of introducing weak isospin breaking effects independently of the coupling
constant~$g_0$~{\footnote{Though it will be argued in section
\ref{sub:radcor-mPGBs} that there might be cases where the power counting for
spurions and for coupling constants have to be related.}} in addition to the
breaking due to~$g_0 \neq 0$ itself, a distinction already mentioned in
{\cite{Bagan:1998vu}}. In addition, the spurion formalism provides a
power counting associated with these isospin breaking effects. This will be of
importance when we study fermions and the mass-splittings within the doublets.

\subsection{Bosons} \label{sub:bosons}

At this stage, we want to show that the bosonic sector of this model is
identical, at tree-level, to that of the SM with the Higgs particle removed.
For this purpose, we use field redefinitions as in section
\ref{sub:field-redef}: we stress again that even though these redefinitions
are suitable for the discussion of the physical fields, the procedure does not
involve gauge-fixing. We once again give these redefinitions as they appear
after injection of the solution to the constraints applied on the spurions.
The~$W$ fields are defined through
\begin{eqnarray}
  W_{\mu} & = & \frac{\mathi}{g_1} \Sigma_{0, 1} \nabla_{\mu} \Sigma_{0,
  1}^{\dag},
\end{eqnarray}
in which case the lagrangian does not depend on the Goldstone bosons anymore:
this is the consequence of the Higgs mechanism. This is sufficient for the
$W^{\pm}$ fields, but we still have to work out the mixing of the neutral
fields: the mass matrix in the~$b^0_{\mu}, W_{\mu}^3$ basis is indeed
\begin{eqnarray}
  & \frac{g_1^2 f_0^2}{4}  \left(\begin{array}{cc}
    0 & 0\\
    0 & 1
  \end{array}\right), & 
\end{eqnarray}
while the kinetic term is in the same basis
\begin{eqnarray}
  &  & - \frac{1}{4}  \left(\begin{array}{cc}
    1 + \left( \frac{g_0}{g_1} \right)^2 & \frac{g_0}{g_1}\\
    \frac{g_0}{g_1} & 1
  \end{array}\right) . 
\end{eqnarray}
This is rewritten in canonical form using the definitions
\begin{eqnarray}
  W^{\pm}_{\mu} & = & \frac{\mathi \sqrt{2}}{g_1}  \left\langle \tau^{\pm} \Sigma_{0,
  1} \nabla_{\mu} \Sigma_{0, 1}^{\dag} \right\rangle,  \label{eq:def-W+-mu}\\
  A_{\mu} & = & s \frac{\mathi}{g_1}  \left\langle \tau^3 \Sigma_{0, 1}
  \nabla_{\mu} \Sigma_{0, 1}^{\dag} \right\rangle + \frac{1}{c} b^0_{\mu}, \\
  Z_{\mu} & = & c \frac{\mathi}{g_1}  \left\langle \tau^3 \Sigma_{0, 1}
  \nabla_{\mu} \Sigma_{0, 1}^{\dag} \right\rangle,  \label{eq:def-Zmu}
\end{eqnarray}
with
\begin{eqnarray}
  c & = & \frac{1}{\sqrt{1 + \left( \frac{g_0}{g_1} \right)^2}}, \\
  s & = & \sqrt{1 - c^2} . 
\end{eqnarray}
The above definitions are useful in order to show that the lagrangian does not
involve the Goldstone fields anymore when written with the appropriate
variables. In order to relate this to the usual diagonalization in the SM, one
may rewrite these field redefinitions in the particular case where~$\pi^a = 0$~{\footnote{To be
more general, one may assume that this condition of setting the Goldstone
modes to zero is achieved by gauge-fixing as one would do to define the
unitary gauge in the standard manner for the SM.}} for~$a =
1, 2, 3$ in
\begin{eqnarray}
  \Sigma_{0, 1} & = & \mathe^{\mathi \frac{\pi^a \tau^a}{f_0}},
\end{eqnarray}
to find
\begin{eqnarray}
  \left. A_{\mu} \right|_{\pi^a = 0} & = & cb_{\mu}^0 + sG^3_{1 \mu}, \\
  \left. Z_{\mu} \right|_{\pi^a = 0} & = & - sb_{\mu}^0 + cG^3_{1 \mu} . 
\end{eqnarray}
One recognizes the SM tree-level formulas after performing the replacement
$g_0 \longmapsto g'$ and~$g_1 \longmapsto g$. The value of the electric charge
is as expected
\begin{eqnarray}
  e & = & \frac{g_0 g_1}{\sqrt{g_0^2 + g_1^2}} . 
\end{eqnarray}
The kinetic term for the Goldstone bosons yields the following mass term
\begin{eqnarray}
  \frac{f^2_0}{4}  \left\langle \nabla_{\mu} \Sigma_{0, 1} \nabla^{\mu}
  \Sigma_{0, 1}^{\dag} \right\rangle & = & M_W^2  \left\langle W_{\mu}^+
  W_{\mu}^- \right\rangle \nonumber\\
&+& \frac{1}{2} M_Z^2  \left\langle Z_{\mu} Z^{\mu}
  \right\rangle,
\end{eqnarray}
with the SM-like definitions
\begin{eqnarray}
  M_W^2 & = & \frac{g_1^2}{4} f_0^2, \\
  M_Z^2 & = & \frac{g_1^2 + g_0^2}{4} f_0^2 . 
\end{eqnarray}
\subsection{Fermions and~$\mathrm{U} \left( 1 \right)_{B - L}$}
\label{sub:U1-fermions}

As already mentioned in~(\ref{eq:Snatural-minimal}), the natural symmetry of
this model involves a~$\mathrm{U} \left( 1 \right)_{B - L}$ group which is
relevant for fermions. This is the subject of this section: within this
minimal case, we want to describe the introduction of fermions charged under
$\mathrm{U} \left( 1 \right)_{B - L}$. We denote the~$\mathrm{U} \left( 1
\right)_{B - L}$ connection by~$B_{\mu}$, transforming as
\begin{eqnarray}
  B_{\mu} & \longmapsto & B_{\mu} - \partial_{\mu} \beta^0 . 
\end{eqnarray}
At this stage,~$B_{\mu}$ is merely an external source~(as are~$L_{0 \mu}$ and
$R_{1 \mu}$) and this is why the corresponding kinetic term does not appear in
the lagrangian. The appropriate power counting is
\begin{eqnarray}
  B_{\mu} & = & \mathcal{O} \left( p^1 \epsilon^0 \right), \\
  B_{\mu \nu} & = & \mathcal{O} \left( p^2 \epsilon^0 \right),
\end{eqnarray}
as can be guessed from~(\ref{eq:pow-count-sources}).

\begin{figure*}\centering
\includegraphics{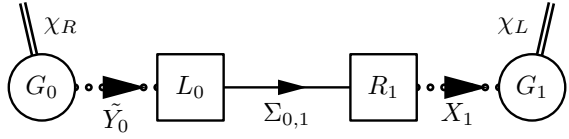}
\caption{Introduction of fermions in the model of section~\ref{sub:one-complex-spurion}.}
\label{fig:one-link-fermions}
\end{figure*}

We will be considering elementary chiral fermion doublets~$\chi_L$ and
$\chi_R$, transforming with respect to the natural symmetry group
$S_{\tmop{natural}}$~(\ref{eq:Snatural-minimal}) as
\begin{eqnarray}
  \chi_L & \longmapsto & G_1 \mathe^{- \mathi \frac{B - L}{2} \beta^0} \chi_L,
  \label{eq:trsf-chiL}\\
  \chi_R & \longmapsto & G_0 \mathe^{- \mathi \frac{B - L}{2} \beta^0} \chi_R,
  \label{eq:trsf-chiR}
\end{eqnarray}
where we have introduced the two four-component projections
$\chi_L, \chi_R$ of the Dirac spinor doublet~$\chi$ which has eight complex
components. The definitions are as usual
\begin{eqnarray}
  \chi_{L, R} & = & \frac{1 \mp \gamma_5}{2} \chi . 
\end{eqnarray}
The transformation properties of these elementary fermions, except for the
$\mathrm{U} \left( 1 \right)$ transformation, are schematized in figure
\ref{fig:one-link-fermions}.

Another possibility would be to consider fermions transforming under~$L_0$ and
$R_1$ instead of~$G_0$ and~$G_1$. Such fields would then have to be
interpreted as bound states resulting from the dynamics of the techni-theory.
Hence, they should be considered as composite as opposed to the elementary
fermions we have just introduced above and which transform under the weak
gauge groups.

Note that, as was noticed in~{\cite{Wudka:1994ny,Nyffeler:1999ap}} for
small-momentum power counting purposes, the following counting for chiral
fermions has to be imposed
\begin{eqnarray}
  \chi_L, \chi_R & = & \mathcal{O} \left( p^{1 / 2} \right) . 
  \label{eq:pow-count-chi}
\end{eqnarray}
This can be deduced for instance by requiring that the kinetic terms of these
fermions appear at the same order as the kinetic terms for the bosons, thereby
allowing them to be part of the effective theory as dynamical fields at the
same level, or alternatively by inspection of the normalization of states.
Both these conditions imply the result~(\ref{eq:pow-count-chi}), with the
proviso that we consider massless or light fermions, again in connection with
naturalness.

We have introduced the additional~$\mathrm{U} \left( 1 \right)_{B - L}$
symmetry, but it turns out we will be interested in the case where the
corresponding gauge fields is identified with the one associated with the
$\mathrm{U} \left( 1 \right)_{\tau^3}$ of section
\ref{sub:one-complex-spurion}. As should be clear from what we have done up to
now, we are going to impose this identification through our usual constraint of
covariant constancy on a spurion. For this purpose, we introduce another
non-propagating field: a complex doublet transforming as~{\footnote{This
complex doublet is not to be confused with the Higgs doublet of the SM, which
would transform under~$G_1$. As we shall also see in section
\ref{sub:fermion-masses}, it does not have the same impact on fermion
masses.}}
\begin{eqnarray}
  \phi & \longmapsto & G_0 \mathe^{\mathi \frac{\beta^0}{2}} \phi . 
\end{eqnarray}
Our choice for the~$\mathrm{U} \left( 1 \right)$ charge of this doublet is
done in order to have the proper normalization for the~$B - L$ charges once
the constraints are solved and the ensuing identification between~$\mathrm{U}
\left( 1 \right)$ connections is performed. The condition of covariant
constancy to be imposed reads explicitly
\begin{eqnarray}
  D_{\mu} \phi \hspace{.2em} = \hspace{.2em} \partial_{\mu} \phi - \mathi \left( g_0 G_{0 \mu}
  - \frac{B_{\mu}}{2} \right) \phi & = & 0 .  \label{eq:Dmuphi=0}
\end{eqnarray}
We next assume that the choice of gauge described in section
\ref{sub:one-complex-spurion} has already been performed and the constraint on
$\tilde{Y}_0$ has already been solved for. The result is then that we obtain
(\ref{eq:Ytilde0-const}),~(\ref{eq:id-L0mu1,2}) and~(\ref{eq:id-L0mu3}). In
order to diagonalize~$\tilde{Y}_0$, we have used a~$G_0$ transformation as
specified in~(\ref{eq:first-G0-transf}). Now, solving~(\ref{eq:Dmuphi=0}) with
our usual method of choosing the simplest gauge appears impossible: it seems
{\tmem{a priori}} that the problem is overdetermined. Thanks to the constraint
themselves, which restrict the gauge configurations to belong to the
$\mathrm{U} \left( 1 \right)$ subgroup of~$\tmop{SU} \left( 2 \right)$, it is
in fact possible to find a solution, resulting in
\begin{eqnarray}
  \left. \phi \right|_{\tmop{const.}} & = & \left(\begin{array}{c}
    \zeta\\
    0
  \end{array}\right),  \label{eq:phi|constraints}
\end{eqnarray}
and
\begin{eqnarray}
  \left. B_{\mu} \right|_{\tmop{const.}} & = & g_0 G_{0 \mu}^3 \hspace{.2em} =
  \hspace{.2em} g_0 b^0_{\mu},  \label{eq:Bmu|const=}
\end{eqnarray}
where~$\zeta$ is a real constant to which we apply for the sake of simplicity
the same power counting as to the other spurions: it is counted as
$\mathcal{O} \left( \epsilon \right)$~{\footnote{The same remark as in
footnote \ref{foot:pow-count-Ytilde0} applies.\label{foot:pow-count-phi}}}. We
will also see that~(\ref{eq:G0mu-1,2=0}) still holds.

To show this, we use the fact that a complex doublet can always be decomposed in
terms of a real function~$\zeta$ and an~$\tmop{SU} \left( 2 \right)$ matrix
$U_Z$ as
\begin{eqnarray}
  \phi & = & U_Z  \left(\begin{array}{c}
    \zeta\\
    0
  \end{array}\right) . 
\end{eqnarray}
We then perform the following~$\tmop{SU} \left( 2 \right)_{G_0} \times
\mathrm{U} \left( 1 \right)_{B - L}$ transformation
\begin{eqnarray}
  G_0 & = & \mathe^{- \mathi f'  \frac{\tau^3}{2}} U_Z^{\dag},
  \label{eq:G0-f}\\
  \beta^0 & = & f',  \label{eq:beta0-f}
\end{eqnarray}
with~$f'$ a gauge function to be solved for later. When this transformation is
applied,~$\phi$ reduces to the real function~$\zeta$ in its upper component,
and zero in the lower component. We then write out explicitly the components
of the constraint equation~(\ref{eq:Dmuphi=0}) to find that~$\zeta$ is indeed
a constant, which is the result~(\ref{eq:phi|constraints}). We also get in
this gauge
\begin{eqnarray}
  G_{0 \mu}'^{1, 2} & = & 0, \\
  g_0 G_{0 \mu}'^3 & = & B_{\mu} .  \label{eq:Bmu=}
\end{eqnarray}
Now we see that this gauge is related to the one used in section
\ref{sub:one-complex-spurion} by
\begin{eqnarray}
  G'_{0 \mu} & = & VG_{0 \mu} V^{\dag} + \frac{\mathi}{g_0} V \partial_{\mu}
  V^{\dag},
\end{eqnarray}
where the gauge transformation~$V$ is given by
\begin{eqnarray}
  V & = & \mathe^{- \mathi f'  \frac{\tau^3}{2}} U_Z^{\dag} G_Y^{\dag}
  \mathe^{\mathi f \frac{\tau^3}{2}} . 
\end{eqnarray}
However, since both~$G'_{0 \mu}$ and~$G_{0 \mu}$ point in the third direction,
the transformation~$V$ only involves the exponential of a function times
$\tau^3$: it is a~$\mathrm{U} \left( 1 \right)$ transformation. Therefore, by
appropriately choosing~$f'$ depending on the function~$f$, we can set this
function to be zero, that is
\begin{eqnarray}
  V & = & 1 . 
\end{eqnarray}
We are then left with only one free gauge function~$f$.

In summary, due to the constraints themselves, we have been able to diagonalize
both spurions in the same gauge, giving
\begin{eqnarray}
  G_{0 \mu}' & = & G_{0 \mu} . 
\end{eqnarray}
Injecting this into~(\ref{eq:Bmu=}), we get the desired result
(\ref{eq:Bmu|const=}).

To summarize, we make a list of useful results from this section~and section
\ref{sub:one-complex-spurion}: we have shown that there exists a gauge in
which the spurions reduce to the following constants, assumed for simplicity
to be~$\mathcal{O} \left( \epsilon \right)$
\begin{eqnarray}
  \left. X_1 \right|_{\tmop{const.}} & = & \xi_1  \mathbbm{1}_{2 \times
  2}, \\
  \left. \tilde{Y}_0 \right|_{\tmop{const.}} & = & \mathe^{\mathi
  \varphi_Y}  \left(\begin{array}{cc}
    \eta_{0 1} & 0\\
    0 & \eta_{0 2}
  \end{array}\right), \\
  \left. \phi \right|_{\tmop{const.}} & = & \left(\begin{array}{c}
    \zeta\\
    0
  \end{array}\right) . 
\end{eqnarray}
This result is of paramount importance since this is the standard gauge we will be using
each time we want to inject the solution to the constraints in order to see the physical content of the terms. This gauge is
reached by fixing nine gauge functions among the thirteen available for
generic~$\tmop{SU} \left( 2 \right)_{G_0} \times \tmop{SU} \left( 2
\right)_{L_0} \times \tmop{SU} \left( 2 \right)_{R_1} \times \tmop{SU} \left(
2 \right)_{G_1} \times \mathrm{U} \left( 1 \right)_{B - L}$ transformations.
We are then left with an~$\tmop{SU} \left( 2 \right) \times \mathrm{U} \left(
1 \right)_Y$ invariance. The~$\mathrm{U} \left( 1 \right)_Y$ degree of freedom
is associated with the gauge function~$f$. In this gauge, the following
relations between connections hold
\begin{eqnarray}
  \left. R^a_{1 \mu} \right|_{\tmop{const.}} & = & g_1 G^a_{1 \mu},
  \hspace{.2em} \tmop{for} \hspace{.2em} a = 1, 2, 3, \\
  \left. L^{1, 2}_{0 \mu} \right|_{\tmop{const.}} & = & G_{0 \mu}^{1, 2}
  \hspace{.2em} = \hspace{.2em} 0, \\
  \left. \hspace{.2em} L^3_{0 \mu} \right|_{\tmop{const.}} & = & \left. \left.
  g_0 G_{0 \mu}^3 \right|_{\tmop{const.}} \hspace{.2em} = \hspace{.2em} B_{\mu}
  \right|_{\tmop{const.}} \nonumber\\
&=& g_0 b_{\mu}^0,
\end{eqnarray}
where~$G_{1 \mu}^a$ and~$b_{\mu}^0$ are the only gauge fields which are
non-zero.~$b_{\mu}^0$ transforms as a~$\mathrm{U} \left( 1 \right)$ connection
\begin{eqnarray}
  b_{\mu}^0 & \longmapsto & b_{\mu}^0 - \frac{1}{g_0} \partial_{\mu} f . 
\end{eqnarray}
Although for notational convenience we count each power of spurion amplitudes
$\xi_1, \eta_{0 1}, \eta_{0 2}$ and~$\zeta$ as~$\mathcal{O} \left( \epsilon
\right)$, it should be stressed once more that the final counting of various
spurions can be consistently determined only taking into account loops in a
full analysis of various symmetry breaking effects. This is however beyond the
scope of this paper.

\subsection{Fermion couplings}

With the notations of the preceding section, we find the following possible
terms involving fermions at order~$\mathcal{O} \left( p^2 \epsilon^0 \right)$
\begin{eqnarray}
  \mathcal{L}^{\left( 2, 0 \right)}_{\tmop{fermions}} & = & \mathi
  \overline{\chi_L} \gamma^{\mu} D_{\mu} \chi_L + \mathi \overline{\chi_R}
  \gamma^{\mu} D_{\mu} \chi_R \nonumber\\
&+& \text{four-fermion interactions} . 
  \label{eq:L20-fermions}
\end{eqnarray}
We want to show that the couplings with vector fields at lowest-order are
automatically identical to those in the SM. We start from
\begin{eqnarray}
  D_{\mu} \chi_L & = & \partial_{\mu} \chi_L \nonumber\\
&-& \mathi \left( g_1 G_{1 \mu} +
  \frac{B - L}{2} B_{\mu} \right) \chi_L, \\
  D_{\mu} \chi_R & = & \partial_{\mu} \chi_R \nonumber\\
&-& \mathi \left( g_0 G_{0 \mu} +
  \frac{B - L}{2} B_{\mu} \right) \chi_R,  \label{eq:Dmu-chiR}
\end{eqnarray}
which become upon application of the constraints and in the standard gauge
\begin{eqnarray}
  \nabla_{\mu} \chi_L & = & \left. D_{\mu} \chi_L \right|_{\tmop{const.}}
  \nonumber\\
  & = & \partial_{\mu} \chi_L \nonumber\\
&-& \mathi \left( g_1 G_{1 \mu} + g_0  \frac{B -
  L}{2} b^0_{\mu} \right) \chi_L , \\
  \nabla_{\mu} \chi_R & = & \left. D_{\mu} \chi_R \right|_{\tmop{const.}}
  \nonumber\\
  & = & \partial_{\mu} \chi_R \nonumber\\
&-& \mathi g_0 b^0_{\mu}  \left( \frac{\tau^3}{2}
  + \frac{B - L}{2} \right) \chi_R . 
\end{eqnarray}
Using unitary rotations involving the~$\tmop{SU} \left( 2 \right)$
matrix-valued field~$U_1$ as defined in~(\ref{eq:decomp-X1}), we next perform
the following fields redefinitions for fermions
\begin{eqnarray}
  \psi_L & = & \Sigma_{0, 1} U_1 \chi_L,  \label{eq:psiL-def}\\
  \psi_R & = & \chi_R,  \label{eq:psiR-def}
\end{eqnarray}
which are inoffensive from the point of view of anomalies as long as we have
three quarks for one lepton, since the trace of~$B - L$ over fermions is then
equal to zero. Denoting
\begin{eqnarray}
  \psi & = & \psi_L + \psi_R,
\end{eqnarray}
and using the field redefinitions we have introduced for vector fields in
(\ref{eq:def-W+-mu}-\ref{eq:def-Zmu}), we find that~(\ref{eq:L20-fermions})
becomes upon application of the constraints in the standard gauge
\begin{eqnarray}
  \left. \mathcal{L}^{\left( 2, 0 \right)}_{\tmop{fermions}}
  \right|_{\tmop{const.}} & = & \mathi \overline{\psi} \gamma^{\mu}
  \partial_{\mu} \psi + e \overline{\psi} \gamma^{\mu} Q \psi A_{\mu}
  \nonumber\\
  &  & + \frac{e}{cs}  \overline{\psi} \gamma^{\mu}  \left\{ \frac{\tau^3}{2}
  \frac{\left( 1 - \gamma_5 \right)}{2} - s^2 Q \right\} \psi Z_{\mu}
  \nonumber\\
  &  & + \frac{1}{\sqrt{2}} \frac{e}{s}  \overline{\psi} \gamma^{\mu}
  \tau^{\mp}  \frac{\left( 1 - \gamma_5 \right)}{2} \psi W_{\mu}^{\pm}
  \hspace{0.25em},
\end{eqnarray}
with
\begin{eqnarray}
  Q & = & \frac{\tau^3}{2} + \frac{B - L}{2} . 
\end{eqnarray}
This is the desired SM-like result: anomalous couplings would come from terms
such as
\begin{eqnarray}
  \mathi \overline{\chi_L} \gamma^{\mu} X_1^{\dag} \Sigma_{0, 1}^{\dag} \left(
  D_{\mu} \Sigma_{0, 1} \right) X_1 \chi_L & = & \mathcal{O} \left( p^2
  \epsilon^2 \right),  \label{eq:anomal-interac}\\
  \mathi \overline{\chi_R} \gamma^{\mu}  \tilde{Y}_0 \Sigma_{0, 1} \left(
  D_{\mu} \Sigma_{0, 1}^{\dag} \right)  \tilde{Y}^{\dag}_0 \chi_R & = &
  \mathcal{O} \left( p^2 \epsilon^2 \right),
\end{eqnarray}
and are therefore of higher order in the spurion expansion. This is
automatically obtained in our spurion formalism without any additional
assumptions and it represents a new result compared to the literature~{\cite{Appelquist:1985rr,Casalbuoni:1987vq,Casalbuoni:1989xm,Peccei:1990kr,Bagan:1998vu}}. It should be stressed that such terms would result in a
violation of the universality of left-handed couplings, and would also allow
couplings of the right-handed fermions to the~$W^\pm$. It is therefore very
satisfactory to find that they are automatically suppressed as a by-product of
our spurion formalism, which we have introduced in relation with other
considerations pertaining to naturalness, and which in addition allows us to
describe the breaking of weak isospin.

Anomalous magnetic moment terms will also be suppressed in the spurion
expansion, and are suppressed by an unknown dimensionful scale, as
are the four-fermion interactions.

Since the lowest-order couplings of the fermions are as in the SM and since we
have recovered the same tree-level relations as in the SM for the bosonic
sector {\emdash}in particular, one may notice that the custodial symmetry
{\cite{Sikivie:1980hm}} is implemented by the~$\tmop{SU}( 2 )_{L_0}$
group{\emdash} it should then be no surprise that an explicit calculation
gives zero tree-level values for the~$S, T, U$ parameters
{\cite{Peskin:1992sw}} in this model.

\subsection{Fermion masses} \label{sub:fermion-masses}

In order to construct fermion mass terms invariant under the whole symmetry~$S_{\tmop{natural}}$, and given the transformation properties of the fermions
(\ref{eq:trsf-chiL}) and~(\ref{eq:trsf-chiR}), spurions are necessary. In this
section, we concentrate on the lowest-order terms~$\mathcal{O} \left( p^1
\epsilon^2 \right)$, i.e. quadratic forms without derivatives, with two
spurion insertions. One may conceive a joint power counting for momenta and
spurions such that the kinetic term~$\mathcal{O} \left( p^2 \epsilon^0
\right)$ and the mass term~$\mathcal{O} \left( p^1 \epsilon^2 \right)$ would
appear at the same order. We first consider quarks, and use~$i, j$ as
generation indices. The most general~$\mathcal{O} \left( p^1 \epsilon^2
\right)$ mass term reads
\begin{eqnarray}
  \mathcal{L}^{\left( 1, 2 \right)}_{\tmop{quarks}} & = & - m_{1 i j} 
  \overline{\chi_L}_i X_1^{\dag} \Sigma_{0, 1}^{\dag}  \tilde{Y}_0^{\dag}
  \chi_{R j}\nonumber\\
&-& m^{\ast}_{1 i j}  \overline{\chi_R}_j  \tilde{Y}_0 \Sigma_{0,
  1} X_1 \chi_{L i} \nonumber\\
  & - &  m_{2 i j}  \overline{\chi_L}_i X_1^{\dag} \Sigma_{0, 1}^{\dag} 
  \tilde{Y}_0^{c \dag} \chi_{R j} \nonumber\\
&-& m^{\ast}_{2 i j}  \overline{\chi_R}_j 
  \tilde{Y}_0^c \Sigma_{0, 1} X_1 \chi_{L i} . 
  \label{eq:minimal-fermion-mass}
\end{eqnarray}
Using the notations
\begin{eqnarray}
  m^u_{i j} & = & \xi_1  \left( \eta_{0 1} \mathe^{- \mathi \varphi_Y} m_{1 i
  j} + \eta_{0 2} \mathe^{\mathi \varphi_Y} m_{2 i j} \right), \\
  m^d_{i j} & = & \xi_1  \left( \eta_{0 2} \mathe^{- \mathi \varphi_Y} m_{1 i
  j} + \eta_{0 1} \mathe^{\mathi \varphi_Y} m_{2 i j} \right), \\
  \psi_i & \longrightarrow & \left(\begin{array}{c}
    u_i\\
    d_i
  \end{array}\right),
\end{eqnarray}
equation~(\ref{eq:minimal-fermion-mass}) becomes, when the constraints are
solved
\begin{eqnarray}
  \left. \mathcal{L}^{\left( 1, 2 \right)}_{\tmop{quarks}}
  \right|_{\tmop{const.}} & = & -  m^u_{i j}  \overline{u_L}_i u_{R
  j} - m^{u \ast}_{i j}  \overline{u_R}_j u_{L i} \nonumber\\
&-& m^d_{i j} 
  \overline{d_L}_i d_{R j} - m^{d \ast}_{i j}  \overline{d_R}_j d_{L i}.
    \label{eq:quark-masses-SM}
\end{eqnarray}
In the generic power counting, both the masses and the mass-splittings within
doublets will be counted as~$\mathcal{O} \left( p^0 \epsilon^2 \right)$. From
the notation in~(\ref{eq:quark-masses-SM}) we see that the masses and
splittings for different doublets are independent of each other, allowing us
to account for the masses of the fermions in a satisfactory way without the
presence of physical scalars in the spectrum. We remark that the freedom is
the same as in the SM, concerning both the magnitude and phases of the
coefficients: the additional parameters~$\xi_1$, $\eta_{0 1}$ and $\eta_{0 2}$ set
the scale of the masses, whereas the phase~$\varphi_Y$ can be absorbed in a
redefinition of the matrices~$m_1$ and~$m_2$. Despite the fact that here the
origin of the fermion masses is not the standard Higgs mechanism, the pattern
of CKM mixing and $C P$ violation is identical to the SM: no predictivity is lost
or gained at this level.

In the lepton sector, we can also write down terms analogous to
those of equation~(\ref{eq:minimal-fermion-mass}). In addition, one can
construct new operators involving the leptons and which are completely invariant
under~$S_{\tmop{natural}}$. These are the first terms we encounter involving
the spurion~$\phi$
\begin{eqnarray}
  N_{R i} & = & \phi^{\dag} \chi_{R i} \hspace{.2em} = \hspace{.2em} \mathcal{O} \left( p^{1 /
  2} \epsilon^1 \right), \\
  N_{L i} & = & \phi^{\dag}  \tilde{Y}_0 \Sigma_{0, 1} X_1 \chi_{L i} \hspace{.2em} =
  \hspace{.2em} \mathcal{O} \left( p^{1 / 2} \epsilon^3 \right),
\end{eqnarray}
and a similar term with the replacement~$\tilde{Y}_0 \longmapsto
\tilde{Y}_0^c$. We have given the simplest power counting for these operators,
assuming a common power counting for all the spurions. However, the fact that
the spurion description is common to all the symmetry-reducing mechanisms does
not imply that the spurions must have the same power counting. In particular,
one may entertain the view that the~$\phi$ spurion should be counted at a
different level. In the remainder of this section, we will therefore give the powers of~$\zeta, \xi_1$ and~$\eta_0$ (where we have generically used~$\eta_0$ for both~$\eta_{0 1}$ and
$\eta_{0 2}$) which appear in the operators, instead of the powers of
$\epsilon$.

With the~$N_{L, R}$ operators, we can write down Lorentz-invariant terms
violating lepton number. We recall that, up to now, we have not encountered any
interactions of the right-handed neutrinos. To elucidate the physical
content of the new terms involving~$N_R$ and~$N_L$, we
use the following notation for the lepton sector
\begin{eqnarray}
  \psi_i & \longrightarrow & \left(\begin{array}{c}
    \nu_i\\
    e_i
  \end{array}\right) .  \label{eq:lepton-doublets}
\end{eqnarray}
Using the definitions~(\ref{eq:psiL-def}) and~(\ref{eq:psiR-def}) for the
fermion fields in the unitary gauge and solving the constraints then leads to
\begin{eqnarray}
  \left. N_{R i} \right|_{\tmop{const.}} & = & \zeta
    \nu_{R i}
,  \label{eq:NR}\\
  \left. N_{L i} \right|_{\tmop{const.}} & = & \mathe^{\mathi \varphi_Y}
  \zeta \xi_1 \eta_{0 1}     \nu_{L i}
 .  \label{eq:NL}
\end{eqnarray}
These operators project out the neutrino components of the doublet, up to
multiplications by constant factors. From~(\ref{eq:NR}) and~(\ref{eq:NL}), we
deduce that the right-handed Majorana masses will be proportional to
$\zeta^2$, while the left-handed Majorana masses will contain factors of
$\zeta^2 \xi_1^2 \eta_0^2$. We can also construct additional Dirac mass-terms
of order~$\zeta^2 \xi_1 \eta_0$ for neutrinos, whereas the ones analogous to the mass matrix
for the quarks~(\ref{eq:minimal-fermion-mass}) are proportional to~$\xi_1
\eta_0$. All the new terms are absent in the case of quarks due to
non-invariance with respect to~$\mathrm{U} \left( 1 \right)_{B - L}$: only for
leptons do they respect the full~$S_{\tmop{natural}}$ symmetry, opening the
possibility for lepton number violation driven by the neutrino sector. This
would in fact be a place to evaluate the orders of magnitude of the various
spurions when confronted to data on lepton flavor violation; however, we will
leave this question open at this stage.

In addition to this, we have learned how the~$\phi$ spurion is related to
naturalness: it introduces lepton number violation. This is indeed the only
physical consequence of the~$\phi$ spurion at this level.

To summarize, we have seen that the minimal model for EWSB, which involves the
dynamical Higgs mechanism, and no physical Higgs boson, fits quite nicely in
our framework: the~$\mathrm{U} \left( 1 \right)_Y$ gauge group is introduced
naturally within the spurion formalism, which gives us a possibility to
properly account for the masses of the fermions while suppressing non-standard
terms. With this firmer basis for the neglect of unwanted terms, we may then
embark on a more thorough study of the effective theory expansion for this
minimal case, along the lines of~{\cite{Nyffeler:1999ap}}. This is however
outside the scope of this paper.

\section{Two examples of non-minimal Higgs-less models}
\label{sec:EWSB-extended-cases}

The fact that all particles that are considered light must be included in our
effective lagrangian implies that there is no single effective lagrangian once
one allows for new particles not too far above the weak scale: we will give
examples of such cases in this section. In fact we will first study in this
section a case were there are vector resonances in the low-energy spectrum
in addition to the minimal particle content, and then one where there is a
triplet of PGBs: barring supersymmetry, the only known way to protect physical
scalar masses from radiative corrections is for them to be PGBs. This last
scenario still deserves the name of Higgs-less model, as these scalars are not
related to the SM Higgs.

\subsection{Case with~$W'$ and~$Z'$} \label{sub:WZprimes}

The previous section~was a showcase to introduce our method of using linear
mooses equipped with spurions as toy-models of EWSB. We now want to give some
insight into the next simplest case which involves an additional gauge
multiplet~($K = 1$). This model will then include excited vectors~$W'$ and~$Z'$. The idea is to discuss the simplest possibility that gives corrections
to the SM-like relations we have seen in section~\ref{sub:EWSB-minimal-case}:
corrections appear in this case at tree-level due to these excited vectors,
which are assumed to be heavier than the~$W^\pm$ and~$Z$, but still lighter
than the next resonances. The generalization to models with more and more
excited vectors is always possible, and may be motivated by considerations of
tree-level unitarity~{\cite{SekharChivukula:2001hz,Csaki:2003dt}}, which we
will not get into here.

\begin{figure*}\centering
\includegraphics{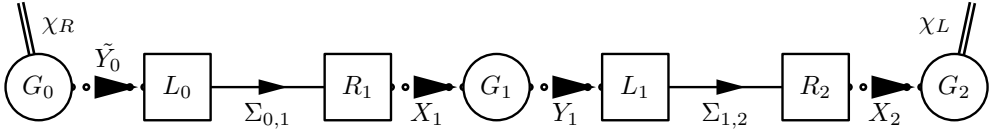}
\caption{A model for EWSB with one internal site.}
\label{fig:WZprimes}
\end{figure*}

Our model is depicted in Fig.~\ref{fig:WZprimes}, with the elementary
fermions already included. The transformation properties of the fermion
doublets are thus the following
\begin{eqnarray}
  \chi_L & \longmapsto & G_2 \mathe^{- \mathi \frac{B - L}{2} \beta^0} \chi_L,
  \\
  \chi_R & \longmapsto & G_0 \mathe^{- \mathi \frac{B - L}{2} \beta^0} \chi_R,
\end{eqnarray}
and the definitions are as before for all the fields.

We will adopt the view that the internal gauge group, which has a particular
status since it describes the low-energy sector of the techni-theory, can be
assumed to have a stronger gauge coupling than the two weak gauge fields
without incoherences
\begin{eqnarray}
  \frac{g_0}{g_1}, \frac{g_2}{g_1} & \ll & 1.  \label{eq:small-gs}
\end{eqnarray}
We will then give the final results as an expansion in~$1 / g_1$ so that it is
clear how we recover the limit of section~\ref{sub:EWSB-minimal-case} when
$g_1$ goes to infinity, as one would expect. Our model is akin to that of
{\cite{Casalbuoni:1987vq,Casalbuoni:1995yb}}, but without reference to the
`hidden symmetry' assumption~{\cite{Bando:1988br}}. Furthermore, our spurion
formalism enables us to describe the fermion sector~(masses and couplings to
vector bosons) without additional assumptions: indeed, we have seen in section
\ref{sub:EWSB-minimal-case} that anomalous couplings are suppressed by powers
of~$\epsilon$, and this result will still hold here.

We focus for simplification on the case where~$f_1 = f_0 = f$ at tree-level,
in order to avoid unreadable formulas.

The natural symmetry of this model is
\begin{eqnarray}
  S_{\tmop{natural}} & = & \tmop{SU} \left( 2 \right)^7 \times \mathrm{U}
  \left( 1 \right)_{B - L},
\end{eqnarray}
which gets reduced via the constraints on the spurions, to
\begin{eqnarray}
  S_{\tmop{reduced}} & = & \tmop{SU} \left( 2 \right)^2 \times \mathrm{U}
  \left( 1 \right)_Y . 
\end{eqnarray}

\subsubsection{Physical fields}

From the above definitions, we find that the~$\mathcal{O} \left( p^2
\epsilon^0 \right)$ lagrangian is
\begin{eqnarray}
  \mathcal{L}^{\left( 2, 0 \right)} & = & \frac{f^2}{4}  \left\langle D_{\mu}
  \Sigma_{0, 1} D^{\mu} \Sigma_{0, 1}^{\dag} \right\rangle + \frac{f^2}{4} 
  \left\langle D_{\mu} \Sigma_{1, 2} D^{\mu} \Sigma_{1, 2}^{\dag}
  \right\rangle \nonumber\\
  &   -& \frac{1}{2}  \left\langle G_{0 \mu \nu} G_0^{\mu \nu} \right\rangle
  - \frac{1}{2}  \left\langle G_{1 \mu \nu} G_1^{\mu \nu} \right\rangle -
  \frac{1}{2}  \left\langle G_{2 \mu \nu} G_2^{\mu \nu} \right\rangle
  \nonumber\\
  &  +& \mathi \overline{\chi_R} \gamma^{\mu} D_{\mu} \chi_L + \mathi
  \overline{\chi_R} \gamma^{\mu} D_{\mu} \chi_R \nonumber\\
&+& \text{four-fermion
  interactions} . 
\end{eqnarray}
The solution to the constraints
\begin{eqnarray}
  D_{\mu} X_1 & = & 0, \\
  D_{\mu} Y_1 & = & 0,
\end{eqnarray}
is as in section~\ref{sub:SU2-case}. Solving the constraints
\begin{eqnarray}
  D_{\mu} X_2 & = & 0, \\
  D_{\mu} \tilde{Y}_0 & = & 0, \\
  D_{\mu} \phi & = & 0,
\end{eqnarray}
where
\begin{eqnarray}
  \phi & \longmapsto & G_0 \mathe^{\mathi \frac{\beta^0}{2}} \phi,
\end{eqnarray}
as before, proceeds in analogy with sections \ref{sub:one-complex-spurion} and
\ref{sub:U1-fermions}. The lowest-order mass term for quarks is of order
$\mathcal{O} \left( p^1 \epsilon^4 \right)$
\begin{eqnarray}
  \mathcal{L}^{\left( 1, 4 \right)}_{\tmop{quarks}} & = & - m_{1 i j} 
  \overline{\chi_L}_i X_2^{\dag} \Sigma_{1, 2}^{\dag} Y_1^{\dag} X_1^{\dag}
  \Sigma_{0, 1}^{\dag}  \tilde{Y}_0^{\dag} \chi_{R j}\nonumber\\
&-& m_{1 i j}^{\ast} 
  \overline{\chi_R}_j  \tilde{Y}_0 \Sigma_{0, 1} X_1 Y_1 \Sigma_{1, 2} X_2
  \chi_{L i} \nonumber\\
  & - &  m_{2 i j}  \overline{\chi_L}_i X_2^{\dag} \Sigma_{1, 2}^{\dag}
  Y_1^{\dag} X_1^{\dag} \Sigma_{0, 1}^{\dag}  \tilde{Y}_0^{c \dag} \chi_{R j}
  \nonumber\\
&-& m_{2 i j}^{\ast}  \overline{\chi_R}_j  \tilde{Y}_0^c \Sigma_{0, 1} X_1 Y_1
  \Sigma_{1, 2} X_2 \chi_{L i} . 
\end{eqnarray}
The masses and splittings are thus of order~$\mathcal{O} \left( p^0 \epsilon^4
\right)$: in contrast with the result~$\mathcal{O} \left( p^0 \epsilon^2
\right)$ we have found in the minimal case, cf. section
\ref{sub:fermion-masses}. The power counting for spurions actually depends on
the total number of spurions involved along the chain, i.e. on the length~$K$
of the moose. This has to be taken into account when comparing the
power counting between~$\epsilon$ and~$p$ on physical grounds.

Performing the proper field redefinitions, we find that this model yields
$W^{\pm}$ and~$Z^0$ vector bosons as well as a massless photon. There are in
addition~$W'^{\pm}$ and~$Z'$ resonances, which also couple directly to
fermions. We will resort to the formalism of oblique parameters in order to
make contact with the literature, however, we will use the definitions of
these parameters in terms of observables~{\cite{Altarelli:1991zd,Sanchez-Colon:1998xg}} as opposed to two-point functions. The field
redefinitions are as follows {\emdash}once again, as they appear after solving
the constraints on the spurions{\emdash}
\begin{eqnarray}
  g_1 W_{1 \mu} & = & \mathi \Sigma_{0, 1} \nabla_{\mu} \Sigma_{0, 1}^{\dag},
  \\
  g_2 W_{2 \mu} & = & \mathi \Sigma_{0, 1} \Sigma_{1, 2} \nabla_{\mu} \left(
  \Sigma_{1, 2}^{\dag} \Sigma_{0, 1}^{\dag} \right) . 
\end{eqnarray}
We then use an orthogonal transformation for the charged fields
\begin{eqnarray}
  \left(\begin{array}{c}
    W^{\pm}_{\mu}\\
    W'^{\pm}_{\mu}
  \end{array}\right) & = & \left(\begin{array}{cc}
    \cos \gamma & - \sin \gamma\\
    \sin \gamma & \cos \gamma
  \end{array}\right)  \left(\begin{array}{c}
    W^{\pm}_{1 \mu}\\
    W^{\pm}_{2 \mu}
  \end{array}\right),
\end{eqnarray}
where~$\cos \gamma$ is found to be
\begin{eqnarray}
  \cos \gamma & = & \frac{1}{\sqrt{1 + \frac{\left( 2 g_1^2 - g_2^2 + \sqrt{4
  g_1^4 + g_2^4} \right)^2}{4 g_1 g_2}}} . 
\end{eqnarray}
The matrix~$N$ for the neutral fields involves rescalings, and is defined as
\begin{eqnarray}
  \left(\begin{array}{c}
    b^0_{\mu}\\
    W_{2 \mu}\\
    W_{1 \mu}
  \end{array}\right) & = & N \left(\begin{array}{c}
    A_{\mu}\\
    Z_{\mu}\\
    Z'_{\mu}
  \end{array}\right) . 
\end{eqnarray}
The entries of this matrix reappear constantly in calculations, but their full
expressions are rather lengthy, so we will not write out the~$N_{i j}$'s
explicitly in the equations, but rather perform the expansion in~$1 / g_1$.
One finds the following tree-level expressions for the masses of the~$W^{\pm}$
and~$Z^0$
\begin{eqnarray}
  M_W^2 & = & \frac{g_2^2}{8} f^2 - \frac{1}{32}  \frac{g_2^4}{g_1^2} f^2
  +\mathcal{O} \left( \frac{1}{g_1^4} \right), \\
  M_Z^2 & = & \frac{g_2^2 + g_0^2}{8} f^2 \nonumber\\
&-& \frac{1}{32}  \frac{\left( g_0^2 -
  g_2^2 \right)^2}{g_1^2} f^2 +\mathcal{O} \left( \frac{1}{g_1^4} \right),
\end{eqnarray}
and for the masses of the~$W'^{\pm}$ and~$Z'$
\begin{eqnarray}
  M_{W'}^2 & = & \frac{g_1^2}{2} f^2 + \frac{g_2^2}{8} f^2 +\mathcal{O} \left(
  \frac{1}{g_1^2} \right), \\
  M_{Z'}^2 & = & \frac{g_1^2}{2} f^2 + \frac{g_2^2 + g_0^2}{8} f^2
  +\mathcal{O} \left( \frac{1}{g_1^2} \right) . 
\end{eqnarray}
\subsubsection{Remarks} \label{sub:remarks-WZprimes}

The electric charge is given by
\begin{eqnarray}
  e & = & \frac{g_0 g_2}{\sqrt{g_0^2 + g_2^2}} - \frac{1}{2 g_1^2} 
  \frac{g_0^3 g_2^3}{\left( g_2^2 + g_0^2 \right)^{\frac{3}{2}}} +\mathcal{O}
  \left( \frac{1}{g_1^4} \right),
\end{eqnarray}
so that in this case, we are not in a position to make contact with the SM by
a simple replacement as in section~\ref{sub:EWSB-minimal-case}, but we need to
study observables and check whether they obey the same relations as in the SM,
or whether there are corrections relative to the SM at leading order. Details
about the determination of the oblique parameters are given in appendix
\ref{sub:STU} for illustration purposes. We only discuss here the~$S$
parameter at tree-level. We get
\begin{eqnarray}
  \alpha S & = & \frac{1}{g_1^2}  \frac{g_0^2 g_2^2}{g_2^2 + g_0^2}
  +\mathcal{O} \left( \frac{1}{g_1^4} \right) . 
\end{eqnarray}
The sign and magnitude of this contribution is as expected from the result
(\ref{eq:L10}), using the relation~{\cite{Espriu:1992vm}} which gives the
value of~$S$ in the electroweak model once the value of~$L_{10}$ for the
corresponding non-linear sigma model~(obtained from the moose after integration of
the resonances) is known
\begin{eqnarray}
  S & = & - 16 \mathpi L_{10} . 
\end{eqnarray}
Of course, the~$S$ parameter would have been further suppressed had we chosen
$f_1 \ll f_0$ as already mentioned in section~\ref{sub:other-aspects}. Indeed,
the suppression by powers of ratios of coupling-constants cannot be taken too
far as it implies that~$g_1$ is not small, and is thus threatening the consistency
of the loop expansion as well as casting a doubt on the meaning of the
power counting for~$g_1$. We emphasize that this is a weak point of the
formulation even though in practice, we shall not be dealing with the loop
expansion for this model.

We further point out that the decoupling of the~$W'$ and~$Z'$ for tree-level
relations is achieved in the limit~$g_1 \longrightarrow + \infty$, as can be
deduced from the formulas of appendix \ref{sub:STU}, in which case one obtains
a shorter moose {\emdash}that of section~\ref{sub:EWSB-minimal-case}. One may
wish to consider the general case with additional resonances, but we will not
pursue further in that direction as far as this paper is concerned.

As a conclusion to this section, we stress the fact that we have limited
ourselves to the tree-level in the effective theory formalism. This formalism
is to be carried out as a loop expansion involving renormalization of
counter-terms of higher and higher orders: there are thus more and more
constants to determine each time we ask for more precision. As long as there
is no way to estimate the values of these constants at a given scale
{\footnote{Some authors have recently revived the idea that there may be a
way to perform calculations for models of EWSB with strong dynamics, using
the correspondence between this case and the five-dimensional models with warped
geometry where the gauge symmetries get broken on branes~{\cite{Csaki:2003zu,Nomura:2003du}}.\label{foot:calculability}}}, other than experiment, it is
difficult to say whether or not there really is a conflict with experiment.
Here we have included the first excited vector states in our effective
lagrangian, which in practice means that we assume their effects to be larger
than those of loops. We believe that this is a possibility, but the reader
should be aware of the fact that limiting oneself to tree diagrams while at
the same time considering the limit~(\ref{eq:small-gs}) is not obviously safe.
Note that the considerations following~(\ref{eq:coeff-matrice-masse}) should
be borne in mind as well. Given these limitations, we do not push the
phenomenological discussion of this model further, but instead, turn to
another possibility: that of having PGBs as the first additional states in the
theory.

\subsection{Case with a triplet of PGBs} \label{sub:closed-moose}

\begin{figure*}\centering
\includegraphics{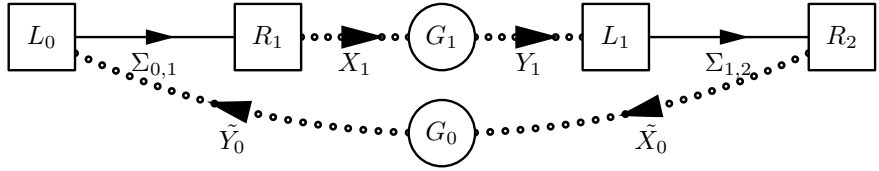}
\caption{The simplest moose model for EWSB with scalars in the spectrum.}
\label{fig:EWSB-3PGBs}
\end{figure*}

We now describe the case where the linear moose is extended by connecting the
two ends via spurions and a gauge group. Since we are interested in
applications to the electroweak sector, this additional gauge group should
only be the~$\mathrm{U} \left( 1 \right)$ subgroup of~$\tmop{SU} \left( 2
\right)$, which is selected by the spurions. We will consider the simplest
case, that with one internal site~($K = 1$) as shown in figure
\ref{fig:EWSB-3PGBs}. The effective theory for this model is constructed along
our standard line of reasoning, and we will find that the spectrum consists in
the~$W^{\pm}$,~$Z^0$, the photon, and a triplet of pseudo-scalars. Note
that, as was done in \ref{sub:WZprimes}, one may consider a longer chain with
additional gauge groups with larger gauge couplings, yielding~$W'$ and~$Z'$
resonances: the discussion of the~$S, T, U$ parameters would then be similar.

The natural symmetry of the model is in this case
\begin{eqnarray}
  S_{\tmop{natural}} & = & \tmop{SU} \left( 2 \right)^6 \times \mathrm{U}
  \left( 1 \right)_{B - L},
\end{eqnarray}
while the reduced symmetry, obtained after applying the constraints, is
\begin{eqnarray}
  S_{\tmop{reduced}} & = & \tmop{SU} \left( 2 \right) \times \mathrm{U} \left(
  1 \right)_Y . 
\end{eqnarray}
\subsubsection{Complex spurions in a closed moose}
\label{sub:two-complex-spurions}

We introduce the spurions~$\tilde{Y}_0, \tilde{X}_0, \phi$ with the following
transformation properties
\begin{eqnarray}
  \tilde{Y}_0 & \longmapsto & G_0  \tilde{Y}_0 L_0^{\dag}, \\
  \phi & \longmapsto & G_0 \mathe^{\mathi \frac{\beta^0}{2}} \phi, \\
  \tilde{X}_0 & \longmapsto & R_2  \tilde{X}_0 G_0^{\dag},
\end{eqnarray}
where~$\text{$\tilde{Y}_0, \tilde{X}_0$}$ are two-by-two complex matrices and
$\phi$ is a complex doublet. At this stage, the complex doublet~$\phi$ is not
necessary as we have not yet introduced fermions. Still, we want to show how
the constraints on these three spurions are solved at the same time: we have
seen in section~\ref{sub:U1-fermions} how the constraints
\begin{eqnarray}
  D_{\mu} \tilde{Y}_0 & = & 0,  \label{eq:constraint-Ytilde0}\\
  D_{\mu} \phi & = & 0,  \label{eq:constraint-phi}
\end{eqnarray}
could be solved simultaneously, resulting in
\begin{eqnarray}
  \left. G^{1, 2}_{0 \mu} \right|_{\tmop{const.}} & = & 0,
  \label{eq:first-rel-connection}\\
  \left. L_{0 \mu}^{1, 2} \right|_{\tmop{const.}} & = & 0, \\
  \left. L_{0 \mu}^3 \right|_{\tmop{const.}} & = & \left. \left. 
  \hspace{.2em} B_{\mu} \right|_{\tmop{const.}} \hspace{.2em} = \hspace{.2em} g_0 G_{0
  \mu}^3 \right|_{\tmop{const.}} \nonumber\\
&=& g_0
  b_{\mu}^0,  \label{eq:last-rel-connection}
\end{eqnarray}
after an appropriate choice of gauge. Solving for both spurion constraints
(\ref{eq:constraint-Ytilde0}) and~(\ref{eq:constraint-phi}) using the same
$G_0$ transformation was possible thanks to to the restrictions implied by the
constraints themselves: a~$\mathrm{U} \left( 1 \right)$ transformation was all
that was required to connect the two configurations, and we had enough freedom
to choose the~$f'$ gauge function for our purposes. We now want to solve in
addition the constraint on the~$\tilde{X}_0$ spurion.

Let us recall that both spurions~$\tilde{Y}_0, \tilde{X}_0$ are assumed to be
generic two-by-two matrices, except that their entries must be functions with
small modulus, to be counted as order~$\epsilon$. Using the same procedure as
in section~\ref{sub:U1-fermions} we shall see that one can solve the constraint
to be applied on~$\tilde{X}_0$
\begin{eqnarray}
  D_{\mu} \tilde{X}_0 & = & 0 .  \label{eq:constraint-Xtilde0}
\end{eqnarray}
in the same gauge. The result will turn out to be the expected one: the
spurion is constant and can be written
\begin{eqnarray}
  \left. \tilde{X}_0 \right|_{\tmop{const.}} & = & \mathe^{\mathi
  \varphi_X}  \left(\begin{array}{cc}
    \xi_{0 1} & 0\\
    0 & \xi_{0 2}
  \end{array}\right),
\end{eqnarray}
and the connections are identified, giving us the following
\begin{eqnarray}
  \left. R_{2 \mu}^{1, 2} \right|_{\tmop{const.}} & = & 0, \\
  \left. R_{2 \mu}^3 \right|_{\tmop{const.}} & = & g_0 G_{0 \mu}^3,
\end{eqnarray}
while we also recover~(\ref{eq:first-rel-connection}).

In order to show this, we assume that the constraint
(\ref{eq:constraint-Ytilde0}) and~(\ref{eq:constraint-phi}) on~$\tilde{Y}_0$
and~$\phi$ have been solved as in sections \ref{sub:one-complex-spurion} and
\ref{sub:U1-fermions} respectively. This involves the gauge function~$f'$ of
(\ref{eq:G0-f}) and~(\ref{eq:beta0-f}), which is now fixed with respect to
$f$. We use the following decomposition for~$\tilde{X}_0$
\begin{eqnarray}
  \tilde{X}_0 & = & \mathe^{\mathi \varphi_X} R_X^{\dag} D_X G_X,
  \label{eq:decomp-Xtilde0}
\end{eqnarray}
where~$\varphi_X$ is real,~$R_X, G_X$ are elements of~$\tmop{SU} \left( 2
\right)$ and~$D_X$ is diagonal and real. The same remark concerning the number
of parameters as made after~(\ref{eq:decomp-Ytilde0}) applies. We then perform
the following~$\tmop{SU} \left( 2 \right)_{G_0}$ and~$\tmop{SU} \left( 2
\right)_{R_2}$ transformations
\begin{eqnarray}
  G_0 & = & \mathe^{- \mathi f''  \frac{\tau^3}{2}} G_X,
  \label{eq:gauge-G0}\\
  R_2 & = & \mathe^{- \mathi f''  \frac{\tau^3}{2}} R_X . 
  \label{eq:gauge-RK+1}
\end{eqnarray}
We have again introduced a gauge function~$f''$, to be solved for later. In
this gauge, we have
\begin{eqnarray}
  \tilde{X}_0 & = & \mathe^{\mathi \varphi_X} D_X,
\end{eqnarray}
independently of our choice for~$f''$, due to the remark following
(\ref{eq:decomp-Ytilde0}). We can then derive, along the same lines as for
$\tilde{Y}_0$, that~$D_X$
\begin{eqnarray}
  D_X & = & \left(\begin{array}{cc}
    \xi_{0 1} & 0\\
    0 & \xi_{0 2}
  \end{array}\right),
\end{eqnarray}
is a constant matrix. As usual, we impose the following power counting for
the two constants appearing in this matrix
\begin{eqnarray}
  \xi_{0 1}, \xi_{0 2} & = & \mathcal{O} \left( \epsilon \right) . 
\end{eqnarray}
In the gauge specified by~(\ref{eq:gauge-G0}) and~(\ref{eq:gauge-RK+1}), we
find, writing the covariant constancy equation component by component
\begin{eqnarray}
  G_{0 \mu}''^{1, 2} & = & 0 . 
\end{eqnarray}
However~$G''_{0 \mu}$ is related via
\begin{eqnarray}
  G''_{0 \mu} & = & WG_{0 \mu} W^{\dag} + \frac{\mathi}{g_0} W \partial_{\mu}
  W^{\dag},
\end{eqnarray}
with the gauge transformation~$W \in \tmop{SU}\left( 2 \right)$, to the connection~$G_{0 \mu}$ we have found when solving the constraints for
$\tilde{Y}_0$ and~$\phi$. The expression for~$W$ indeed reads
\begin{eqnarray}
  W & = & \mathe^{- \mathi f''  \frac{\tau^3}{2}} G_X G_Y^{\dag}
  \mathe^{\mathi f \frac{\tau^3}{2}} . 
\end{eqnarray}
However, since both~$G_{0 \mu}$ and~$G''_{0 \mu}$ point in the third
direction, we can deduce that~$W$ only involves a~$\mathrm{U} \left( 1
\right)_{\tau^3}$ transformation. Therefore, we can choose the function~$f''$
so as to have
\begin{eqnarray}
  W & = & 1 . 
\end{eqnarray}
Thus, due to the constraints~(\ref{eq:constraint-Xtilde0}) and~(\ref{eq:constraint-Ytilde0}), we have been able to diagonalize all three
spurions in the same gauge, giving
\begin{eqnarray}
  G''_{0 \mu} & = & G_{0 \mu}' \hspace{.2em} = \hspace{.2em} G_{0 \mu} . 
\end{eqnarray}
We also find that~$\varphi_X$ is a constant,
and that
\begin{eqnarray}
  R_{2 \mu}^a & = & g_0 G^a_{0 \mu}, \hspace{.2em} \tmop{for} \hspace{.2em} a = 1, 2,
  3. 
\end{eqnarray}
In summary, we have found a gauge in which the spurions are rewritten as the
following constant matrices
\begin{eqnarray}
  \left. \tilde{X}_0 \right|_{\tmop{const.}} & = & \mathe^{\mathi
  \varphi_X}  \left(\begin{array}{cc}
    \xi_{0 1} & 0\\
    0 & \xi_{0 2}
  \end{array}\right), \\
  \left. \tilde{Y}_0 \right|_{\tmop{const.}} & = & \mathe^{\mathi
  \varphi_Y}  \left(\begin{array}{cc}
    \eta_{0 1} & 0\\
    0 & \eta_{0 2}
  \end{array}\right),\\
 \left. \phi \right|_{\tmop{const.}} & = & \left(\begin{array}{c}
    \zeta\\
    0
  \end{array}\right) . 
 \end{eqnarray}
due to the constraints. In addition, the following connections are identified
\begin{eqnarray}
  \left. R^{1, 2}_{2 \mu} \right|_{\tmop{const.}} & = & \left. \left.
  L_{0 \mu}^{1, 2} \right|_{\tmop{const.}} \hspace{.2em} = \hspace{.2em} g_0 G^{1, 2}_{0
  \mu} \right|_{\tmop{const.}} \nonumber\\
&=& 0, \\
  \left. R^3_{2 \mu} \right|_{\tmop{const.}} & = & \left. \left. \left.
  L_{0 \mu}^3 \right|_{\tmop{const.}} \hspace{.2em} = \hspace{.2em} B_{\mu}
  \right|_{\tmop{const.}} \hspace{.2em} = \hspace{.2em} g_0 G_{0 \mu}^3
  \right|_{\tmop{const.}}  \nonumber\\
  & = & g_0 b_{\mu}^0,  \label{eq:closed-U1-id}
\end{eqnarray}
This proves the result announced above. Again,
the only invariance left from the original~$\tmop{SU} \left( 2 \right)_{L_0}
\times \tmop{SU} \left( 2 \right)_{G_0} \times \mathrm{U} \left( 1 \right)_{B
- L}$ is the~$\mathrm{U} \left( 1 \right)_Y$ degree of freedom, under which
\begin{eqnarray}
  b_{\mu}^0 & \longmapsto & b_{\mu}^0 - \frac{1}{g_0} \partial_{\mu} f,
\end{eqnarray}
with the identification~(\ref{eq:closed-U1-id}).

As for the constraints on the real spurions~$X_1, Y_1$
\begin{eqnarray}
  D_{\mu} X_1 & = & 0, \\
  D_{\mu} Y_1 & = & 0,
\end{eqnarray}
they are solved independently, as in section~\ref{sub:SU2-case}, giving the
following results in the standard gauge
\begin{eqnarray}
  X_1 |_{\tmop{const.}} & = & \xi_1  \mathbbm{1}_{2 \times 2}, \\
  Y_1 |_{\tmop{const.}} & = & \eta_1  \mathbbm{1}_{2 \times 2},
\end{eqnarray}
where~$\xi_1$ and~$\eta_1$ are real constants, used as expansion parameters.
In this gauge, we have the following identification of connections
\begin{eqnarray}
  \left. R_{1 \mu} \right|_{\tmop{const.}} & = & \left. L_{1 \mu}
  \right|_{\tmop{const.}} \hspace{.2em} = \hspace{.2em} g_1 G_{1 \mu} . 
\end{eqnarray}

\subsubsection{Bosons}

Going back to the unconstrained lagrangian, we find that the~$\mathcal{O}
\left( p^2 \epsilon^0 \right)$ lagrangian for the bosonic sector of this model
is

\begin{eqnarray}
  \mathcal{L}^{\left( 2, 0 \right)}_{\tmop{bosons}} & = & \frac{f^2_0}{4} 
  \left\langle D_{\mu} \Sigma_{0, 1} D^{\mu} \Sigma_{0, 1}^{\dag}
  \right\rangle + \frac{f_1^2}{4}  \left\langle D_{\mu} \Sigma_{1, 2} D^{\mu}
  \Sigma_{1, 2}^{\dag} \right\rangle \nonumber\\
    & -& \frac{1}{2}  \left\langle G_{1 \mu \nu} G_1^{\mu \nu} \right\rangle
  - \frac{1}{2}  \left\langle G_{0 \mu \nu} G_0^{\mu \nu} \right\rangle . 
  \label{eq:L-closed-moose}
\end{eqnarray}

Solving the constraints on the spurions,~(\ref{eq:L-closed-moose}) becomes in
the standard gauge used in section~\ref{sub:two-complex-spurions}

\begin{eqnarray}
  \left. \mathcal{L}^{\left( 2, 0 \right)}_{\tmop{bosons}}
  \right|_{\tmop{const.}} & = & \frac{f^2_0}{4}  \left\langle
  \nabla_{\mu} \Sigma_{0, 1} \nabla^{\mu} \Sigma_{0, 1}^{\dag} \right\rangle \nonumber\\
&+&
  \frac{f_1^2}{4}  \left\langle \nabla_{\mu} \Sigma_{1, 2} \nabla^{\mu}
  \Sigma_{1, 2}^{\dag} \right\rangle \nonumber\\
  & - &  \frac{1}{2}  \left\langle G_{1 \mu \nu} G_1^{\mu \nu} \right\rangle
  - \frac{1}{4} b^0_{\mu \nu} b^{0 \mu \nu},
\end{eqnarray}

where
\begin{eqnarray}
  \nabla_{\mu} \Sigma_{0, 1} & = & \left. D_{\mu} \Sigma_{0, 1}
  \right|_{\tmop{const.}} \nonumber\\
  & = & \partial_{\mu} \Sigma_{0, 1} - \mathi g_0 b^0_{\mu}  \frac{\tau^3}{2}
  \Sigma_{0, 1} \nonumber\\
&+& \mathi g_1 \Sigma_{0, 1} G_{1 \mu}, \\
  \nabla_{\mu} \Sigma_{1, 2} & = & \left. D_{\mu} \Sigma_{1, 2}
  \right|_{\tmop{const.}} \nonumber\\
  & = & \partial_{\mu} \Sigma_{1, 21} - \mathi g_1 G_{1 \mu} \Sigma_{1, 2} \nonumber\\
&+&
  \mathi g_0 b^0_{\mu} \Sigma_{1, 2} \frac{\tau^3}{2} . 
\end{eqnarray}

We also find plaquette terms~{\footnote{Note that one could in principle use
the same spurion formalism to order the different possible plaquette terms in
little Higgs models~{\cite{Arkani-Hamed:2002qx}} according to an assumed
pattern for the identification of symmetries.}}, which are the terms of lowest
order in~$\epsilon$ among those which carry no power of~$p$. They are given by
terms of the form
\begin{eqnarray}
  &  & \left\langle \Sigma_{0, 1} X_1 Y_1 \Sigma_{1, 2}  \tilde{X}_0 
  \tilde{Y}_0 \right\rangle\nonumber\\
&+& \left\langle \tilde{Y}_0^{\dag}  \tilde{X}_0^{\dag} \Sigma_{1, 2}^{\dag}
  Y_1^{\dag} X_1^{\dag} \Sigma_{0, 1}^{\dag} \right\rangle,
  \label{eq:plaquette}
\end{eqnarray}
and similar ones when one replaces~$\tilde{X}_0$ and~$\tilde{Y}_0$
respectively by their conjugates~$\tilde{X}_0^c$ and~$\tilde{Y}_0^c$. There
are four such terms, which give a tree-level mass-squared to the PGBs. We find
therefore that the mass-squared of the PGBs is automatically counted as a
small parameter of order~$\mathcal{O} \left( p^0 \epsilon^4 \right)$.

We only briefly describe the field redefinitions as they are analogous to
those of section~\ref{sub:field-redef}: after injecting the solution to the
constraints on spurions, we define the Goldstone boson fields remaining in the
spectrum through
\begin{eqnarray}
  \left. U \right|_{\tmop{const.}} & = & \Sigma_{0, 1} \Sigma_{1, 2},
\end{eqnarray}
and the vector fields according to
\begin{eqnarray}
  \left. g_1 W_{1 \mu} \right|_{\tmop{const.}} & = &  \mathi
  \Sigma_{0, 1} \nabla_{\mu} \Sigma_{0, 1}^{\dag} \nonumber\\
&-& \left.\mathi
  \frac{\alpha_1}{g_1} U \nabla_{\mu} U^{\dag} \right|_{\tmop{const.}},
\end{eqnarray}
where
\begin{eqnarray}
  \alpha_1 & = & \frac{f_1^2}{f_0^2 + f_1^2} . 
\end{eqnarray}
We then perform the change of variables
\begin{eqnarray}
  \left\{ G_{1 \mu}^a, b^0_{\mu}, \Sigma_{0, 1}, \Sigma_{1, 2} \right\} &
  \longrightarrow & \left\{ W_{1 \mu}^a, b^0_{\mu}, U, \Sigma_{1, 2} \right\},
\end{eqnarray}
to find that~$\Sigma_{1, 2}$ does not appear anymore, due to the symmetries of
the lagrangian. There remains one triplet of PGBs {\emdash}collected in the
unitary matrix~$U$ {\emdash} with decay constant~$f_{\pi}$, which can be
inferred from section~\ref{sub:field-redef} or \ref{sub:K=1-moose} to be
\begin{eqnarray}
  f_{\pi}^2 & = & \frac{f_0^2 f_1^2}{f_0^2 + f_1^2} . 
\end{eqnarray}
The mixing of~$b^0_{\mu}$ and~$W_{1 \mu}^3$ is identical to the case of
section~\ref{sub:EWSB-minimal-case}, with the replacement~$f_0^2 \longmapsto
f_0^2 + f_1^2$.

\subsubsection{Radiative corrections to the masses of PGBs}
\label{sub:radcor-mPGBs}

The plaquette terms~(\ref{eq:plaquette}) do not contribute to the
mass-splitting within the PGB triplet. On the other hand, such a splitting
arises from electroweak radiative corrections. A study of such corrections was first
performed for the pions in the context of low-energy QCD in
{\cite{Das:1967it}}, and reformulated in the case of effective theories in
{\cite{Ecker:1989te,Moussallam:1997xx}}. A consequence of the WSRs in this
context is the softening of divergences in the one-loop corrections to the PGB
masses~{\footnote{In fact, in the SM, if weak corrections to the pion masses
are considered in addition to the electromagnetic ones, the divergence is not
quadratic, but only logarithmic even without the first Weinberg sum rule
{\cite{Knecht:1998sp}}.}}: this is then another interpretation of the fact
that the moose models yield softer corrections to PGB masses~{\footnote{See
also~{\cite{Lane:2002pe}} in this respect.}}, which has been the starting
point of the little Higgs discussion~{\cite{Arkani-Hamed:2001nc}}, later
generalized to a broader class of models that do not necessarily have a moose
representation~{\cite{Arkani-Hamed:2002qy}}~{\footnote{There are also
connections with Randall-Sundrum~{\cite{Randall:1999ee}} type models where
EWSB is dictated by boundary conditions~{\cite{Csaki:2003zu,Contino:2003ve,Nomura:2003du,Davoudiasl:2003me}}, as can be guessed from
{\cite{Barbieri:2003pr}}: see footnote \ref{foot:calculability}. In
particular, the softness of corrections to two-point functions may well be one
aspect of the general asymptotic softness property studied in
{\cite{SekharChivukula:2001hz,Csaki:2003dt}}.}}. For simplicity we will
consider the limit where the constant factors in front of the plaquette terms
(\ref{eq:plaquette}) appearing in the lagrangian are set to zero: the PGBs
then have zero tree-level masses.

One may calculate the contribution of electroweak loops to the masses of the
PGBs directly via Feynman diagrams. However, we wish to explicitly show the
implications of WSRs on this contribution and will therefore resort to the
formulas given in~{\cite{Peskin:1980gc}}, based on Dashen's results
{\cite{Dashen:1971et}}. This gives the correction to first order in~$g_0^2$ as
a convolution of the left-right two-point function studied in section
\ref{sec:generalized-WSRs}, and of the correlator~$\left\langle 0 \left|
Tb^0_{\mu} \left( x \right) b^0_{\nu} \left( 0 \right) \right| 0
\right\rangle$
\begin{eqnarray}
&&  \left. \left( m_{\pi^{\pm}}^2 - m_{\pi^0}^2 \right)_{} \right|_{\tmop{loop}}\nonumber\\
  & = & 4 \mathi \frac{g_0^2}{f_{\pi}^2}  \int \mathd^D x 
\left\langle 0
  \left| TJ_{L_0}^{3 \mu} \left( x \right) b^0_{\mu} \left( x \right) J^{3
  \nu}_{R_2} \left( 0 \right) b^0_{\nu} \left( 0 \right) \right| 0
  \right\rangle \bignone \nonumber\\
  \bignone & = & - \frac{g_0^2}{f_{\pi}^2}  \int \mathd^D x \left\langle 0
  \left| Tb^0_{\mu} \left( x \right) b^0_{\nu} \left( 0 \right) \right| 0
  \right\rangle \nonumber\\
 &\times & \int \frac{\mathd^D q}{\left( 2 \mathpi \right)^D} \mathe^{-
  \mathi q \cdot x}  \left( \eta^{\mu \nu} q^2 - q^{\mu} q^{\nu} \right)
 \Pi_{L R} \left( - q^2 \right) . 
\end{eqnarray}
However,~$b^0_{\mu}$ is not a physical field, and we have to rewrite it in
terms of~$A_{\mu}$ and~$Z_{\mu}$: this is the reason why the formula is only
the first term in an expansion. The effect of the full diagonalization is to
replace the mass of the third component of the triplet~$W_{1 \mu}^a$ by
$M_Z^2$. When this is done, we are in a position to express the electroweak
PGB mass shift as
\begin{eqnarray}
&&  \left. \left( m_{\pi^{\pm}}^2 - m_{\pi^0}^2 \right)_{} \right|_{\tmop{loop}}\nonumber\\
  & = & \mathi \frac{\left( D - 1 \right) e^2}{f_{\pi}^2}  \int \bignone
  \frac{\mathd^D q}{\left( 2 \mathpi \right)^D} \nonumber\\
&\times&
 \left( 1 + \left( \frac{s}{c}
  \right)^2  \frac{q^2}{q^2 - M_Z^2} \right)
  \Pi_{L R} \left( - q^2 \right),
\end{eqnarray}
with~$\Pi_{L R}$ given by~(\ref{eq:PiLR-1-WSR}). The constants~$e, c, s$ are
defined as in section~\ref{sub:bosons}. This relation shows that the mass
squared is only logarithmically divergent, and not quadratically divergent,
due to the first Weinberg sum rule. This is true independently of the
regularization method used. Note that, if we were to add at least another
internal site to the chain, and thus one multiplet of resonances, we would get
a convergent integral at this order. Finally, we find here
\begin{eqnarray}
&&  \left. \left( m_{\pi^{\pm}}^2 - m_{\pi^0}^2 \right) \right|_{\tmop{loop}} \nonumber\\
&  = & - g_0^2 M_W^2  \left( 6 \lambda + \frac{1}{16 \mathpi^2}  \right. 
  \left.\left( 3 \ln
  \left( \frac{M_Z^2}{\mu^2} \right) + 2 \right) \right) \nonumber\\
&+&\mathcal{O} \left( D
  - 4 \right),
\end{eqnarray}
where the remaining logarithmic divergence is contained in the constant
$\lambda$
\begin{eqnarray}
  \lambda & = & \frac{\mu^{D - 4}}{16 \mathpi^2} \left( \frac{1}{D - 4} -
  \frac{1}{2}  \left( \ln 4 \mathpi - \gamma + 1 \right) \right),
\end{eqnarray}
and has to be canceled by the pole term in the appropriate counter-terms
present at higher orders in the effective lagrangian. Such counter-terms will
necessarily involve the spurions, and in fact, we note that we can build terms
which encompass those of~{\cite{Urech:1995hd}} when the constraints on the
spurions are used: the following~$\mathcal{O} \left( p^0
\epsilon^8 \right)$ term will absorb the divergences if our expansion allows it to
appear at the same level as one-loop corrections
\begin{eqnarray}
  & \left\langle \tilde{Y}_0^{\dag}  \tilde{Y}_0 \Sigma_{0, 1} X_1 Y_1
  \Sigma_{1, 2}  \tilde{X}_0  \tilde{X}_0^{\dag} \Sigma_{1, 2}^{\dag}
  Y_1^{\dag} X_1^{\dag} \Sigma_{0, 1}^{\dag} \right\rangle . & 
  \label{eq:Urech-like}
\end{eqnarray}
There are altogether sixteen such possible terms when one replaces the tilded
spurions by their conjugates, though not all depend on the PGBs due to the
relations
\begin{eqnarray}
  \tilde{X}_0  \tilde{X}_0^{c \dag} & = & \det X_0  \mathbbm{1}_{2 \times 2},
  \\
  \tilde{Y}_0  \tilde{Y}_0^{c \dag} & = & \det Y_0  \mathbbm{1}_{2 \times 2} .
\end{eqnarray}
The fact that these counter-terms have to appear at one-loop to absorb the
divergences to the masses of the PGBs then allows us to match the counting for
the spurion expansion with that of the momentum expansion in this particular
case where PGBs remain in the spectrum. We find that the correspondence between
the two expansion parameters should then read, for our particular~$K = 1$ case
\begin{eqnarray}
  \epsilon & = & \mathcal{O} \left( p^{1 / 2} \right) .  \label{eq:epsilon-p}
\end{eqnarray}
Note that, from the plaquette terms~(\ref{eq:plaquette}), this automatically gives in addition
\begin{eqnarray}
  m_{\tmop{PGBs}}^2 & = & \mathcal{O} \left( p^2 \right),
\end{eqnarray}
and we already mention that, from the results in section
\ref{sub:fermions-closed}, it would also imply for the fermions
\begin{eqnarray}
  m_{\tmop{fermions}} & = & \mathcal{O} \left( p^1 \right) . 
\end{eqnarray}
Both results are quite satisfactory from the point of view of reproducing the
poles of the corresponding particles without resummation. Still, for the sake
of generality, we shall refrain from using assumption~(\ref{eq:epsilon-p}) in
the remainder of this paper.

\subsubsection{Fermions} \label{sub:fermions-closed}

\begin{figure*}\centering
\includegraphics{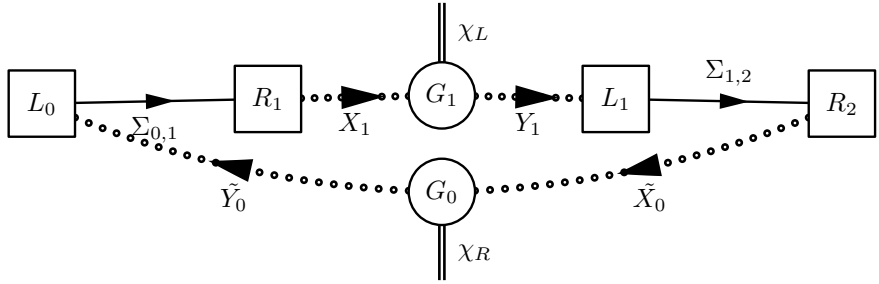}
\caption{Coupling fermions to the moose of section~\ref{sub:two-complex-spurions}.}
\label{fig:closed-moose+fermions}
\end{figure*}

We now briefly describe the introduction of fermions in this model: the
procedure is similar to that of section~\ref{sub:EWSB-minimal-case}. The model
is summarized in Fig.~\ref{fig:closed-moose+fermions}.

The elementary chiral fermions transform as
\begin{eqnarray}
  \chi_L & \longmapsto & G_1 \mathe^{- \mathi \frac{B - L}{2} \beta^0} \chi_L,
  \\
  \chi_R & \longmapsto & G_0 \mathe^{- \mathi \frac{B - L}{2} \beta^0} \chi_R,
\end{eqnarray}
and the identification of~$\mathrm{U} \left( 1 \right)$ connections is
performed as detailed in section~\ref{sub:two-complex-spurions} using the
$\phi$ spurion. We find the following~$\mathcal{O} \left( p^2 \epsilon^0
\right)$ lagrangian for fermions
\begin{eqnarray}
  \mathcal{L}^{( 2, 0 )}_{\tmop{fermions}} & = & \mathi \overline{\chi_L}
  \gamma^{\mu} D_{\mu} \chi_L + \mathi \overline{\chi_R} \gamma^{\mu} D_{\mu}
  \chi_R \nonumber\\
&+& \text{four-fermion interactions},
\end{eqnarray}
giving the same leading-order couplings as in the SM.

The following~$\mathcal{O} \left( p^1 \epsilon^2 \right)$ terms yield masses
for the quarks
\begin{eqnarray}
  \mathcal{L}_{\tmop{quarks}}^{( 1, 2 )} & = & - m_{1 i j} 
  \overline{\chi_L}_i X_1^{\dag} \Sigma_{0, 1}^{\dag}  \tilde{Y}_0 ^{\dag}
  \chi_{R j} \nonumber\\
&-& m_{1 i j}^{\ast}  \overline{\chi_R}_j  \tilde{Y}_0 \Sigma_{0,
  1} X_1 \chi_{L i} \nonumber\\
  & - &  m_{2 i j}  \overline{\chi_L}_i X_1^{\dag} \Sigma_{0, 1}^{\dag} 
  \tilde{Y}^{c \dag}_0 \chi_{R j} \nonumber\\
&-& m_{2 i j}^{\ast}  \overline{\chi_R}_j 
  \tilde{Y}_0^c \Sigma_{0, 1} X_1 \chi_{L i} \nonumber\\
  & - &  m_{3 i j}  \overline{\chi_L}_i Y_1 \Sigma_{1, 2}  \tilde{X}_0
  \chi_{R j} \nonumber\\
&-& m_{3 i j}^{\ast}  \overline{\chi_R}_j  \tilde{X}_0^{\dag}
  \Sigma^{\dag}_{1, 2} Y_1^{\dag} \chi_{L i} \nonumber\\
  & - &  m_{3 i j}  \overline{\chi_L}_i Y_1 \Sigma_{1, 2}  \tilde{X}_0^c
  \chi_{R j} \nonumber\\
&-& m_{4 i j}^{\ast}  \overline{\chi_R}_j  \tilde{X}_0^{c \dag}
  \Sigma^{\dag}_{1, 2} Y_1^{\dag} \chi_{L i},  \label{eq:Yukawa-closed}
\end{eqnarray}
and there are again enough free parameters available per fermion doublet for
the masses of the two states to be independent at tree-level. The additional
parameters compared to the previous cases are not relevant for the masses and
mixing, only for the Yukawa terms~(see below). The usual term yielding
Majorana masses for right-handed neutrinos are also present at this level. The
power countings are as in section~\ref{sub:fermion-masses}.

Note in addition that, whatever the counting for~$\epsilon$ is, we have the
result that the masses of the fermions are counted at the same level as that
of the PGBs
\begin{eqnarray}
  m^2_{\tmop{fermions}}, m^2_{\tmop{PGBs}} & = & \mathcal{O} \left( \epsilon^4
  \right) .
\end{eqnarray}
On the other hand, the scalars remaining in the spectrum in this model bear
little resemblance to the physical Higgs boson of the SM. In order to study
this, one has to define the unitary gauge fermion fields through
\begin{eqnarray}
  \psi_L & = & \Sigma_{0, 1} U_1 \chi_L, \\
  \psi_R & = & \chi_R,
\end{eqnarray}
yielding the same couplings of fermions to vector bosons as in section
\ref{sub:EWSB-minimal-case}. These rewritings, together with those involving the vector bosons, yield a
term
\begin{eqnarray}
  &  & \mathi \frac{f_{\pi}^2}{f_0^2}  \overline{\psi_L} \gamma^{\mu}  \left(
  U \nabla_{\mu} U^{\dag} \right) \psi_L .  \label{eq:der-coupling}
\end{eqnarray}
which involves couplings of the fermions with the PGBs alone and with the
PGBs and the neutral vector fields. The term linear in the PGB fields in~(\ref{eq:der-coupling}) can be removed by
a subsequent redefinition
\begin{eqnarray}
  \Psi_L & = & \mathe^{- \mathi \frac{f_{\pi}}{f_0^2} \pi^a \tau^a} \psi_L, \\
  \Psi_R & = & \psi_R,
\end{eqnarray}
where
\begin{eqnarray}
  U & = & \mathe^{\mathi \frac{\pi^a \tau^a}{f_{\pi}}} . 
\end{eqnarray}
All the other terms involving interactions of the left-handed fermions with
the PGBs are then modified, in particular the Yukawa terms. The fact that we
have a triplet of PGBs and not a single boson {\emdash}such as the physical
Higgs of the SM{\emdash} and the remark following~(\ref{eq:Yukawa-closed}) on
the number of terms imply that there is no simple relation between the masses
and Yukawa couplings of fermions to the PGBs. Still, one would for instance
expect general order-of-magnitude relations to hold: the Yukawa couplings of
the third generation will be larger, except if particular cancellations occur.

We conclude this section~by recalling the salient points encountered in the
discussion of this model: we have presented the first approximation in the
low-energy expansion for a model of EWSB where the only light particles not
yet discovered at accelerators are a triplet of PGBs. We have discussed the
way this model is derived from the moose idea, in the framework of effective
theories, with emphasis on the spurion formalism. We have also described the
consequences of the Weinberg sum rules derived in section~\ref{sec:generalized-WSRs} on the calculation of radiative corrections to the
masses of the PGBs. We have pointed out that the terms we can build with our
spurions do in particular allow us to renormalize the divergence in these
masses: there is a connection with the study of Urech~{\cite{Urech:1995hd}},
provided we identified the~$\epsilon$ power counting with the proper factors
in momentum expansion~(in fact with powers of gauge coupling constants). It is
reassuring to find that this power counting in turn yields quite sensible
results for the power counting of the masses of scalars and fermions. The
three scalars remaining in the spectrum have Yukawa interactions with the
fermions; however the connection between the coupling constants and the masses
is not so direct as would be in the SM for instance.

\section{Conclusion} \label{sec:concl}

In this paper, we have proposed to restore naturalness in Higgs-less EWSB
effective theories within a systematic spurion formulation. The spurion
formalism starts with a symmetry group~$S_{\tmop{natural}}$, larger than the
symmetry group~$S_{\tmop{reduced}}$ acting on the low-energy degrees of
freedom. Both the restriction of the space of gauge connections and the
coupling of Goldstone bosons to gauge fields follow from the condition of
covariant constancy imposed on the spurions. This constraint effectively
reduces the symmetry to~$S_{\tmop{reduced}}$.

The spurions in fact play a double role. Their first use is to select the
vacuum alignment of gauge fields~$S_{\tmop{reduced}} \subset
S_{\tmop{natural}}$. This alignment is maintained in the limit of vanishing
spurions. This effect of spurions is therefore already visible in the
$\mathcal{O} \left( p^2 \epsilon^0 \right)$ part of the lagrangian although
the latter does not explicitly involve any spurions. The second role of the
spurions is to provide expansion parameters: when the constraints are solved
in the standard gauge, the spurions reduce to constants, which we assume to be
small.

To begin with a simpler example before focusing on EWSB, we have first used
our spurion formalism in connection with a low-energy description of~$K$
vector resonances coupled to the Goldstone bosons of a global chiral symmetry.
The couplings at lowest order in the spurion expansion are identical to those
obtained in the dimensional deconstruction approach. We therefore have an
independent and more general bottom-up approach to justify this choice of
leading-order interactions. We have shown that a set of~$K$ generalized
Weinberg sum-rules follow at tree-level. Corrections will only occur at higher
orders in the spurion expansion.

Turning to EWSB, we have identified the set of spurions necessary to reduce
the natural symmetry to the one that is gauged:~$\tmop{SU} \left( 2 \right)
\times \mathrm{U} \left( 1 \right)_Y$. Known difficulties usually associated
with Higgs-less theories at leading order are relegated to higher orders in
the spurion expansion. This concerns~$\mathcal{O} \left( p^2 \right)$
contributions to the~$S$ parameter and non-standard couplings of the fermions
to gauge fields~(including non-universal couplings of the left-handed fermions
and couplings of right-handed fermions to the~$W^{\pm}$). Another well-known
difficulty of Higgs-less theories is to account for mass-splittings within
fermion doublets at the same chiral order as the masses themselves. This finds
a solution within the spurion formalism: as shown in section
\ref{sub:EWSB-minimal-case}, the full CKM structure can be recovered. Also,
due to the presence of the spurion~$\phi$, whose leading couplings only appear
in the neutrino sector, lepton number violation is introduced. One should
stress the unifying aspect of the spurion approach, which offers a
simultaneous solution to seemingly unrelated problems. Indeed, the scope of
the formalism was further illustrated in section~\ref{sec:EWSB-extended-cases}
with two extensions of the minimal effective theory of EWSB based on larger
natural symmetries and therefore on a larger set of light states protected by
them. These protected states involve either a tower of excited vector bosons
or PGBs which remain in the spectrum. Such scalars are kept naturally light
but share little resemblance with the SM Higgs boson. Examination of the
spurion contributions to the tree-level as well as to the radiative masses of
the PGBs suggests that the parameters~$\xi$ and~$\eta_0$ descending from the
spurions might be of chiral order~$\epsilon =\mathcal{O} \left( p^{1 / 2}
\right)$. This in turn would imply that the leading-order contribution to the
(Dirac) fermion masses arising from spurions be counted as
$m_{\tmop{fermions}} =\mathcal{O} \left( p^1 \right)$, as one would expect in
a low-energy expansion.

On the other hand, higher orders in the spurion expansion certainly require a
more complete analysis. First, the assumption of a common power counting,
which we have been using for simplicity, is by no means granted. For instance,
in the generic minimal EWSB effective theory of section
\ref{sub:EWSB-minimal-case}, the spurion~$\phi$ which allows for lepton number
violation need not carry the same power counting as the real spurion~$X_1$
responsible for the identification of two~$\tmop{SU} \left( 2 \right)$ groups
or as the complex spurion~$\tilde{Y}_0$ responsible for the selection of the
$\mathrm{U} \left( 1 \right)_{\tau^3}$ group and for the introduction of
isospin breaking. Furthermore, one needs to understand better how the spurion
expansion fits in the whole low-energy expansion including loops. This
requires a study of the way the powers of coupling constants and the constants
descending from spurions feed back into each other in the renormalization
procedure. A complete loop-level investigation going beyond the partial
results of section~\ref{sub:radcor-mPGBs} would be in order.

Insight into the magnitude of the spurions relative to each other and relative
to loops could be gained from the phenomenology of flavor violation, in
particular in the lepton sector~(using information from rare~$K$ decays and
from neutrino experiments for example). There obviously remains a lot of work
to be done for the completion of this program. Needless to say, our formalism
may also well be relevant outside the scope of Higgs-less EWSB.

\begin{acknowledgement}
{\em Acknowledgements.}
We would like to thank Andreas Nyffeler for sharing his opinions and knowledge with us, and Bachir Moussallam for his interest and continuous
encouragements. We also benefited from discussing some of the ideas exposed in
this paper with Marc Knecht, Hagop Sazdjian and Mike Pennington.

This work was supported in part by the European Community EURIDICE network
under contract HPRN-CT-2002-00311.\\
\end{acknowledgement}

\appendix \section{Resonance corrections to electroweak relations}
\label{sub:STU}

In this appendix, we give details about calculations for the oblique
parameters in the model of section~\ref{sub:WZprimes}:  evaluating the
Fermi constant from muon decay,  a contribution from~$W'$ exchange has to be taken into account, to find
\begin{eqnarray}
  G_{\mu} & = & \frac{g_2^2}{4 \sqrt{2}}  \left( \frac{\cos^2 \gamma}{M_W^2} +
  \frac{\sin^2 \gamma}{M_{W'}^2} \right) \nonumber\\
  & = & \frac{\sqrt{2}}{f_1^2} . 
\end{eqnarray}
The effective angle measured in low-energy~$\nu N$ scattering is then
\begin{eqnarray}
  s_{\ast}^2 \left( 0 \right) & = & - \frac{g_0}{g_2}  \left( \frac{\frac{N_{1
  2} N_{2 2}}{M_Z^2} + \frac{N_{1 3} N_{2 3}}{M_{Z'}^2}}{\frac{N^2_{2
  2}}{M_Z^2} + \frac{N^2_{2 3}}{M_{Z'}^2}} \right) \nonumber\\
  & = & \frac{g_0^2}{g_2^2 + g_0^2} +\frac{1}{2 g_1^2} \frac{g_0^2 g_2^2 
  \left( g_2^2 - g_0^2 \right)}{\left( g_2^2 + g_0^2 \right)^2} \nonumber\\
&+& \mathcal{O}
  \left( \frac{1}{g_1^4} \right) . 
\end{eqnarray}
Turning to the comparison between~$G_{\mu}$ and the four-fermion terms
generated by~$Z$ and~$Z'$ exchange, one finds at tree-level
\begin{eqnarray}
  \rho_{\ast} \left( 0 \right) & = & \frac{g_2^2}{4 \sqrt{2} G_{\mu}}  \left(
  \frac{N_{2 2}^2}{M_Z^2} + \frac{N_{2 3}^2}{M_{Z'}^2} \right) \nonumber\\
  & = & 1 . 
\end{eqnarray}
This can be understood as the consequence of the custodial symmetry, as noted
in~{\cite{Casalbuoni:1996qt,Chivukula:2003wj}}, in relation with
{\cite{Georgi:1978wk}}. However this symmetry, embodied here by~$\tmop{SU}\left(2\right)_{L_0}$, does
not enforce~$T = 0$, nor~$\rho = 1$, which are different in such cases
{\cite{Chang:2003un}}. Indeed, to define~$\rho$, we first have to define the
angle appearing in the~$Z^0$ couplings
\begin{eqnarray}
  s_f^2 & = & \frac{1}{4}  \left( 1 - \frac{g_V}{g_A} \right) \nonumber\\
  & = & - \frac{g_0 N_{1 2}}{g_2 N_{2 2}} \nonumber\\
  & = & \frac{g_0^2}{g_2^2 + g_0^2} +\frac{1}{2 g_1^2} \frac{g_0^2 g_2^2 
  \left( g_2^2 - g_0^2 \right)}{\left( g_2^2 + g_0^2 \right)^2} \nonumber\\
&+&\mathcal{O}
  \left( \frac{1}{g_1^4} \right),
\end{eqnarray}
where~$g_V$ and~$g_A$ are the vector and axial couplings of charged leptons to
the~$Z^0$. We then find
\begin{eqnarray}
  \rho & = & \frac{1}{1 - s_f^2}  \frac{M_W^2}{M_Z^2} \nonumber\\
  & = & 1 - \frac{1}{4}  \frac{g_0^2}{g_1^2} +\mathcal{O} \left(
  \frac{1}{g_1^4} \right),
\end{eqnarray}
and we note that~$s_{\ast}^2 \left( 0 \right)$ and~$s_f^2$ are different
\begin{eqnarray}
  s_{\ast}^2 \left( 0 \right) - s_f^2 & = & \frac{1}{8}  \frac{g_0^2
  g_2^2}{g_1^4}  \frac{g_0^2 - g_2^2}{g_0^2 + g_2^2} +\mathcal{O} \left(
  \frac{1}{g_1^6} \right) . 
\end{eqnarray}
Turning now to the~$S, T, U$ parameters, we get, from the definition
\begin{eqnarray}
  \Gamma \left( Z^0 \longrightarrow l^+ l^- \right) & = & \frac{G_{\mu}
  M_Z^3}{24 \sqrt{2} \mathpi}  \left( 1 - 4 s_f^2 + 8 s_f^4 \right)  \nonumber\\
&\times& \left( 1
  + \alpha T \right),
\end{eqnarray}
the following negative contribution to~$T$
\begin{eqnarray}
  \alpha T & = & \frac{g_2^2 N_{2 2}^2}{4 \sqrt{2} G_{\mu} M_Z^2} - 1
  \nonumber\\
  & = & - \frac{1}{16 g_1^4}  \left( g_0^2 - g_2^2 \right)^2 +\mathcal{O}
  \left( \frac{1}{g_1^6} \right) . 
\end{eqnarray}
Defining the~$S$ and~$U$ parameters requires the introduction of~$s_0^2$
through
\begin{eqnarray}
  s_0^2  \left( 1 - s_0^2 \right) & = & \frac{\mathpi \alpha}{\sqrt{2} G_{\mu}
  M_Z^2},
\end{eqnarray}
in order to extract~$S$ from the definition
\begin{eqnarray}
  s_f^2 & = & s_0^2 - \frac{s_0^2  \left( 1 - s_0 \right)^2}{1 - 2 s_0^2}
  \alpha T + \frac{1}{4 \left( 1 - 2 s_0^2 \right)} \alpha S . 
\end{eqnarray}
The value of the~$S$ parameter at tree-level has already been given in section
\ref{sub:remarks-WZprimes}.

Using the following definition for~$U$
\begin{eqnarray}
  \frac{1}{1 - s_0^2}  \frac{M_W^2}{M_Z^2} & = & 1 + \frac{1 - s_0^2}{1 - 2
  s_0^2} \alpha T - \frac{1}{2 \left( 1 - 2 s_0^2 \right)} \alpha S \nonumber\\
&+& 
  \frac{1}{4 s_0^2} \alpha U,
\end{eqnarray}
yields a positive value for~$U$
\begin{eqnarray}
  \alpha U & = & \frac{1}{4 g_1^4} g_0^2 g_2^2 +\mathcal{O} \left(
  \frac{1}{g_1^6} \right) . 
\end{eqnarray}
From all these expressions, one sees how the decoupling of the~$W'$ and~$Z'$
is achieved in the limit~$g_1\longrightarrow+\infty$, in which one obtains
a shorter moose {\emdash}that of section~\ref{sub:EWSB-minimal-case}.

\bibliographystyle{epjc-bib}
\bibliography{../../biblio/biblio.bib}

\end{document}